# ChatGPT Chemistry Assistant for Text Mining and Prediction of MOF Synthesis

Zhiling Zheng,[†,‡,§] Oufan Zhang,[†] Christian Borgs,[§,◊] Jennifer T. Chayes, [§,◊,††,‡‡,§§] Omar M. Yaghi[†,‡,§,‖,]*

[†] Department of Chemistry, University of California, Berkeley, California 94720, United States

[‡] Kavli Energy Nanoscience Institute, University of California, Berkeley, California 94720, United States

[§] Bakar Institute of Digital Materials for the Planet, College of Computing, Data Science, and Society, University of California, Berkeley, California 94720, United States

[◊] Department of Electrical Engineering and Computer Sciences, University of California, Berkeley, California 94720, United States

[††] Department of Mathematics, University of California, Berkeley, California 94720, United States

[‡‡] Department of Statistics, University of California, Berkeley, California 94720, United States

[§§] School of Information, University of California, Berkeley, California 94720, United States

[‖] KACST–UC Berkeley Center of Excellence for Nanomaterials for Clean Energy Applications, King Abdulaziz City for Science and Technology, Riyadh 11442, Saudi Arabia

**ABSTRACT:** We use prompt engineering to guide ChatGPT in the automation of text mining of metal-organic frameworks (MOFs) synthesis conditions from diverse formats and styles of the scientific literature. This effectively mitigates ChatGPT's tendency to hallucinate information—an issue that previously made the use of Large Language Models (LLMs) in scientific fields challenging. Our approach involves the development of a workflow implementing three different processes for text mining, programmed by ChatGPT itself. All of them enable parsing, searching, filtering, classification, summarization, and data unification with different tradeoffs between labor, speed, and accuracy. We deploy this system to extract 26,257 distinct synthesis parameters pertaining to approximately 800 MOFs sourced from peer-reviewed research articles. This process incorporates our ChemPrompt Engineering strategy to instruct ChatGPT in text mining, resulting in impressive precision, recall, and F1 scores of 90-99%. Furthermore, with the dataset built by text mining, we constructed a machine-learning model with over 87% accuracy in predicting MOF experimental crystallization outcomes and preliminarily identifying important factors in MOF crystallization. We also developed a reliable data-grounded MOF chatbot to answer questions on chemical reactions and synthesis procedures. Given that the process of using ChatGPT reliably mines and tabulates diverse MOF synthesis information in a unified format, while using only narrative language requiring no coding expertise, we anticipate that our ChatGPT Chemistry Assistant will be very useful across various other chemistry subdisciplines.

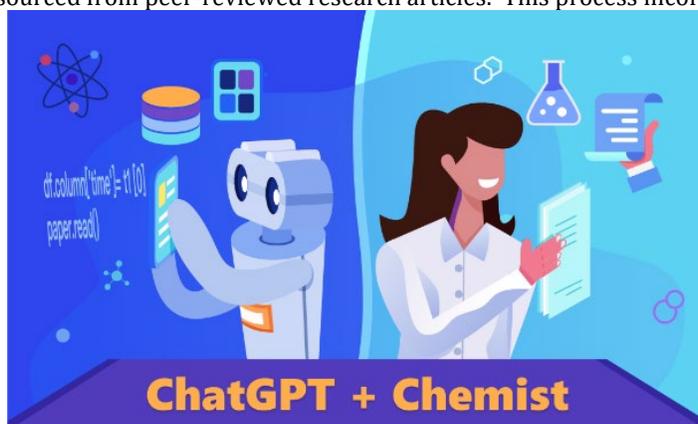

# INTRODUCTION

The dream of chemists is to create matter in the hope of advancing human knowledge for the betterment of society.[1, 2] As we stand on the precipice of the age of Artificial General Intelligence (AGI), the potential for synergy between AI and chemistry is vast and promising.[3, 4] The idea of creating AI-powered chemistry assistants offers unprecedented opportunities to revolutionize the landscape of chemistry research by applying knowledge across various disciplines, efficiently processing labor-intensive and time-consuming tasks, such as literature searches, compound screening and data analysis. AI-powered chemistry may ultimately transcend the limits of human cognition.[5-8]

Identifying chemical information for compounds, including ideal synthesis conditions and physical and chemical properties, has been a critical endeavor in chemistry research. The comprehensive summary of chemical information from literature reports, such as publications and patents, and their subsequent storage in an organized database format is the next logical and necessary step toward discovery of materials.[9] The challenge lies in efficiently mining the vast amount of available literature to obtain valuable information and insights. Traditionally, specialized natural language processing (NLP) models have been employed to address this issue.[10-14] However, these approaches can be labor-intensive and necessitate expertise in coding, computer science, and data science. Furthermore, they are less generalizable, requiring rewriting the program when the target changes. The advent of large language models (LLMs), such as GPT-3, GPT-3.5 and GPT-4, has the potential to fundamentally transform this process and revolutionize the routine of chemistry research in the next decade.[9, 15-18]

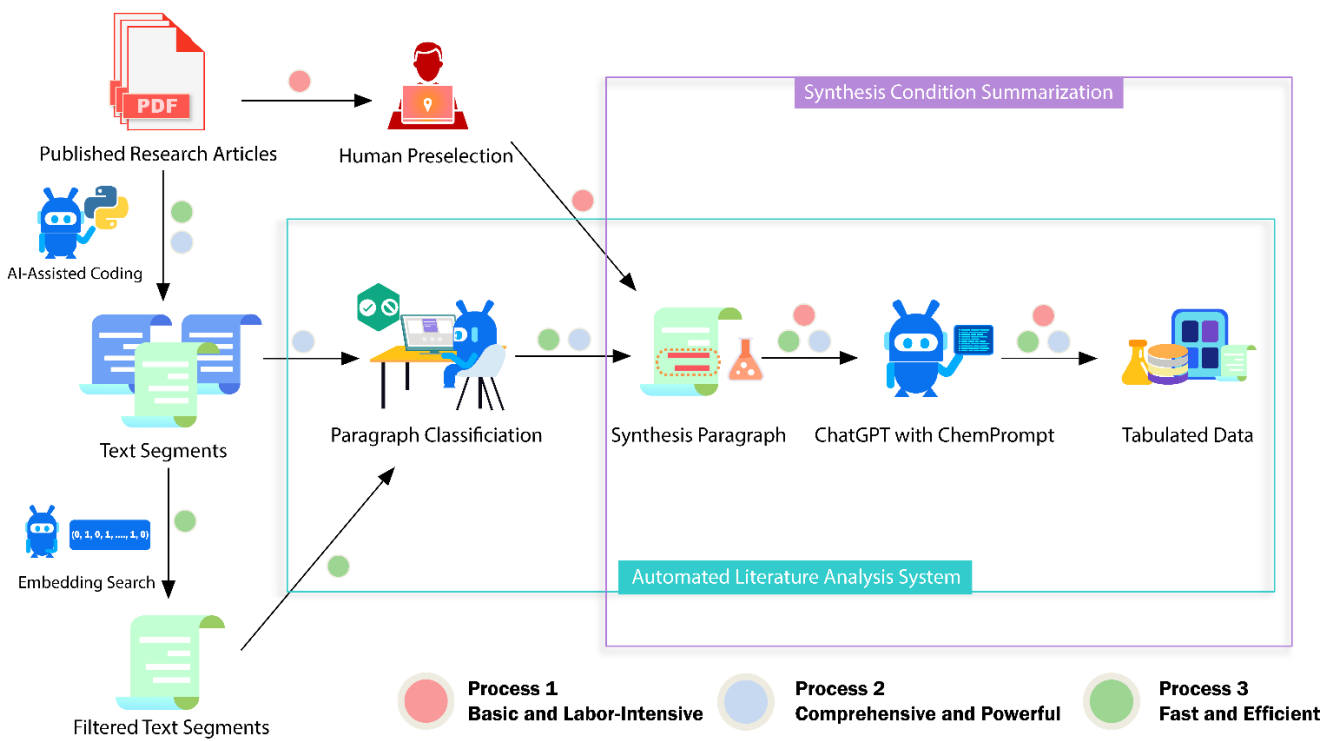

**Figure 1.** Schematics of ChatGPT Chemistry Assistant workflow having three different processes employing ChatGPT and ChemPrompt for efficient text mining and summarization of MOF synthesis conditions from a diverse set of published research articles. Each process is distinctively labeled with red, blue, and green dots respectively. To illustrate, Process 1 initiates with "Published Research Articles", proceeds to "Human Preselection", moves onto the "Synthesis Paragraph", integrates "ChatGPT with ChemPrompt", and culminates in "Tabulated Data". Steps shared among multiple processes are indicated with corresponding color-coded dots. The two-snakes logo of Python is included to indicate the use of the Python programming language, with the logo's credit attributed to the Python Software Foundation (PSF).

Herein, we demonstrate that LLMs, including ChatGPT based on the GPT-3.5 and GPT-4 model, can act as chemistry assistants to collaborate with human researchers, facilitating text mining and data analysis to accelerate the research process. To harness the power of what we termed as the ChatGPT Chemistry Assistant (CCA), we provide a comprehensive guide on ChatGPT prompt engineering for chemistry-related tasks, making it accessible to researchers regardless of their familiarity with machine learning, thus bridging the gap between chemists and computer scientists. In this report, we present (1) A novel approach to using ChatGPT for text mining the synthesis conditions of metal-organic frameworks (MOFs), which can be easily generalizable to other contexts requiring minimal coding knowledge and operating primarily on verbal instructions. (2) Assessment of ChatGPT's intelligence in literature text mining through accuracy evaluation and its ability for data refinement. (3) Utilization of the chemical synthesis reaction dataset obtained from text mining to train a model capable of predicting reaction results as crystalline powder or single crystals. Furthermore, we demonstrate that the CCA chatbot can be tuned to specialize in answering questions related to MOF synthesis based on literature conditions, with minimal hallucinations. This study underscores the transformative potential of ChatGPT and other LLMs in the realm of chemistry research, offering new avenues for collaboration and accelerating scientific discovery.

## MATERIALS AND METHODS

**Design Considerations for ChatGPT-Based Text Mining.** In curating research papers for ChatGPT to read and extract information, it is imperative to account for the diversity in MOF synthesis conditions, such as variations in metal sources, linkers, solvents, and equipment, as well as the different writing styles employed. Notably, the absence of a standardized format for reporting MOF synthesis conditions leads to variable reporting templates by research groups and journals. Indeed, by incorporating a broad spectrum of narrative styles, we can examine ChatGPT's robustness in processing information from heterogeneous sources. On the other hand, it is essential to recognize that the challenge of establishing unambiguous criteria to identify MOF compounds in the literature may lead to the inadvertent inclusion of some non-MOF compounds reported in earlier publications that are non-porous inorganic complexes and amorphous coordination polymers (included in some MOF datasets). As such, maintaining a balance between quality and quantity is vital, and prioritizing the selection of high-quality and well-cited papers, rather than incorporating all associated papers indiscriminately can ensure that the text mining of MOF synthesis conditions yields reliable and accurate data.

Moreover, papers discussing post-synthetic modifications, catalytic reactions of MOFs, and MOF composites are not directly pertinent to our objective of identifying MOF synthesis conditions. Hence, such papers have been excluded. Another consideration is that MOFs can be synthesized as both microcrystalline powders and single crystals, both of which should be regarded as valid candidates for our dataset. Utilizing the above-mentioned selection criteria, we narrowed our selection to 228 papers from an extensive pool of MOF papers, retrieved from Web of Science, Cambridge Structure Database MOF subset,[19] and the CoreMOF database.[20, 21] This sample represents a diverse range of MOF synthesis conditions and narrative styles.

To enable ChatGPT to process each paper, we devised three different approaches analogous to human paper reading: (1) locating potential sections containing synthesis conditions within the document, (2) confirming the presence of synthesis conditions in the identified sections, and (3) extracting synthesis parameters one by one. For our ChatGPT Chemistry Assistant, these steps are accomplished through filtering, classification, and summarization (Figure 1).

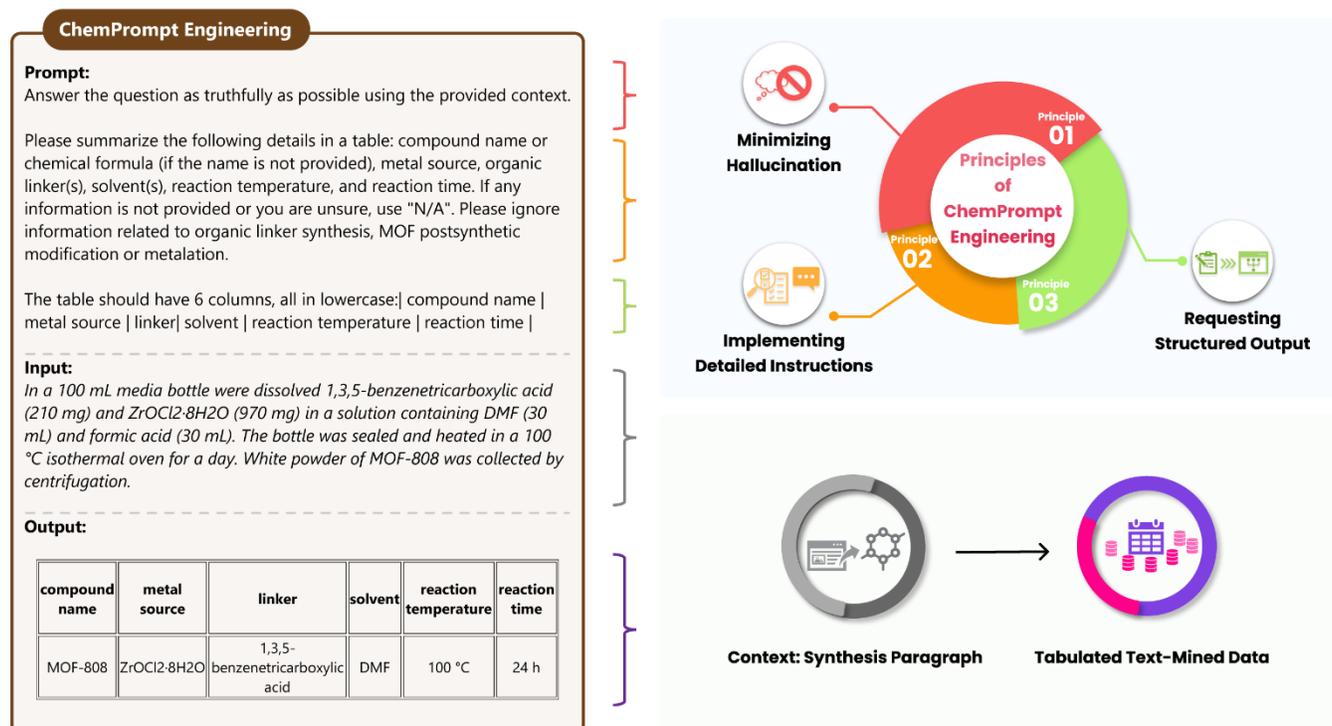

**Figure 2.** Illustration of a carefully designed ChemPrompt (shown on the left), encapsulating all three fundamental principles of ChemPrompt Engineering (shown on the right). The prompt guides ChatGPT to systematically extract and summarize synthesis conditions from a specified section in a research article, organizing the data into a well-structured table.

In Process 1, we developed prompts to guide ChatGPT in summarizing text from designated experimental sections contained in those papers. To replace the need for human intervention to obtain synthesis sections, in Process 2, we designed a method for ChatGPT to categorize text inputs as either "experimental section" or "non-experimental section", enabling it to generate experimental sections for summarization. In Process 3, we further devised a technique to swiftly eliminate irrelevant paper sections, such as references, titles, and acknowledgments, which are unlikely to encompass comprehensive synthesis conditions. This accelerates processing speed for the later classification task. As such, in Process 1, ChatGPT is solely responsible for summarizing and tabulating synthesis conditions and requires one or more paragraphs of experimental text as input, while Process 2 and 3 can be considered as an "automated paper reading system". While Process 2 entails a thorough examination of the entire paper to scrutinize each section, the more efficient Process 3 rapidly scans the entire paper, removing the least relevant portions, thereby reducing the number of paragraphs that ChatGPT must meticulously analyze.

**Prompt Engineering.** In the realm of chemistry-related tasks, ChatGPT's performance can be significantly enhanced by employing prompt engineering (PE)—a meticulous approach to designing prompts that steer ChatGPT towards generating precise and pertinent information. We propose three fundamental principles in prompt engineering for chemistry-focused applications, denoted as ChemPrompt Engineering:

*(1) Minimizing Hallucination*, which entails the formulation of prompts to avoid eliciting fabricated or misleading content from ChatGPT. This is particularly important in the field of chemistry, where the accuracy of information can have significant implications on research outcomes and safety. For instance, when asked to provide synthesis conditions for MOFs without any additional prompt or context, ChatGPT may recognize that MOF-99999 does not exist but will generate fabricated conditions for existing compounds with names like MOF-41, MOF-419, and MOF-519. We should note that with additional prompts followed after the question, it is possible to minimize hallucination and enforce ChatGPT to answer the questions based on its knowledge (Table 1 and Table 2). Furthermore, we demonstrate that with well-designed prompts and context, hallucination occurrences can be minimized (Supporting Information, Section S2.1). We note that this should be the first and foremost principle to follow when designing prompts for ChatGPT to perform in handling text and questions relevant to chemical information.

(2) *Implementing Detailed Instructions*, whereby explicit directions are provided in the prompt to assist ChatGPT in understanding the context and desired response format. By incorporating detailed guidance and context into the prompts, we can facilitate a more focused and accurate response from ChatGPT. In chemistry-related tasks, this approach narrows down the potential answer space and reduces the likelihood of irrelevant or ambiguous responses. For example, we can specify not to include any organic linker synthesis conditions and focus solely on MOF synthesis (Supporting Information, Figure S8). In this case, we found that ChatGPT can recognize the features of organic linker synthesis and differentiate them from MOF synthesis. With proper prompts, information from organic linker synthesis will not be included. Additionally, instructions can provide step-by-step guidance, which has proven effective when multiple tasks are included in one prompt (Supporting Information, Section S2.2).

**Table 1.** Assessment of hallucination in ChatGPT response without prompt engineering.

| Query | ChatGPT Response [a] |
|---|---|
| Which metal is used in the synthesis of MOF-5? | Zinc (Correct) |
| Which metal is used in the synthesis of MOF-519? | Zirconium (Incorrect) |
| What is the linker used in the synthesis of MOF-99999? | I don't know (Correct) |
| What is the linker used in the synthesis of MOF-419? | Terephthalic acid (Incorrect) |
| What is the linker used in the synthesis of ZIF-8? | 2-methylimidazole (Correct) |

**Table 2.** Improvements in ChatGPT response accuracy utilizing a basic prompt engineering strategy.

| Initial Query | Guided Prompt | ChatGPT Response [a] |
|---|---|---|
| Which metal is used in the synthesis of MOF-5? | If you're uncertain, please reply with 'I do not know'. | Zinc (Correct) |
| Which metal is used in the synthesis of MOF-519? | | I don't know (Correct) |
| What is the linker used in the synthesis of MOF-99999? | | I don't know (Correct) |
| What is the linker used in the synthesis of MOF-419? | | I don't know (Correct) |
| What is the linker used in the synthesis of ZIF-8? | | 2-methylimidazole (Correct) |

[a] Responses are representative answers selected from a series of 100 repeated queries, followed by parenthetical indications of their correctness, which is based on the established facts concerning the respective compounds referenced in the queries.

(3) *Requesting Structured Output*, which includes the incorporation of an organized and well-defined response template or instruction to facilitate data extraction. We emphasize that this principle is particularly valuable in the context of chemistry, where data can often be complex and multifaceted. Structured output enables the efficient extraction and interpretation of critical information, which in turn can significantly contribute to the advancement of research and knowledge in the field. Take synthesis condition extraction as an example, without clear instructions on the formatted output, ChatGPT can generate a table, list-like bullet points, or a paragraph, with the order of parameters such as reaction temperature, reaction time, and solvent volume not being uniform, making it challenging for later sorting and storage of the data. This can be easily improved by explicitly asking it to generate a table and providing a fixed header to start with prompt (Supporting Information, Section S2.3). By incorporating these principles, the resulting prompt can ensure that ChatGPT yields accurate and reliable results, ultimately enhancing its utility in tackling complex chemistry-related tasks (Figure 2). We further employ the idea of interactive prompt refinement, in which we start with asking ChatGPT to write a prompt to instruct itself by giving it preliminary descriptions and information (Supporting Information, Figure S15). Through conversation, we add more specific details and considerations to the prompt, testing it with some texts, and once we obtain output, we provide feedback to ChatGPT and ask it to improve the quality of the prompt (Supporting Information, Section S2.4).

As there has been almost no literature systematically discussing prompt engineering in Chemistry, and the fact that this field is relatively new, we provide a comprehensive step-by-step ChemPrompt Engineering guide for beginners to start with, including numerous chemistry-related examples in the Supporting Information, Section S2. At present, everyone is at the same starting point, and no one possesses exclusive expertise in this area. It is our hope that this work will stimulate the development of more powerful prompt engineering skills and help every chemist quickly understand the art of ChemPrompt Engineering, thereby advancing the field of chemistry at large.

**Process 1: Synthesis Conditions Summarization**. One revolutionary aspect of ChatGPT is its specialized domain knowledge due to its extensive pre-trained text corpus, which enables an understanding of chemical nomenclature and reaction conditions.[18] In contrast to traditional NLP methods, ChatGPT requires no additional training for named entity recognition, and can readily identify inorganic metal sources, organic linkers, solvents, and other compounds within a given experimental text. Another notable feature is ChatGPT's ability to recognize and associate compound abbreviations (e.g., DMF) with their full names (*N,N*-dimethylformamide) within the context of MOF synthesis (Supporting Information, Figure S5). This capability is crucial as the use of different abbreviations for the same compound can inflate the number of "unique compounds" in the dataset post text mining, leading to redundancy without providing new information. This challenge is difficult to address using traditional NLP methods or packages, as no model can inherently discern that DMF and *N,N*-dimethylformamide are the same compound without a manually curated dictionary of chemical abbreviations. Although ChatGPT may not cover all abbreviations, its proficiency in identifying and associating the most common ones such as DEF, DI water, EtOH, and $CH_3CN$ with their full names enhances data consistency and reduces redundancy. This, in turn, facilitates data retrieval and analysis, ensuring that different names of the same compound are treated as a single entity with its unique chemical identity and information.

Our first goal is to develop a ChatGPT-based AI assistant that demonstrates high performance in converting a given experimental section paragraph into a table containing all synthesis parameters (Supporting Information, Figure S22). To design the prompt for this purpose, we incorporate the three principles discussed earlier into ChemPrompt Engineering (Figure 2). The rationale for using tabulation as the output for synthesis condition summarization is that the tabular format simplifies subsequent data sorting, analysis, and storage. In terms of the choice of 11 synthesis parameters, we include those deemed most important and non-negligible for each MOF synthesis. Specifically, these parameters encompass metal sources and quantities, dictating metal centers in the framework and their relative concentrations; the linker and its quantity, which affect connectivity and pore size within the MOF; the modulator and its quantity or volume, which can fine-tune the MOF's structure by impacting the nucleation and growth of the MOF in the reaction; the solvent and its volume, which can influence both the crystallization process and the final MOF structure; and the reaction temperature and duration, which are vital parameters governing the kinetics and thermodynamics of MOF formation in each synthesis. In our prompt, we also account for the fact that some papers may report multiple synthesis conditions for the same compound and instruct ChatGPT to use multiple rows to include each variation. For multiple units of the same synthesis parameters, such as when molarity mass and weight mass are both reported, we encourage ChatGPT to include them in the same cell, separated by a comma, which can be later streamlined depending on the needs. If any information is not provided in the sections, e.g., most MOF reactions may not involve the use of modulators and some papers may not specify the reaction time, we expect ChatGPT to answer "N/A" for that parameter. Importantly, to eliminate non-MOF synthesis conditions such as organic linker synthesis, post-synthetic modification, or catalysis reactions, which are not helpful for studying MOF synthesis reactions, we simply add one line of narrative instruction, asking ChatGPT to ignore these types of reactions and focus solely on MOF synthesis parameters. Notably, this natural language-based instruction is highly convenient, requiring no complex and laborious rule-based code to identify unwanted cases and filter them out, and is friendly to researchers without coding experience.

The finalized prompts for Process 1 consist of three parts: (i) a request for ChatGPT to summarize and tabulate the reaction conditions, and only use the text or information provided by humans, which adheres to Principle 1 to minimize hallucination; (ii) a specification of the output table's structure, enumerating expectations and handling instructions, which follows Principles 2 and 3 for detailed instructions and structured output requests; and (iii) the context, consisting of MOF synthesis reaction condition paragraphs from experimental sections or supporting information in research articles. Note that parts (i) and (ii) are fixed prompts, while part (iii) is considered as "input." The combined prompt results in a single question-and-answer interaction, allowing ChatGPT to generate a summarization of the given synthesis conditions as output.

**Process 2: Synthesis Paragraph Classification.** The next question to be answered is, "if ChatGPT is given an entire research article, can it correctly locate the sections of experimental sections?" The objective of Process 2 is to accept an entire research paper as input and selectively forward paragraphs containing chemical experiment details to the next assistant for summarization. However, locating the experimental synthesis section within a research paper is a complex task, as simple techniques such as keyword searches often prove insufficient. For instance, the synthesis of MOFs may be embedded within the supporting information or combined with organic linker synthesis. In earlier publications, synthesis information might appear as a footnote. Furthermore, different journals or research groups utilize varying section titles, including "Experimental," "Methods," "General Methods and Materials," "Experimental methods," "Synthesis and Characterization," "Synthetic Procedures," "Methods Summary," and more. Manually enumerating each case is labor-intensive, especially when synthesis paragraphs may be dispersed with non-MOF synthesis, characterization conditions, or instrument details. Even a human might take considerable time to identify the correct section.

To address this challenge and enable ChatGPT to accurately discern synthesis details within a lengthy research paper, we draw inspiration from the human process. A chemistry Ph.D. student, when asked to locate the MOF synthesis section in a new research paper, would typically start with the first paragraph and ask themselves if it contains synthesis parameters. They would then draw upon prior knowledge from previously read papers to determine if the section is experimental. This process is repeated paragraph by paragraph until the end of the supporting information is reached, with no guarantee that additional synthesis details will not be encountered later. To train ChatGPT similarly, we prompt it to read paper sections incrementally, focusing on one or two paragraphs at a time. Using a few-shot prompt strategy, we provided ChatGPT with a couple of example cases of both synthesis and non-synthesis paragraphs and asked it to classify the sections it reads as either "Yes" (synthesis paragraph) or "No" (non-synthesis paragraph). The ChatGPT Chemistry Assistant would then continue processing the research paper section by section, passing only the paragraphs labeled as "Yes" to the following assistant for summarization.

This few-shot prompt strategy is more convenient than traditional approaches, which require researchers to manually identify and label a large number of paragraphs as "Synthesis Paragraphs" and train their models accordingly. In fact, ChatGPT can even perform such classification using a zero-shot prompt strategy with detailed descriptions of what a "Synthesis Paragraph" should look like and contain. However, we have found that providing four or five short examples in a few-shot prompt strategy enables ChatGPT to identify the features of synthesis paragraphs more effectively, streamlining the classification process (Supporting Information, Figure S24).

The finalized prompt for Process 2 comprises three parts: (i) a request for ChatGPT to determine whether the provided context includes a comprehensive MOF synthesis, answering only with "Yes" or "No"; (ii) some example contexts labeled as "Yes" and other labeled as "No"; (iii) the context to be classified, consisting of one or more research article paragraphs. Similar to Process 1's prompt, parts (i) and (ii) are fixed, while part (iii) is replaced with independent sections from the paper to be classified. The entire research article is parsed into sections of 100-500 words, which are iteratively incorporated into the prompt and sent separately to ChatGPT for a "Yes" or "No" response. Each prompt represents a one-time conversation, and ChatGPT cannot view answers from previous prompts, preventing potential bias in its decision-making for the current prompt.

**Process 3: Text Embeddings for Search and Filtering.** Text embeddings are high-dimensional vector representations of text that capture semantic information, enabling quantification of the relatedness of textual content.[22, 23] The distance between these vectors in the embedded space correlates with the semantic similarity between corresponding text strings, with smaller distances indicating greater relatedness.[24, 25] While Process 2 can automatically read and summarize papers, it must evaluate every section to identify synthesis paragraphs. To expedite this process, we developed Process 3, which filters sections least likely to contain synthesis parameters using OpenAI embeddings before exposing the article to classification assistant in Process 2. To achieve this, we employed a two-step approach to construct Process 3: first, parsing all papers and converting each segment into embeddings; and second, calculating and ranking the similarity scores of each segment based on their relevance to a predefined prompt encapsulating synthesis parameter.

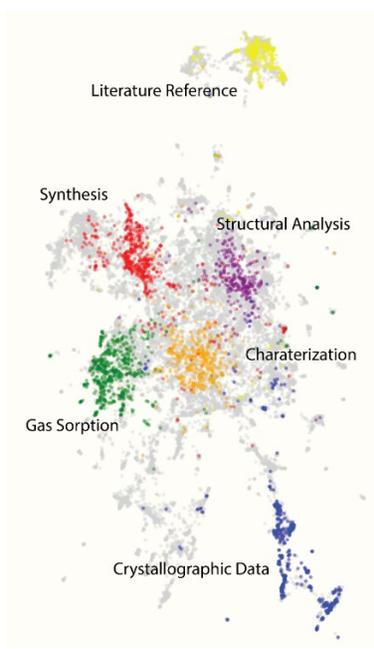

**Figure 3.** Two-dimensional visualization of 18,248 text segment embeddings, with each point representing a text segment from the research articles selected. Color coding denotes thematic categories: red for "synthesis", green for "gas sorption", yellow for "literature reference", blue for "crystallographic data", purple for "structural analysis", orange for "characterization", and grey for other text segments not emphasized in this study.

In particular, we partitioned the 228 research articles into 18,248 individual text segments (Supporting Information, Figure S30–S32). Each segment was converted into a 1536-dimensional text embedding using OpenAI's *text-embedding-ada-002*, a simple but efficient model for this process (Supporting Information, Figure S33–S35). These vectors were stored for future use. To identify segments most and least likely to contain synthesis parameters, we employed interactive prompt refinement strategy (Supporting Information, Section S2.4), consulting with ChatGPT to optimize the prompt. The prompt used in Process 3, unlike previous prompts, served as a text segment for search and similarity comparison rather than instructing ChatGPT (Supporting Information, Figure S25). Next, the embeddings of all 18,248 text segments were compared with the prompt's embedding, and a relevance score was assigned to each segment based on the cosine similarity between the two embeddings. Highly relevant segments were passed on to classification assistant for further processing, while low similarity segments were filtered out (Figure 1).

To evaluate the effectiveness of this approach, we conducted a visual exploration of our embedding data (Figure 3). By reducing the vectors' dimensionality, we observed distinct clusters corresponding to different topics. Notably, we identified distinct clusters related to topics like "gas sorption", "literature reference", "characterization", "structural analysis" and "crystallographic data", which were separate from the "synthesis" cluster. This observation strongly supports the efficiency of our embedding-based filtering strategy. However, this strategy, while effective at filtering out less relevant text and passing segments of mid to high relevance to the subsequent classification assistant, cannot directly search for synthesis paragraphs to feed to the summarization assistant, thus bypassing the classification assistant. In other words, the searching-to-classifying-to-summarizing pipeline cannot be simplified to a searching-to-summarizing pathway due to the inherent search limitations of the embeddings. As shown in Figure 3, embeddings alone may not accurately identify all relevant "synthesis" sections, particularly when they contain additional information such as characterization and sorption data. The presence of these elements in a synthesis section can reduce its similarity score and its proximity to the center of the "synthesis" cluster. Points between the "synthesis" and "characterization" or "crystallographic data" clusters may not have the highest similarity scores and could be missed. However, by filtering only the lowest scores, mid-relevance points are retained and passed to the classification assistant, which can more accurately classify ambiguous content.

**ChatGPT-Assisted Python Code Generation and Data Processing**. Rather than relying on singular, time-consuming conversations with web-based ChatGPT to process textual data from a multitude of research articles, OpenAI's *GPT-3.5-turbo*, which is identical to the one underpinning the ChatGPT product, facilitates a more efficient approach, as it incorporates an Application Programming Interface (API), enabling batch processing of text from an extensive array of articles. This is achieved through iterative context and prompt submissions to ChatGPT, followed by the collection of its responses (Supporting Information, Section S3.4).

Specifically, our approach involves having ChatGPT to create Python scripts for parsing academic papers, generating prompts, executing text processing through Processes 1, 2, and 3, and collating the responses into cleaned, tabulated data (Supporting Information, Figures S28–S39). Traditionally, such a process could necessitate substantial coding experience and be time-consuming. However, we leverage the code generation capabilities of ChatGPT to establish Processes 1, 2, and 3 for batch processing using OpenAI's APIs, namely, *gpt-3.5-turbo* and *text-embedding-ada-002*. In essence, researchers only need to express their requirements for each model in natural language - specifying inputs and desired outputs - and ChatGPT will generate the appropriate Python code (Supporting Information, Section S3.5). This code can be copied, pasted, and executed in the relevant environment. Notably, even in the event of an error, ChatGPT, especially when equipped with the GPT-4 model, can assist in code revision. We note that while coding assistance from ChatGPT may not be necessary for those with coding experience, it does provide an accessible platform for individuals lacking such experience to engage in the process. Given the simplicity and straightforwardness of the logic involved in Processes 1, 2, and 3, ChatGPT-generated Python code exhibits minimal errors and significantly accelerates the programming process.

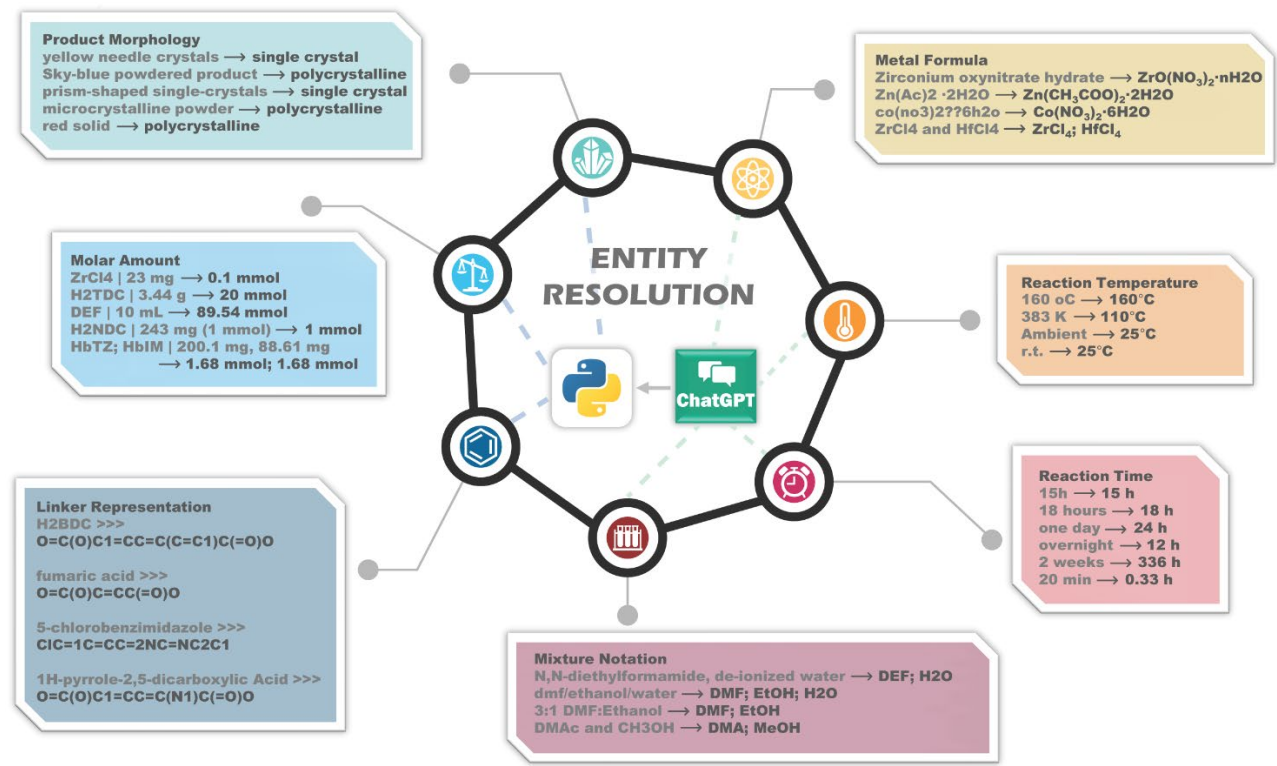

**Figure 4.** Schematic representation of the diverse data unification tasks managed either directly by ChatGPT or through Python code written by ChatGPT. The figure distinguishes between simpler tasks handled directly by ChatGPT, such as standardizing chemical notation, and converting time and temperature units in reactions. More complex tasks, such as matching linker abbreviations to their full names, converting these to SMILES codes, classifying product morphology, and calculating metal amounts, are accomplished via Python code generated by ChatGPT. The Python logo displayed is credited to PSF.

ChatGPT also aids in entity resolution post text mining (Figure 4). This step involves standardizing data formats including units, notation, and compound representations. For each task, we designed a specific prompt for ChatGPT to handle data directly or a specialized Python code generated by ChatGPT. More details on designing prompts to handle different synthesis parameters are available in a cookbook style in Supporting Information, Section S4. In simpler cases, ChatGPT can directly handle conversions such as time and reaction temperature. For complex calculations, we take advantage of ChatGPT in generating Python code. For instance, to calculate the molar mass of each metal source, ChatGPT can generate the appropriate Python code based on the given compound formulas. For harmonizing notation of compound pairs or mixtures, ChatGPT can standardize different notations to a unified format, facilitating subsequent data processing.

To standardize compound representations, we employ the Simplified Molecular Input Line Entry System (SMILES). We faced challenges with some synthesis procedures, where only abbreviations were provided. To overcome this, we designed prompts for ChatGPT to search for the full names of given abbreviations. We then created a dictionary linking each unique PubChem Compound identification number (CID) or Chemical Abstracts Service (CAS) number to multiple full names and abbreviations and generated the corresponding SMILES code. We note that for complicated linkers or those with missing full names, inappropriate nomenclature or non-existent CID or CAS numbers,[26-33] manual intervention was occasionally necessary to generate SMILES codes for such chemicals (Supporting Information, Figure S50–S54). However, most straightforward cases were handled efficiently by ChatGPT's generated Python code. As a result, we achieved uniformly formatted data, ready for subsequent evaluation and utilization.

## RESULTS AND DISCUSSION

**Evaluation of Text Mining Performance.** We began our performance analysis by first evaluating the execution time consumption for each process (Figure 5a). As previously outlined, the ChatGPT assistant in Process 1 exclusively accepts pre-selected experimental sections for summarization. Consequently, Process 1 requires human intervention for the identification and extraction of the synthesis section from a paper to operate autonomously. As illustrated in Figure 5a, this process can vary in duration based on the length and structure of the document and its supporting information file. In our study, the complete selection procedure spanned 12 hours for 228 papers, averaging around 2.5 minutes per paper. This period must be considered as the requisite time for Process 1's execution. For summarization tasks, ChatGPT Chemistry Assistant demonstrated an impressive performance, taking an average of 13 seconds per paper. This is noteworthy considering that certain papers in the dataset contained more than 20 MOF compounds, and human summarization in the traditional way without AI might consume a significantly larger duration. By accelerating the summarization process, we alleviate the burden of repetitive work and free up valuable time for researchers.

In contrast, Process 2 operates in a fully automated manner, integrating the classification and result-passing processes to the next assistant for summarization. There is no doubt that it outperforms the manual identification and summarization combination of Process 1 in terms of speed due to ChatGPT's superior text processing capabilities. Lastly, Process 3, as anticipated, is the fastest due to the incorporation of section filtering powered by embedding, reducing the classification tasks, and subsequently enhancing the speed. The efficiency of Process 3 can be further optimized by storing the embeddings locally as a CSV file during the first reading of a paper, which reduces the processing time by 15-20 seconds (28%-37% faster) in subsequent readings. This provides a convenient solution in scenarios necessitating repeated readings for comparison or extraction of diverse information.

To evaluate the accuracy of the three processes in text mining, instead of sampling, we conducted a comprehensive analysis of the entire result dataset. In particular, we manually wrote down the ground truth for all 11 parameters for approximately 800 compounds reported in all papers across the three processes, which was used to judge the text mining output. This involved the grading of nearly 26,000 synthesis parameters by us. Each synthesis parameter was assigned one of three labels: True Positive (TP, correct identification of synthesis parameters by ChatGPT), False Positive (FP, incorrect assignment of a compound to the wrong synthesis parameter or extraction of irrelevant information), and False Negative (FN, failure of ChatGPT to extract some synthesis parameters). Notably, a special rule for assigning labels on modulators, most of which were anticipated to be acid and base, was introduced to accommodate the neutral solvents in a mixed solvent system, due to the inherent challenges in distinguishing between co-solvents and modulators. For instance, in a $DMF:H_2O = 10:1$ solution, the role of $H_2O$ becomes ambiguous. In such situations, we labeled the result as a TP if $H_2O$ was considered either as a solvent or modulator. However, we labeled it as FP or FN if it appeared or was absent in both solvent and modulator columns. Nevertheless, acids and bases were still classified as modulators, and if labeled as solvents, they were graded as FP.

The distribution of TP labels counted for each of the 11 synthesis parameters across all papers is presented in Figure 5b. It should be noted that not all MOF synthesis conditions necessitate reporting of all 11 parameters; for instance, some syntheses do not involve modulators, and in such cases, we asked ChatGPT to assign an "N/A" to the corresponding column and its amount. Subsequently, we computed the precision, recall, and F1 scores for each parameter across all three processes, illustrated in Figure 5c and d. All processes demonstrated commendable performance in identifying compound names, metal source names, linker names, modulator names, and solvent names. However, they encountered difficulties in accurately determining the quantities or volumes of the chemicals involved. Meanwhile, parameters like reaction temperature and reaction time, which usually have fixed patterns (e.g., units such as °C, hours), were accurately identified by all processes, resulting in high recall, precision, and F1 scores. The lowest scores were associated with the recall of solvent volumes. This is because ChatGPT often captured only one volume in mixed solvent systems instead of multiple volumes. Moreover, in some literatures, the stock solution was used for dissolving metals and linkers, and in principle these volumes should be added to the total volume and unfortunately, ChatGPT lacked the ability to report the volume for each portion in these cases.

Nevertheless, it should be noted that our instructions did not intend for ChatGPT to perform arithmetic operations in these cases, as the mathematical reasoning of the large languages models is limited, and the diminishment of the recall scores is unavoidable. In other instances, only one exemplary synthesis condition for MOF was reported, and then for similar MOFs, the paper would only state "following similar procedures". In such cases, while occasionally ChatGPT could duplicate conditions, most of the time it recognized solvents, reaction temperature, and reaction time as "N/A", which was graded as a FN, thus reducing the recall scores across all processes.

Despite these irregularities, which were primarily attributable to informal synthesis reporting styles, the precision, recall, and F1 scores for all three processes remained impressively high, with less than 9.8% of NP and 0 cases of hallucination detected by human evaluators. We further calculated the average and standard deviation of each process on precision, recall, and F1 scores, respectively, as shown in Figure 5c. By considering and averaging precision, recall, and F1 scores across the 11 parameters, given their equal importance in evaluating overall performance of the process, we found that all three processes achieved impressive precision (> 95%), recall (> 90%), and F1 scores (> 92%).

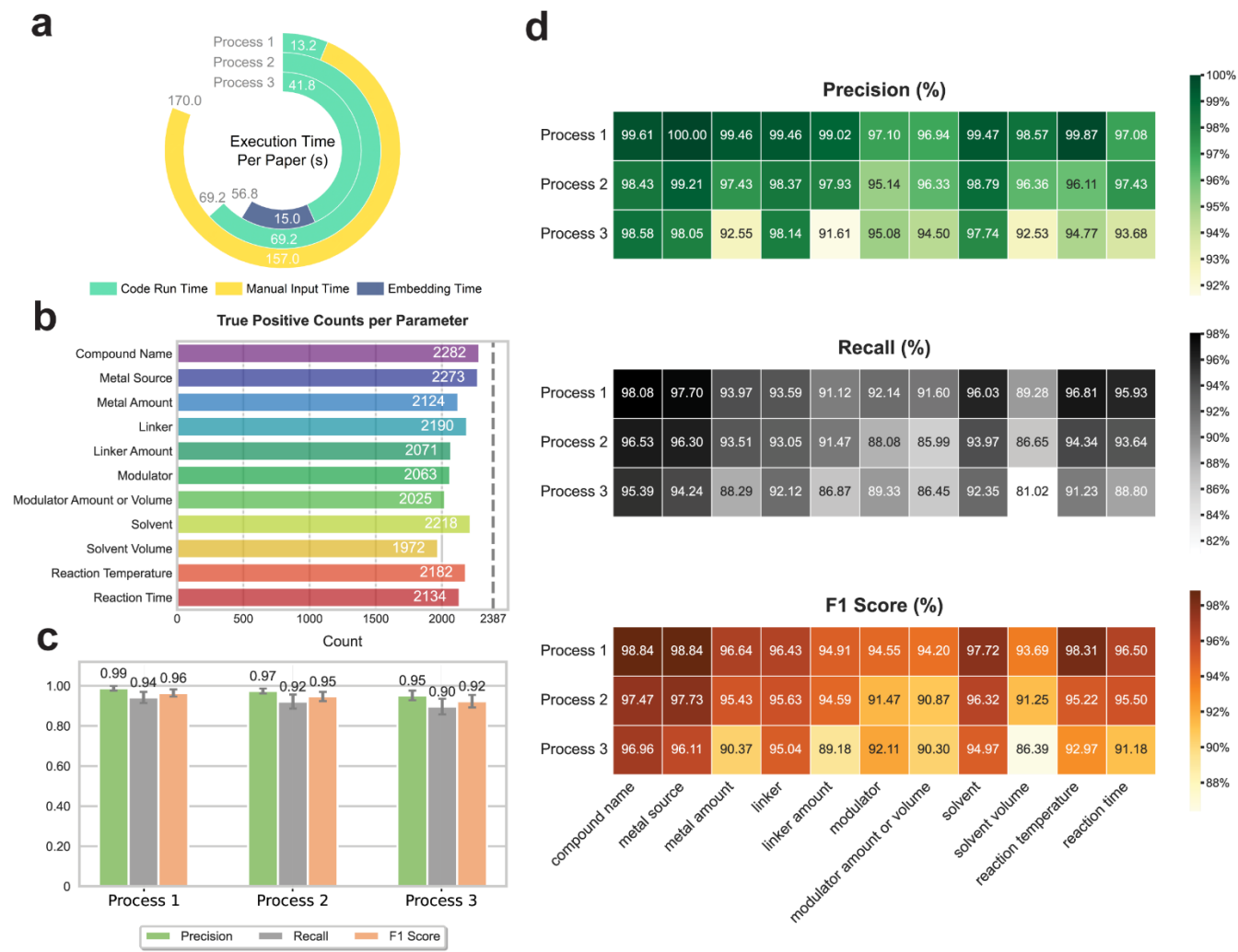

**Figure 5.** Multifaceted performance analysis of ChatGPT-based text mining processes. (a) Comparison of the average execution time required by each process to read and process a single paper, highlighting their relative efficiency. (b) Distribution of true positive counts for each of the 11 synthesis parameters, derived from the cumulative results of Processes 1, 2, and 3 based on a total of 2387 synthesis conditions. Despite minor discrepancies, the counts are closely aligned, demonstrating the assistants' proficiency in effectively extracting the selected parameters. (c) Aggregate average precision, recall, and F1 scores for each process, indicating their overall accuracy and reliability. Standard deviations are represented by grey error bars in the chart. (d) Heatmap illustrating the detailed percentage precision, recall, and F1 scores for each synthesis parameter across the three processes, providing a nuanced understanding of the ChatGPT-based assistants' performance in accurately identifying specific synthesis parameters.

The performance metrics of Process 1 substantiated our hypothesis that ChatGPT excels in summarization tasks. Upon comparing the performance of Processes 2 and 3 — both of which are fully automated paper-reading systems capable of generating datasets from PDFs with a single click — we observed that Process 2, by meticulously examining every paragraph across all papers, ensures high precision and recall by circumventing the omission of any synthesis paragraphs or extraction of incorrect data from irrelevant sections. Conversely, while Process 3's accuracy is marginally lower than that of Process 2, it provides a significant reduction in processing time, thus enabling faster paper reading while maintaining acceptable accuracy, courtesy of its useful filtration process.

To the best of our knowledge, these scores surpass most of other models in text mining in the MOF-related domain.[11, 13, 14, 34, 35] Notably, the entire workflow, established via code and programs generated from ChatGPT, can be assembled by one or two researchers with only basic coding proficiency in a period as brief as a week, whilst maintaining remarkable performance. The successful establishment of this innovative ChatGPT Chemistry Assistant workflow including the ChemPrompt

Engineering system, which harnesses AI for processing chemistry-related tasks, promises to significantly streamline scientific research. It liberates researchers from routine laborious work, enabling them to concentrate on more focused and innovative tasks. Consequently, we anticipate that this approach will catalyze potentially revolutionary shifts in research practices through the integration of AI-powered tools.

**Prediction Modeling of MOF Synthesis Outcomes.** Given the large quantity of synthesis conditions obtained through our ChatGPT-based text mining programs, our aim is to utilize this data to investigate, comprehend, and predict the crystallization conditions of a material of interest. Specifically, our goal was to determine the crystalline state based on synthesis conditions - we seek to discern which synthesis conditions will yield MOFs in the form of single crystals, and which conditions are likely to yield non-single crystal forms of MOFs, such as microcrystalline powder or solids.

With this objective in mind, we identified the need for a label signifying the crystalline state of the resulting MOF for each synthesis condition, thereby forming a target variable for prediction. Fortunately, nearly all research papers in the MOF field consistently include the description of crystal morphological characteristics such as the color and shape of as-synthesized MOFs (e.g. yellow needle crystals, red solid, sky-blue powdered product). This facilitated in re-running our processes with the same synthesis paragraphs as input and modifying the prompt to instruct ChatGPT to extract the description of reaction products, summarizing and categorizing them (Supporting Information, Figure S23 and Figure S47). The final label for each condition will either be Single-Crystal (SC) or Polycrystalline (P), and our objective is to construct a machine learning model capable of accurately predicting whether a given condition will yield SC or P. Furthermore, we recognized that the crystallization process is intrinsically linked with the synthesis method (e.g., vapor diffusion, solvothermal, conventional, microwave-assisted method). Thus, we incorporated an additional synthesis variable, "Synthesis Method", to categorize each synthesis condition into four distinct groups. Extracting the reaction type variable for each synthesis condition can be achieved using the same input but a different few-shot prompt to guide our ChatGPT-based assistants for classification and summarization, subsequently merging this data with the existing dataset. This process parallels the method for obtaining MOF crystalline state outcomes, and both processes can be unified in a single prompt. Moreover, as the name of the MOF is a user-defined term and does not influence the synthesis result, we have excluded this variable for the purposes of prediction modeling.

After unifying and organizing the data to incorporate 11 synthesis parameter variables and 1 synthesis outcome target variable, we designed respective descriptors for each synthesis parameter capable of robustly representing the diversity and complexity in the synthesis conditions and facilitating the transformation of these variables into features suitable for machine learning algorithms. A total of six sets of chemical descriptors were formulated for the metal node(s), linker(s), modulator(s), solvent(s), their respective molar ratios, and the reaction condition(s) - aligning with the extracted synthesis parameters (Supporting Information, Section S5).[36-40] These MOF-tailored, hierarchical descriptors have been previously shown to perform well in various prediction tasks.[13, 41] To distill the most pertinent features and streamline the model, a recursive feature elimination (REF) with 5-fold cross-validation was performed on 80% of the total data. The rest was preserved as a held out set unseen during the learning process for independent evaluation (Figure 6a). This down-selection process reduced the number of descriptors from 70 to 33, thereby preserving comparative model performance on the held out set while removing the non-informative features that can lead to overfitting (Supporting Information Section S5).

Subsequently, we constructed a machine learning model to train for synthesis conditions to predict if a given synthesis condition can yield single crystals. A binary classifier was trained based on a random forest model (Supporting Information, Section S5). The random forest (RF) is an ensemble of decision trees, whose independent predictions are max voted in the classification case to arrive at the more precise prediction.[42] In our study, we trained an RF classifier to predict crystalline states from synthesis parameters, given its ability to work with both continuous and categorical data, its advantage in ranking important features towards prediction, its robustness against noisy data,[43] and its demonstrated efficacy in various chemistry applications such as chemical property estimation,[44-47] spectroscopic analysis,[48-51] and material characterization and discovery.[52]

The dimension-reduced data was randomly divided into different training sizes; for each train test split, optimal hyperparameters, in particular, number of tree estimators and minimum samples required for leaf split, were determined with 5-fold cross validation of the training set. Model performance was gauged in terms of class weighted accuracy, precision, recall, and F1 score over 10 runs on the held out set and test set (Figure 6b and Supporting Information, Figure S64). The model converged to an average accuracy of 87% and an F1 score of 92% on the held out set, indicating a reasonable performance in the presence of the imbalanced classification challenge.

Following the creation of the predictive model, our objective was to apply this model for descriptor analysis to illuminate the factors impacting MOF crystalline outcomes. This aids in discerning which features in the synthesis protocol are more crucial in determining whether a synthesis condition will yield MOF single crystals. Although the random forest model is not inherently interpretable, we probed the relative importance of descriptors used in building the model. One potential measure of a descriptor's importance is the percent decrease in the model's accuracy score when values for that descriptor are randomly shuffled and the model is retrained. We found that among the descriptors involved, the top ten most influential descriptors are key in predicting MOF crystallization outcomes (Figure 6c). In fact, these descriptors broadly align with the chemical intuition and our understanding on MOF crystal growth.[53, 54] For example, the descriptors related to stoichiometry of the MOF synthesis, namely the "modulator to metal ratio", "solvent to metal ratio", and "linker to metal ratio", take precedence in the ranking. These descriptors reflect the vital role of precise stoichiometric control in MOF crystal formation, and they directly impact the crystallization process, playing critical roles in determining the quality and morphology of the MOF crystals.

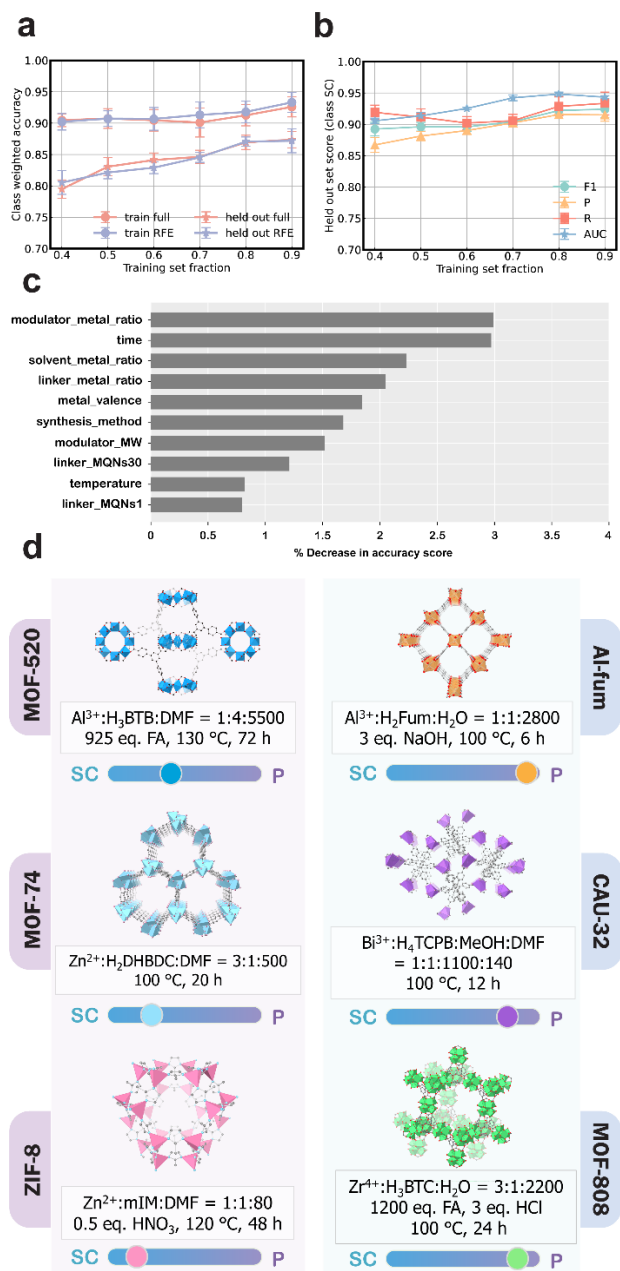

**Figure 6.** Performance of the classification models in predicting the crystalline state of MOFs from synthesis. (a) Learning curves of the classifier model with 1σ standard deviation error bars. (b) Model performance evaluation through the F1 Score, Precision, Recall, and Area Under the Curve metrics. The training set fraction was in ratio to the data excluding the held out set. (c) The ten most significant descriptors of the trained random forest model, determined by accuracy score increase. (d) Six examples of MOFs, MOF-520, MOF-74, ZIF-8, Al-fum, CAU-32, and MOF-808, along with their synthesis conditions derived from the literature.[55-60] Circle positions on the bar represent the likelihood of resulting in single-crystal or polycrystalline states predicted by the model. The model's predictions for these six examples aligned with actual experimental results.

Following closely is the descriptor "time", and it highlights the significant role of reaction duration in the crystallization process. Additionally, the "metal valence" descriptor emphasizes the key role of the nature and reactivity of the metal ions used in MOF synthesis. The valence directly influences the secondary building units (SBUs) and the final crystalline state of the MOF. In the meantime, descriptors related to the molecular and the linker can impact the kinetics of the synthesis, influencing the orderliness of crystal growth. Together, this result provides a greater understanding of the crucial factors affecting the crystallization of MOFs and will aid in the design and optimization of synthesis conditions for the targeted preparation of single-crystal or polycrystalline MOFs (Figure 6d).

**Interrogating the Synthesis Dataset via a Chatbot.** Having utilized text mining techniques to construct a comprehensive MOF Synthesis Dataset, our aim was to leverage this resource to its fullest potential. To enhance data accessibility and aid in the interpretation of its intricate contents, we embarked on a journey to convert this dataset into an interactive and user-friendly dialogue system, which effectively converts the dataset to dialogue. The resulting chatbot is part of the umbrella concept of ChatGPT Chemistry Assistant thus serving as a reliable and fact-based assistant in chemistry, proficient in addressing a broad spectrum of queries pertaining to chemical reactions, in particular MOF synthesis. Unlike typical and more general web-based ChatGPT provided by OpenAI, which may suffer from limitations such as the inability to access the most recent data and a propensity for hallucinatory errors. This chatbot is grounded firmly in the factual data contained within the MOF synthesis dataset from text mining and is engineered to ensure that responses during conversations are based on accurate information and synthesis conditions derived from text mining the literature (Supporting Information, Section S6).

In particular, to construct the chemistry chatbot, our initial step was the creation of distinct entries corresponding to each MOF we identified from the text mining, which encompasses a comprehensive array of synthesis parameters, such as the reaction time, temperature, metal, and linker, among others, using the dataset we have. Recognizing the value of bibliographic context, we compiled a list of paper information, such as authors, DOI, and publication years, collated from Web of Science, into each section (Supporting Information, Table S3). Subsequently, we generated embeddings for each of these information cards of different compounds, thereby constructing an embedding dataset (Figure 7). When a user asks a question, if it is the first query, the system first navigates to the embedding dataset to locate the most relevant information card using the question's embedding, which is based on a similarity score calculation and is similar to the foundation of Process 3 in text mining. The information of the highest-ranking entry is then dispatched to the prompt engineering module of MOF chatbot, guiding it to construct responses centered solely around the given synthesis information.

To mitigate the possibility of hallucination, the chatbot is programmed to refrain from addressing queries that fall outside the scope of the dataset. Instead, it encourages the user to rephrase the question (Supporting Information, Figure S69). It's worth noting that, following the initial query, the chatbot 'memorizes' the conversation context by being presented with the context of prior interactions between user and itself. This includes the synthesis context and paper information identified from the initial query, ensuring that the answers to subsequent queries are also based on factual information from the dataset. Consequently, this strategy guarantees that responses to ensuing queries are contextually accurate, being grounded in the facts outlined in the synthesis dataset and corresponding paper information (Figure 7 and Supporting Information, Figures S71–S74).

By virtue of its design, the chatbot addresses the challenge of enhancing data accessibility and interpretation. It accomplishes this by delivering synthesis parameters and procedures in a clear and comprehensible manner. Furthermore, it ensures data integrity and traceability by providing DOI links to the original papers, guiding users directly to the source of information. This functionality proves particularly beneficial for newcomers to the field. By leveraging ChatGPT's general knowledge base, they can receive guided instructions through the synthesis process, even when faced with a procedure in a journal that is ambiguously or vaguely described. In this case, the user can consult ChatGPT to "chat with the paper" for a more precise explanation, thereby simplifying the learning process and facilitating a more efficient understanding of complex synthesis procedures. This capability fosters independent learning and expedites comprehension of intricate synthesis procedures, reinforcing ChatGPT's role as a valuable assistant in the field of chemistry research.

**Exploring Adaptability and Versatility in Large Language Models.** The adaptability of LLM-based programs, a hallmark feature distinguishing them from traditional NLP programs, lies in their inherent ability to modify search targets or tasks simply by adjusting the input prompt. Whereas traditional NLP models may necessitate a complete overhaul of rules and coding in the event of task modifications, programs powered by ChatGPT and some other LLMs utilize a more intuitive approach. A simple change in narrative language within the prompt can adequately steer the model towards the intended task, obviating the need for elaborate code adjustments.

However, we do recognize limitations within the current workflow, particularly concerning token limitations. Research articles for text mining were parsed into short snippets due to 4096 token limit from *GPT-3.5-turbo*, since longer research articles can extend to 20,000 – 40,000 tokens. This fragmentation may inadvertently result in the undesirable segmentation of synthesis paragraphs or other sections containing pertinent information. To alleviate this, we envision that a large language model that can process higher token memory [61, 62] such as *GPT-4-32K* (OpenAI), or *Claude-v1* (Anthropic) will be very helpful, since each time it reads the entire paper rather than just sections, which can further increase its accuracy by avoiding undesirable segmentation of the synthesis paragraph or other targeted paragraph containing information. Longer reading capabilities will also have the added benefit of reducing the number of tokens used in repeated questions, thus enhancing processing times. As we continue to refine our workflow, we believe that there are further opportunities for improvement. For instance, parts of the fixed prompt could be more concise to save tokens, and the examples in the few-shot prompt can be further optimized to reduce total tokens. Given that each paper may have around 100 segments, such refinements could dramatically reduce time and costs, particularly for classification and summarization tasks, which must process every section with the same fixed prompt, especially for few-shot instructions.

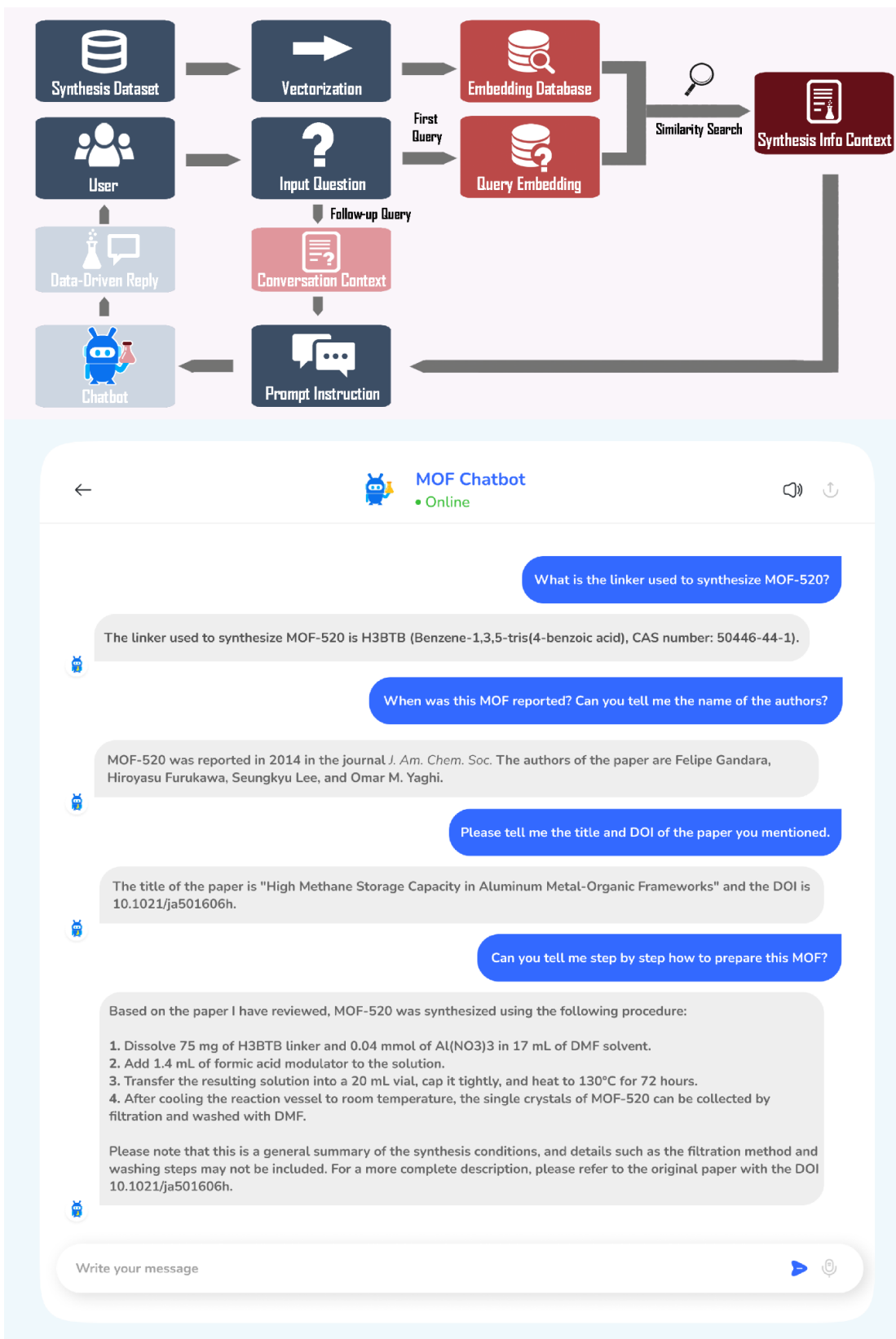

**Figure 7.** Integrated workflow of the MOF chatbot transforming comprehensive synthesis datasets into contextually accurate dialogue systems and demonstration of conversation with the data-driven chatbot. The process ensures enhanced data accessibility, interpretation, and facilitates independent learning in the field of chemistry research.

Furthermore, language versatility, a crucial aspect in the realm of text mining, is seamlessly addressed by LLMs. Traditional NLP models, trained in a specific language, often struggle when the task requires processing text data in another language. For example, if the model is trained on English data, it may require substantial adjustments or even a complete rewrite to process text data in Arabic, Chinese, French, German, French, Japanese, Korean and some other languages. However, with LLMs that can handle multiple languages, such as ChatGPT, we showed that researchers just need to slightly alter the instructions or prompts to achieve the goal, without the necessity of substantial code modifications (Supporting Information, Figure S55–S58).

The adaptable nature of LLMs can further extend versatility in handling diverse tasks. We demonstrated how prompts can be changed to direct ChatGPT to parse and summarize different types of information from the same pool of research articles. For instance, with minor modification of the prompts, we show that our ChatGPT Chemistry Assistants have the potential to be instructed to summarize diverse information such as thermal stability, BET surface area, $CO_2$ uptake, crystal parameters, water stability, and even MOF structure or topology (Supporting Information, Section S4). This adaptability was previously a labor-intensive process, requiring experienced specialists to manually collect or establish training sets for text mining each type of information.[11, 13, 35, 41, 63-66]

Moreover, the utility of this approach can benefit the broader chemistry domain: it is capable of not only facilitating data mining in research papers addressing MOF synthesis but also extending to all chemistry papers with the accorded modifications. By fine-tuning the prompt, the ChatGPT Chemistry Assistant can effectively extract and tabulate data from diverse fields such as organic synthesis, biochemistry preparations, perovskite preparations, polymer synthesis, and more. This capability underscores the versatility of the ChatGPT-based assistant, not only in terms of subject matter but also in the level of detail it can handle. In the event that key parameters for data extraction are not explicitly defined, ChatGPT can be prompted to suggest parameters based on its trained understanding of the text. This level of adaptability and interactivity is unparalleled in traditional NLP models, highlighting a key advantage of the ChatGPT approach. The shift from a code-intensive approach to a natural language instruction approach democratizes the process of data mining, making it accessible even to those with less coding expertise, makes it an innovative and powerful solution for diverse data mining challenges.

## CONCLUDING REMARKS

Our research has successfully demonstrated the potential of LLMs, particularly GPT models, in the domain of chemistry research. We presented a ChatGPT Chemistry Assistant, which includes three different but connected approaches to text mining with ChemPrompt Engineering: Process 3 is capable of conducting search and filtration, Processes 2 and 3 both classify synthesis paragraphs, and Processes 1, 2 and 3 are capable of summarizing synthesis conditions into structured datasets. Enhanced by three fundamental principles of prompt engineering specific to chemistry text processing, coupled with the interactive prompt refinement strategy, the ChatGPT-based assistant have substantially advanced the extraction and analysis of MOF synthesis literature, with precision, recall, and F1 scores exceeding 90%.

We elucidated two crucial insights from the dataset of synthesis conditions. First, the data can be employed to construct predictive models for reaction outcomes, which shed light into the key experimental factors that influence the MOF crystallization process. Second, it is possible to create a MOF chatbot that can provide accurate answers based on text mining, thereby improving access to the synthesis dataset, and achieving a data-to-dialogue transition. This investigation illustrates the potential for rapid advancement inherent to ChatGPT and other LLMs as a proof-of-concept.

On a fundamental level, this study provides guidance on interacting with LLMs to serve as AI assistants for chemists, accelerating research with minimal prerequisite coding expertise and thus bridging the gap between chemistry and the realms of computational and data science more effectively. Through interaction and chatting, the code and design of experiments can be modified, democratizing data mining and enhancing the landscape of scientific research. Our work sets a foundation for further exploration and application of LLMs across various scientific domains, paving the way for a new era of AI-assisted chemistry research.

## ASSOCIATED CONTENT

**Supporting Information**. Detailed instructions and design principles for ChemPrompt Engineering, as well as the specifics of the prompts employed in the ChatGPT Chemistry Assistant for text mining and other chemistry-related tasks. Additional information on the ChatGPT-assisted coding and data processing methods. An extensive explanation of the machine learning models and methods used, as well as the steps involved in setting up the MOF chatbot based on the MOF synthesis condition dataset. This material is available free of charge via the Internet at http://pubs.acs.org.

## AUTHOR INFORMATION

**Corresponding Author**

**Omar M. Yaghi** – *Department of Chemistry; Kavli Energy Nanoscience Institute; and Bakar Institute of Digital Materials for the Planet, College of Computing, Data Science, and Society, University of California, Berkeley, California 94720, United States; UC Berkeley–KACST Joint Center of Excellence for Nanomaterials for Clean Energy Applications, King Abdulaziz City for Science and Technology, Riyadh 11442, Saudi Arabia; orcid.org/0000-0002-5611-3325; Email: yaghi@berkeley.edu*

**Other Authors**

**Zhiling Zheng** – *Department of Chemistry; Kavli Energy Nanoscience Institute; and Bakar Institute of Digital Materials for the Planet, College of Computing, Data Science, and Society, University of California, Berkeley, California 94720, United States;* orcid.org/0000-0001-6090-2258

**Oufan Zhang** – *Department of Chemistry, University of California, Berkeley, California 94720, United States*

**Christian Borgs** – *Bakar Institute of Digital Materials for the Planet, College of Computing, Data Science, and Society; Department of Electrical Engineering and Computer Sciences, University of California, Berkeley, California 94720, United States;* orcid.org/0000-0001-5653-0498

**Jennifer T. Chayes** – *Bakar Institute of Digital Materials for the Planet, College of Computing, Data Science, and Society; Department of Electrical Engineering and Computer Sciences; Department of Mathematics; Department of Statistics; and School of Information, University of California, Berkeley, California 94720, United States;* orcid.org/0000-0003-4020-8618

## ACKNOWLEDGMENTS

Z.Z. extends special gratitude to Jiayi Weng (OpenAI) for valuable discussions on harnessing the potential of ChatGPT. In addition, Z.Z. acknowledges the inspiring guidance and input from Kefan Dong (Stanford University), Long Lian (University of California, Berkeley), and Yifan Deng (Carnegie Mellon University), all of whom contributed to shaping the study's design and enhancing the performance of ChatGPT. We express our appreciation to Dr. Nakul Rampal from the Yaghi Lab for insightful discussions. Our gratitude is also extended for the financial support received from the Defense Advanced Research Projects Agency (DARPA) under contract HR0011-21-C-0020. O.Z. acknowledges funding and extends thanks for the support provided by the National Institute of Health (NIH) under Grant 5R01GM127627-04. Additionally, Z.Z. thanks for the financial support received through a Kavli ENSI Graduate Student Fellowship and the Bakar Institute of Digital Materials for the Planet (BIDMaP). his work is independently developed by the University of California, Berkeley research team and not affiliated, endorsed, or sponsored by OpenAI.

# Supporting Information

# ChatGPT Chemistry Assistant for Text Mining and Prediction of MOF Synthesis

Zhiling Zheng,[†,‡,§] Oufan Zhang,[†] Christian Borgs, [§,◊] Jennifer T. Chayes, [§,◊,††,‡‡,§§]
Omar M. Yaghi[†,‡,§,∥,*]

[†] Department of Chemistry, University of California, Berkeley, California 94720, United States
[‡] Kavli Energy Nanoscience Institute, University of California, Berkeley, California 94720, United States
[§] Bakar Institute of Digital Materials for the Planet, College of Computing, Data Science, and Society, University of California, Berkeley, California 94720, United States
[◊] Department of Electrical Engineering and Computer Sciences, University of California, Berkeley, California 94720, United States
[††] Department of Mathematics, University of California, Berkeley, California 94720, United States
[‡‡] Department of Statistics, University of California, Berkeley, California 94720, United States
[§§] School of Information, University of California, Berkeley, California 94720, United States
[∥] KACST–UC Berkeley Center of Excellence for Nanomaterials for Clean Energy Applications, King Abdulaziz City for Science and Technology, Riyadh 11442, Saudi Arabia
* To whom correspondence should be addressed: yaghi@berkeley.edu

**Table of Contents**

## Section S1. General Information

<u>Large Language Models</u>

Three prominent large language models (LLMs) were involved in this study: GPT-3,[1] ChatGPT (GPT-3.5), and GPT-4. These models are developed and maintained by OpenAI, and although the comprehensive specifics of their training data and architectural design are proprietary, each model is an instantiation of an autoregressive language model that operates on the transformer architecture.[2] For clarity, in this study, we refer to the default GPT-3.5 based chatbot as "ChatGPT", whereas we explicitly denote the GPT-4 based chatbot as "ChatGPT (GPT-4)" when referenced. Both of these are web-based chatbots and accessible through the OpenAI website chat.openai.com.

<u>Application Programming Interface (API)</u>

Two LLM APIs were involved in this study: *text-embedding-ada-002* and *gpt-3.5-turbo*. It should be noted that the model *gpt-3.5-turbo* is essentially the same model as that supporting the default web-based ChatGPT, so we refer to GPT-3.5 API as the ChatGPT API. We note that as of May 2023, access to the GPT-4 API is limited and requires being on a waitlist, and its cost surpasses that of GPT-3.5 significantly. Therefore, our research does not incorporate any usage of the GPT-4 API. In this study, we used *text-embedding-ada-002* for the Text Mining step in Process 3, whereas *gpt-3.5-turbo* served the Text Mining steps in Processes 1 and 2.

<u>Article Retrieval</u>

We obtained 228 papers and their corresponding 225 supporting documents from 32 different journals with the authorization from eight distinct publisher groups: American Chemical Society (ACS), Elsevier, Royal Society of Chemistry (RSC), American Association for the Advancement of Science (AAAS), World Scientific, De Gruyter, Springer Nature, and Wiley. The papers, published between May 2004 and March 2023, were downloaded in PDF format.

<u>Prompt Engineering</u>

The prompts utilized in this study were developed through a process of interactive prompt refinement. For additional details, please refer to Supporting Information Section S2.

<u>Python Code</u>

The majority of the code for text mining, data preprocessing, and chatbot operation was generated by ChatGPT and tested on a Jupyter notebook. Comprehensive code and instructions can be found in Supporting Information Sections S3, S4, and S6. Details of code and data can be found at https://github.com/zach-zhiling-zheng/ChatGPT_Chemistry_Assistant.

<u>Machine Learning Methods</u>

The machine learning model implemented in this study was based on the random forest algorithm, as made available through the scikit-learn library in Python. For additional details, please refer to Supporting Information Section S5.

# Section S2. Chemistry ChatGPT Prompt Engineering (ChemPrompt Engineering)

## S2.1. Principle 1: Minimizing Hallucination

> **Human-ChatGPT Interaction**
>
> **Human:** What is the synthesis condition of MOF-5?
>
> ---
>
> **ChatGPT:** MOF-5 (Metal-organic framework 5) is a type of metal-organic framework that consists of zinc ions coordinated with 1,4-benzenedicarboxylic acid (also known as terephthalic acid) linkers to form a three-dimensional porous structure.
>
> The synthesis conditions for MOF-5 typically involve the following steps:
>
> 1. Dissolving zinc nitrate hexahydrate (Zn(NO3)2·6H2O) in a solvent such as N,N-dimethylformamide (DMF) or N,N-dimethylacetamide (DMA).
> 2. Adding 1,4-benzenedicarboxylic acid (H2BDC) to the solution and stirring for several hours to form a homogeneous mixture.
> 3. Heating the mixture at a high temperature (e.g. 120-140°C) for several hours to promote crystallization.
> 4. Cooling the mixture to room temperature and filtering the resulting solid product.
> 5. Washing the solid product with a suitable solvent (e.g. methanol or DMF) to remove any impurities.
> 6. Drying the product under vacuum to obtain the final MOF-5 material.
>
> The precise conditions for MOF-5 synthesis may vary depending on the specific laboratory protocol and the desired properties of the material.

**Figure S1.** An interaction with ChatGPT showcasing its ability to accurately identify the name of MOF-5 and utilize its knowledge to provide a description of the synthesis conditions, including chemicals, steps, and temperature ranges.

> **Human-ChatGPT Interaction**
>
> **Human:** What is the synthesis condition of MOF-99999?
>
> ---
>
> **ChatGPT:** I'm sorry, but MOF-99999 is not a known or registered MOF. In fact, the highest numbered MOF as of my knowledge cutoff in September 2021 was MOF-901, which was reported in a paper by Zhang et al. in 2017. Without knowing the specific MOF you are referring to, I cannot provide information on its synthesis conditions. If you have any additional information or clarification on the MOF you are interested in, I may be able to help you further.
>
> ---
>
> **Human:** What is the synthesis condition of MOF-419?
>
> ---
>
> **ChatGPT:** MOF-419 is a metal-organic framework (MOF) that is synthesized under solvothermal conditions. Specifically, MOF-419 is typically synthesized by mixing a solution of the metal precursor (e.g. zinc nitrate hexahydrate) and the organic ligand (e.g. 2,5-dihydroxyterephthalic acid) in a solvent (e.g. N,N-dimethylformamide) in a sealed reaction vessel. The mixture is then heated under autogenous pressure at a specific temperature (e.g. 120°C) for a certain duration (e.g. 24 hours) to allow for the formation of the MOF crystals. After the reaction, the resulting MOF-419 crystals can be isolated and purified through various methods such as centrifugation, filtration, or solvent exchange.

**Figure S2.** Two examples of dialogues with ChatGPT that demonstrate its limitations due to hallucinations, where it provides incorrect information about the synthesis conditions of MOFs. In the first example, ChatGPT recognizes that MOF-99999 does not exist but inaccurately describes MOF-901. In the second example, ChatGPT provides incorrect synthesis conditions for MOF-419[3], which was not known at the time of its training.

> **Human-ChatGPT Interaction**
>
> **Human:** What is the synthesis condition of MOF-99999?
>
> ---
>
> **ChatGPT (GPT-4):** As of my knowledge cutoff date in September 2021, MOF-99999 is not a known or documented metal-organic framework (MOF). It is possible that the designation MOF-99999 has been assigned to a new material since then, but I do not have any information on its synthesis or properties.
>
> ---
>
> **Human:** What is the synthesis condition of MOF-419?
>
> ---
>
> **ChatGPT (GPT-4):** As an AI language model, I don't have real-time access to the latest research or data. However, based on the information available up until September 2021, MOF-419 isn't a well-known or extensively studied metal-organic framework (MOF). It's possible that MOF-419 may have been developed or studied more recently, but I cannot provide any specific synthesis conditions for it.

**Figure S3.** A comparison of responses from GPT-4, illustrating its improved handling of hallucination issues compared to the previous GPT-3.5 model. In the case of MOF-99999, GPT-4 correctly identifies that it does not have any information on the compound. However, it still cannot provide useful information on MOF-419, as this compound was published in March 2023, beyond GPT-4's knowledge cutoff date of September 2021.

> **Human-ChatGPT Interaction**
>
> **Human:** <u>Answer the question as truthfully as possible using the provided context. If any information is not provided or you are unsure, use "N/A".</u>
>
> Q: What is the synthesis condition of MOF-99999?
>
> Context: *Section S3: Synthesis of MOF-901*
> *Benzene-1,4-dialdehyde (10.0 mg, 0.0750 mmol) was dissolved in 1 mL of methanol and then dropped into a mixture of 4-aminobenzoic acid (16.0 mg, 0.117 mmol) and titanium isopropoxide (9.0 μL, 0.030 mmol). To increase the solubility, an additional 2 mL of methanol was added to the mixture. The solution was then introduced to a 15 × 1.5 cm Pyrex tube. The tube was flash frozen at 77 K using a liquid nitrogen bath, evacuated for 10 min in order to remove oxygen, and then flamed seal to maintain a vacuum during the reaction process. The reaction mixture was left at 125 °C for 3 d.*
>
> ----
>
> **ChatGPT:** N/A
>
> ----
>
> **Human:** <u>Answer the question as truthfully as possible using the provided context. If any information is not provided or you are unsure, use "N/A".</u>
>
> Q: What is the synthesis condition of MOF-419?
>
> Context: *Synthesis of MOF-419. In a 20 mL scintillation vial, H2BDH (98 mg, 0.5 mmol) was dispersed in DMF (10 mL). The resulting mixture was heated for 30 min in a preheated oven at 100 °C to yield a clear solution. Yttrium (III) nitrate hexahydrate (191 mg, 0.5 mmol) was dissolved in DMF (2 mL) and added to the H2BDH solution. The combined solution was sonicated for 10 min and formic acid (4 mL, 100 mmol) was added. The vial was placed in an oven at 100 °C for 90 min.*
>
> ----
>
> **ChatGPT:** The synthesis condition of MOF-419 involves heating a mixture of H2BDH and DMF at 100 °C for 30 minutes, followed by adding a solution of Yttrium (III) nitrate hexahydrate in DMF, sonication for 10 minutes, addition of formic acid, and heating at 100 °C for 90 minutes. However, the exact synthesis condition of MOF-419 may involve additional steps or conditions that are not mentioned in the provided context.

**Figure S4.** Dialogue demonstrating the performance of ChatGPT when provided with context[3, 4] for answering questions on MOF synthesis conditions. In the first example, ChatGPT correctly identifies that the synthesis condition of MOF-99999 is not available based on the context provided. In the second example, ChatGPT successfully extracts the synthesis conditions of MOF-419 from the most recent information provided in the "context", showcasing its ability to adapt and utilize context-driven information.

> **Human-ChatGPT Interaction**
>
> **Human:** <u>Answer the question as truthfully as possible using the provided context. If any information is not provided or you are unsure, use "N/A".</u>
>
> Q: What is the organic linker used in this reaction? What is the metal source in this reaction?
>
> Context: *Synthesis of MOF-419. In a 20 mL scintillation vial, H2BDH (98 mg, 0.5 mmol) was dispersed in DMF (10 mL). The resulting mixture was heated for 30 min in a preheated oven at 100 °C to yield a clear solution. Yttrium (III) nitrate hexahydrate (191 mg, 0.5 mmol) was dissolved in DMF (2 mL) and added to the H2BDH solution. The combined solution was sonicated for 10 min and formic acid (4 mL, 100 mmol) was added. The vial was placed in an oven at 100 °C for 90 min.*
>
> **ChatGPT:** The organic linker used in this reaction is H2BDH (no information is provided about its full name). The metal source in this reaction is Yttrium (III) nitrate hexahydrate.
>
> **Human:** Can you write the chemical formula of the metal source?
>
> **ChatGPT:** Yes, the chemical formula of Yttrium (III) nitrate hexahydrate is Y(NO3)3·6H2O.
>
> **Human:** What solvent was used?
>
> **ChatGPT:** The solvent used in this reaction is DMF (N,N-Dimethylformamide).
>
> **Human:** What is the reaction time and temperature?
>
> **ChatGPT:** The reaction time is 90 minutes and the reaction temperature is 100 °C.
>
> **Human:** Can you summarize the synthesis conditions into a table?
>
> **ChatGPT:** Yes, here is a summary of the synthesis conditions:
>
> | Compound/Parameter | Amount/Value |
> |---|---|
> | H2BDH | 98 mg, 0.5 mmol |
> | DMF | 12 mL |
> | Yttrium (III) nitrate hexahydrate | 191 mg, 0.5 mmol |
> | Formic acid | 4 mL, 100 mmol |
> | Reaction time | 90 minutes |
> | Reaction temperature | 100 °C |

**Figure S5.** An interaction with ChatGPT showcasing its ability to extract and summarize relevant information from the provided context for MOF-419 synthesis.[3] The AI model identifies the organic linker, metal source, solvent, and reaction conditions, and then presents a summarized table of the synthesis parameters, demonstrating its effectiveness in processing and presenting complex information.

> **Human-ChatGPT Interaction**
>
> **Human:** <u>Answer the question as truthfully as possible using the provided context. If any information is not provided or you are unsure, use "N/A".</u>
>
> Q: Can you summarize the synthesis conditions into a table?
>
> Context:
>
> *MOF-802. H2PZDC (0.27 g, 1.5 mmol) and ZrOCl2 · 8H2O (0.40 g, 1. 3 mmol) in a solvent mixture of DMF/formic acid (50 mL/ 35mL) were placed in a 125 mL screw capped glass jar, which was heated at 130 °C for three days.*
>
> *MOF-808. H3BTC (0.11 g, 0.50 mmol) and ZrOCl2 ·8H2O (0.1 6 g, 0.50 mmol) in a solvent mixture of DMF/formic acid (20 mL/ 20mL) were placed in a 60 mL screw capped glass jar, which was heated at 1 0 0 °C for seven days.*
>
> *DUT-67. DUT-67 was prepared according to published procedures with slight modification (S4). H2TDC (0. 069 g, 0.40 mmol) and ZrOCl2 · 8H2O (0.19 g, 0.60 mmol) in a solvent mixture of DMF/formic acid (20 mL/ 11 mL) were placed in a 60 mL screw capped glass jar, which was heated at 130 °C for three days.*
>
> ---
>
> **ChatGPT:** Yes, a table can be made to summarize the synthesis conditions of the MOFs. The table can include the following information:
>
> | MOF Type | Ligand | Metal Source | Solvent Mixture | Solvent Volume | Temperature | Duration |
> |---|---|---|---|---|---|---|
> | MOF-802 | H2PZDC | ZrOCl2·8H2O | DMF/formic acid | 50 mL/35 mL | 130 °C | 3 days |
> | MOF-808 | H3BTC | ZrOCl2·8H2O | DMF/formic acid | 20 mL/20 mL | 100 °C | 7 days |
> | DUT-67 | H2TDC | ZrOCl2·8H2O | DMF/formic acid | 20 mL/11 mL | 130 °C | 3 days |

**Figure S6.** An interaction with ChatGPT where it successfully extracts and organizes synthesis conditions for three different MOFs (MOF-802, MOF-808, and DUT-67) from the provided context[5]. The AI model presents the information in a concise and easily readable table format, demonstrating its ability to process and structure complex information from multiple sources. Note that ChatGPT can efficiently extract and summarize information from multiple synthesis conditions in a single step, without the need for separate conversations or step-by-step extraction.

> ### Human-ChatGPT Interaction
>
> **Human:** <u>Answer the question as truthfully as possible using the provided context. If any information is not provided or you are unsure, use "N/A".</u>
>
> Q: Can you summarize the synthesis conditions into a table?
>
> Context:
>
> *Optimized Synthesis of MOF-521-H Single Crystals. In a 4 mL scintillation vial, 3.0 mL DMF was added to dissolve Al(NO3)3·9H2O (18.0 mg, 0.048 mmol) and H3BTB (15.8 mg, 0.036 mmol). After adding 0.24 mL formic acid and 30 μL deionized water to the solution, the vial was capped and placed in the preheated 140 °C for 2 days, and colorless needle-shape crystals were obtained. The crystals were washed with MeOH (3 × 20 mL) for 3 days before activation and characterization. Full activation of the MOF was conducted under dynamic vacuum (~10-3 mbar) at 100°C for 12 hours, yielding pure and desolvated product.*
>
> ---
>
> **ChatGPT:** Based on the provided context, here is a summarized table of the synthesis conditions for MOF-521-H single crystals:
>
> | Compound/Reagent | Amount/Volume |
> |---|---|
> | Al(NO3)3·9H2O | 18.0 mg (0.048 mmol) |
> | H3BTB | 15.8 mg (0.036 mmol) |
> | DMF | 3.0 mL |
> | Formic Acid | 0.24 mL |
> | Deionized Water | 30 μL |
> | Preheating Temperature | 140 °C |
> | Preheating Duration | 2 days |

**Figure S7.** An interaction with ChatGPT wherein the model successfully delineates and organizes the synthesis conditions for MOF-521, an unpublished compound as of May 2023, from the provided context. It should be noted that prior to this instance, the synthesis description for this compound was inaccessible to ChatGPT as the compound is from an unpublished manuscript. The AI model exhibits its proficiency in processing and structuring intricate information by presenting it in a concise, reader-friendly table format. Please note that the structured output instructions have not been incorporated in the prompt at this stage, leading to potential variances in formatting elements (such as column headers and layout) when compared to outputs illustrated in Figure S5 and Figure S6. Solutions to these discrepancies will be thoroughly explored in the ensuing Sections S2.2 and S2.3.

## S2.2. Principle 2: Implementing Detailed Instructions

> **Human-ChatGPT Interaction**
>
> **Human:** Answer the question as truthfully as possible using the provided context. If any information is not provided or you are unsure, use "N/A".
>
> <u>Please focus on extracting experimental conditions from only the MOF synthesis and ignore information related to organic linker synthesis.</u>
>
> Q: What is the synthesis condition of the following compound?
>
> Context:
> *Synthesis of (E)-5-(2-carboxyvinyl)-1H-pyrazole-3-carboxylic acid (H2PZVDC):*
> *Step 1: A 100-mL round-bottom flask equipped with a stirring bar was charged with 1 (1.5 g, 8.9 mmol, 1 equiv.) and dry THF (50 mL) under argon atmosphere. The mixture was cooled down to –10 °C using an acetone/ice bath, and 2 (3.5 g, 10.5 mmol, 1.2 equiv.) was added portion-wise.*
> *The reaction was allowed to warm up to room temperature over a period of 15 hours. After concentrating the resulting solution under reduced pressure, a mixture containing E- and Z-isomers was identified via 1H NMR analysis. The desired E-isomer 3 was isolated via column chromatography using acetone/hexane (1:5) as eluent ($R_f$ = 0.1). Yield: 1.3 g, 65%.*
> *$^1$H NMR (400 MHz, CDCl$_3$) δ 10.93 (s, 1H), 7.67 (d, J = 16.0 Hz, 1H), 7.04 (s, 1H), 6.48 (d, J = 16.0 Hz, 1H), 4.41 (q, J = 7.1 Hz, 2H), 3.81 (s, 3H), 1.41 (t, J = 7.1 Hz, 3H).*
>
> 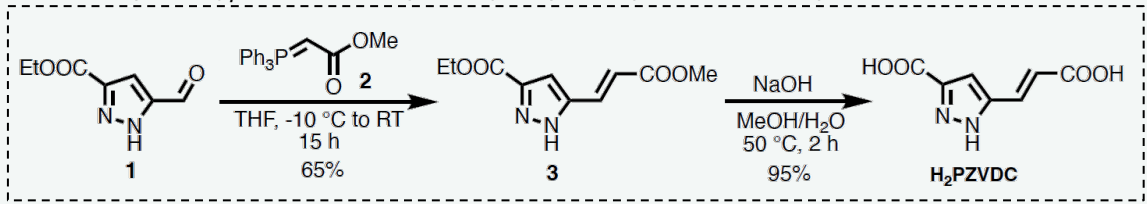
>
> **ChatGPT:** N/A.

**Figure S8.** Engagement of ChatGPT in a context-specific organic synthesis discussion[6], adhering to the explicit instruction to exclude organic linker synthesis, thereby minimizing ambiguity. ChatGPT's response illustrates its capability to discern the relevance of organic linker synthesis. Note that accompanying reaction schematic is not a component of the input but is provided to enhance reader comprehension.

## Human-ChatGPT Interaction

**Human:** Answer the question as truthfully as possible using the provided context.

Please summarize the following details in a table: compound name, metal source, metal amount, organic linker(s), linker amount, reaction temperature, and reaction time. If any information is not provided or you are unsure, use "N/A".

If multiple conditions are provided for the same compound, use multiple rows to represent them. If multiple units or components are provided for the same factor (e.g. g and mol for the weight, multiple linker or metals, multiple temperature and reaction time, mixed solvents, etc.), include them in the same cell and separate by comma.

Context:
*Solvothermal synthesis of MOF-LA2-1:*
*In a 4-mL scintillation vial, the linker H2PZVDC (91.0 mg, 0.5 mmol, 1 equiv.) was dissolved in N,N-dimethylformamide (DMF) (0.6 mL) upon sonication. An aqueous solution of AlCl3·6H2O (2.4 mL, 0.2 M, 1 equiv.) was added dropwise, and the resulting mixture was heated in a 120 °C oven for 24 hours.*

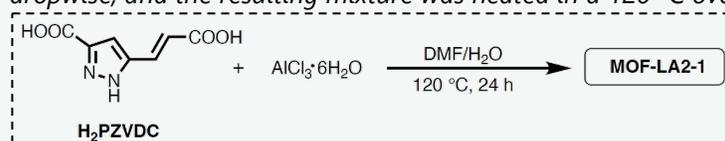

*Green synthesis of MOF-LA2-1:*
*In a 50-mL round-bottom flask, the linker H2PZVDC (364 mg, 2 mmol, 1 equiv.) and NaOH (160 mg, 4 mmol, 2 equiv.) were dissolved in deionized water (10 mL) upon sonication. An aqueous solution of AlCl3·6H2O (6 mL, 0.33 M, 1 equiv.) was added dropwise for 10 minutes, and the reaction mixture was heated to 120 °C and refluxed for 2 hours.*

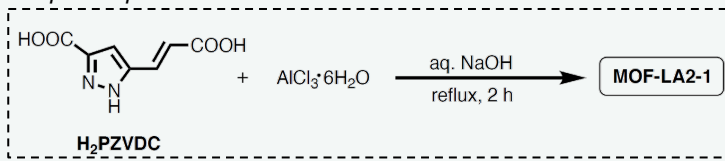

**ChatGPT:**

| Compound Name | Metal Source | Metal Amount | Organic Linker(s) | Linker Amount | Reaction Temperature | Reaction Time |
|---|---|---|---|---|---|---|
| MOF-LA2-1 | AlCl3·6H2O | 2.4 mL, 0.2 M, 1 equiv. | H2PZVDC | 91.0 mg, 0.5 mmol, 1 equiv. | 120 °C | 24 hours |
| MOF-LA2-1 | AlCl3·6H2O | 6 mL, 0.33 M, 1 equiv. | H2PZVDC | 364 mg, 2 mmol, 1 equiv. | 120 °C | 2 hours |

**Figure S9.** An illustration of an interaction with ChatGPT, demonstrating its capability to generate multiple synthesis paragraphs with varying parameters.[6] This figure emphasizes the essential role of prompt engineering in guiding ChatGPT to manage specific scenarios effectively. It is evident that ChatGPT can comprehend detailed instructions and yield a well-organized output table. Note that included reaction schematic is not part of the input but is provided solely for the convenience of the reader.

> **Human-ChatGPT Interaction**
>
> **Human:** Answer the question as truthfully as possible using the provided context.
>
> Question: Describe the solvent(s) in the synthesis of MOF-2.
>
> Context:
> *Synthesis of MOF-2: Slow vapor diffusion at room temperature of triethylamine (0.05 mL) and toluene (5 mL) into a DMF solution (2 mL) containing a mixture of Zn(NO3)2 6H2O (0.073 g, 0.246 mmol) and the acid form of BDC (0.040 g, 0.241 mmol) diluted with toluene (8 mL) yields colorless prism-shaped crystals that were formulated as Zn(BDC)(DMF)(H2O).*
>
> ---
>
> **ChatGPT:**
> The solvent used in the synthesis of MOF-2 is a mixture of DMF (2 mL) and toluene (8 mL), into which triethylamine (0.05 mL) and toluene (5 mL) were slowly vapor diffused at room temperature.
>
> ---
>
> **Human:**
> Task: Summarize the synthesis condition of this MOF into a table.
> The table should have 6 columns, all in lowercase:
> | compound name | metal source | metal amount | linker | linker amount | reaction temperature |
>
> ---
>
> **ChatGPT:**
>
> | compound name | metal source | metal amount | linker | linker amount | reaction temperature |
> |---|---|---|---|---|---|
> | MOF-2 | Zn(NO3)2 6H2O | 0.073 g (0.246 mmol) | BDC (acid form) | 0.040 g (0.241 mmol) | Room temperature |

**Figure S10.** A schematic representation of interaction with ChatGPT, highlighting the effective use of delimiters, such as '|' and ':', in prompt engineering for chemistry-related tasks. The application of these delimiters in separating input and output sections enables the AI model to process complex information, including chemical formulas, reactions, and equations, more efficiently and accurately.

> **Human-ChatGPT Interaction**
>
> **Human:** Answer the question as truthfully as possible using the provided context.
>
> Task:
> 1. Create a new table that includes only the MOF synthesis temperature and reaction time.
> 2. Convert temperatures to °C.
> 3. Convert reaction times to hours.
>
> Context:
>
> | MOF | Ligand | Metal Source | Solvent Mixture | Solvent Volume | Temperature | Duration |
> |---|---|---|---|---|---|---|
> | MOF-808 | H3BTC | ZrOCl2·8H2O | DMF/formic acid | 20 mL/20 mL | 100 °C | 7 days |
> | MFM-300(Cr) | H4BPTC | CrCl3·6H2O | deionised water | 100 mL | 483 K | 3 days |
> | MIL-101 | H2BDC | Cr(NO3)3·9H2O | water | 75 mL | 493 K | 9 h |
>
> **ChatGPT:** Based on the provided context, a new table including only the MOF synthesis temperature and reaction time can be created as follows:
>
> | MOF | Temperature (°C) | Reaction Time (hours) |
> |---|---|---|
> | MOF-808 | 100 | 168 |
> | MFM-300(Cr) | 210 | 72 |
> | MIL-101 | 220 | 9 |
>
> 1. To convert the temperatures to °C, the following conversions can be made:
>
>    483 K = 210 °C
>    493 K = 220 °C
>
> 2. To convert the reaction times to hours, the following conversions can be made:
>
>    7 days = 168 hours
>    3 days = 72 hours
>    9 hours (already in hours)

**Figure S11.** An illustration of an interaction with ChatGPT, underscoring the importance of prompt engineering for guiding the AI model through a multi-step task within a single conversation. This approach optimizes ChatGPT's performance and provides detailed insights into its problem-solving process for each step.

## S2.3. Principle 3: Requesting Structured Output

**Human-ChatGPT Interaction**

**Human:** Task: Please summarize the following details in a table: compound, metal source, metal amount, organic linker(s), linker amount, solvent(s), solvent volume(s), reaction temperature, and reaction time. If any information is not provided or you are unsure, use "N/A".

Output table should have 9 columns: **| Compound name | Metal source | Metal amount | Linker | Linker amount | Solvent | Solvent volume | Reaction temperature | Reaction time |**

Context:
MOF-950. A solvent mixture of H3BTAC (0.033 g, 0.11 mmol) and Zn(NO3)2·6H2O (0.25 g, 0.84 mmol) in 20 mL DMF was placed in a 20 mL screw-capped scintillation vial, which was heated at 85 °C for 3 d.

MOF-905. A solvent mixture of H3BTAC (0.045 g, 0.16 mmol), H2BDC (0.048 g, 0.29 mmol) and Zn(NO3)2·6H2O (0.26 g, 0.87 mmol) in 18 mL DMF and 1.8 mL ethanol was placed in a 20 mL screw-capped scintillation vial, which was heated at 85 °C for 1 d.

MOF-905-Me2. A solvent mixture of H3BTAC (0.045 g, 0.16 mmol), H2BDC-Me2 (0.067 g, 0.34 mmol) and Zn(NO3)2·6H2O (0.26 g, 0.87 mmol) in 18 mL DMF and 1.8 mL ethanol was placed in a 20 mL screw-capped scintillation vial, which was heated at 85 °C for 1 d.

MOF-905-Naph. A solvent mixture of H3BTAC (0.045 g, 0.16 mmol), H2NDC (0.074 g, 0.34 mmol) and Zn(NO3)2·6H2O (0.26 g, 0.87 mmol) in 18 mL DMF and 1.8 mL ethanol was placed in a 20 mL screw-capped scintillation vial, which was heated at 85 °C for 1 d.

MOF-905-NO2. A solvent mixture of H3BTAC (0.045 g, 0.16 mmol), H2BDC-NO2 (0.062 g, 0.29 mmol) and Zn(NO3)2·6H2O (0.26 g, 0.87 mmol) in 18 mL DMF and 1.8 mL ethanol was placed in a 20 mL screw-capped scintillation vial, which was heated at 85 °C for 36 h.

**ChatGPT:**

| Compound name | Metal source | Metal amount | Linker | Linker amount | Solvent | Solvent volume | Reaction temperature | Reaction time |
|---|---|---|---|---|---|---|---|---|
| MOF-950 | Zn(NO3)2·6H2O | 0.84 mmol | H3BTAC | 0.11 mmol | DMF | 20 mL | 85 °C | 3 d |
| MOF-905 | Zn(NO3)2·6H2O | 0.87 mmol | H3BTAC, H2BDC | 0.16 mmol, 0.29 mmol | DMF, ethanol | 18 mL, 1.8 mL | 85 °C | 1 d |
| MOF-905-Me2 | Zn(NO3)2·6H2O | 0.87 mmol | H3BTAC, H2BDC-Me2 | 0.16 mmol, 0.34 mmol | DMF, ethanol | 18 mL, 1.8 mL | 85 °C | 1 d |
| MOF-905-Naph | Zn(NO3)2·6H2O | 0.87 mmol | H3BTAC, H2NDC | 0.16 mmol, 0.34 mmol | DMF, ethanol | 18 mL, 1.8 mL | 85 °C | 1 d |
| MOF-905-NO2 | Zn(NO3)2·6H2O | 0.87 mmol | H3BTAC, H2BDC-NO2 | 0.16 mmol, 0.29 mmol | DMF, ethanol | 18 mL, 1.8 mL | 85 °C | 36 h |

**Figure S12.** An illustration of interaction with ChatGPT, highlighting the necessity for structured output requests. This example showcases a prompt that clarifies tasks and provides headers, thus enabling ChatGPT to understand that the task is 'tabulation'.[7] The specified headers subsequently guide the AI model to generate a table in the desired format.

> **Human-ChatGPT Interaction**
>
> **Human:** Task: Given the following text describing the synthesis of MOFs, please extract the MOF compound name and reaction product morphology. Present the information in a table format with a header and a new row for each compound. If the compound name, or reaction results is not mentioned in the text, use "N/A". Header: **| MOF Compound | Reaction Results |**
>
> Example 1:
> Input: *Synthesis of Zn(2-mbim)2: the mixture of Zn(CH3CO2)2·2(H2O), 2-methylbenzimdazole in DMF was sealed in a 20 mL vial and heated to 120 oC for 2 days, and then cooled to room-temperature. The pale yellow polyhedral crystals were obtained.*
> Output:
> | MOF Compound | Reaction Results |
> | Zn(2-mbim)2 | pale yellow polyhedral crystals|
>
> Example 2:
> Input: *The synthesis of HKUST-1 and MOF-199 was performed in a 20 mL Teflon-lined autoclave. Blue octahedral crystals were obtained for HKUST-1.*
> Output:
> | MOF Compound | Reaction Results |
> | HKUST-1 | Blue octahedral crystals |
> | MOF-199 | N/A |
>
> Example 3:
> Input: *A mixture of H2PylDC (0.2 mol) and NaOH (0.6 mol) was dissolved and heated to 120°C. After refluxing for 6 hours, the white crystalline powder was collected.*
> Output:
> | MOF Compound | Reaction Results |
> | N/A | white crystalline powder |
>
> Please provide the extracted information for the following text:
> *ZIF-21: A solid mixture of cobalt(II) nitrate hexahydrate Co(NO3)2·6H2O and purine was dissolved in DMF in a 20 ml vial. To the solution, 2.0M dimethylamine solution in methanol was added. The vial was tightly capped and heated at 85 ◦C in an isothermal oven for 24 h to yield purple octrahedral crystals.*
>
> ------
>
> **ChatGPT:**
> | MOF Compound | Reaction Results |
> | ZIF-21 | purple octahedral crystals |

**Figure S13.** An illustration of interaction with ChatGPT using few-shot prompts. In contrast to the zero-shot approach, ChatGPT is presented with several example contexts along with their corresponding correct responses. This process aids the model in identifying and extracting specific parameters such as synthesis vessel type and crystal class, tasks that would otherwise be challenging to accomplish without the provision of examples.

> **Human-ChatGPT Interaction**
>
> **Human:** Task: Answer the question as truthfully as possible using the provided context. Please summarize the following details in a table: compound name or chemical formula (if the name is not provided), metal source, metal amount, organic linker(s), linker amount, modulator, modulator amount or volume, solvent(s), solvent volume(s), reaction temperature, and reaction time. If any information is not provided or you are unsure, use "N/A".
>
> The table should have 11 columns, all in lowercase:**| compound name | metal source | metal amount | linker | linker amount | modulator | modulator amount or volume | solvent | solvent volume | reaction temperature | reaction time |**
>
> Context:
>
> *Optimized Synthesis of MOF-521-H Single Crystals. In a 4 mL scintillation vial, 3.0 mL DMF was added to dissolve Al(NO3)3·9H2O (18.0 mg, 0.048 mmol) and H3BTB (15.8 mg, 0.036 mmol). After adding 0.24 mL formic acid and 30 µL deionized water to the solution, the vial was capped and placed in the preheated 140 °C for 2 days, and colorless needle-shape crystals were obtained. The crystals were washed with MeOH (3 × 20 mL) for 3 days before activation and characterization. Full activation of the MOF was conducted under dynamic vacuum (~10-3 mbar) at 100°C for 12 hours, yielding pure and desolvated product.*
>
> ---
>
> **ChatGPT:**
>
> | compound name | metal source | metal amount | linker | linker amount | modulator | modulator amount or volume | solvent | solvent volume | reaction temperature | reaction time |
> |---|---|---|---|---|---|---|---|---|---|---|
> | MOF-521-H | Al(NO3)3·9H2O | 18.0 mg | H3BTB | 15.8 mg | Formic acid | 0.24 mL | DMF | 3.0 mL | 140 °C | 2 days |

**Figure S14.** An illustration of interaction with ChatGPT, demonstrating how specific instructions can direct the AI model to generate a table in a desired format. In contrast to the prompt displayed in Figure S12, the number and nomenclature of headers can be effortlessly altered through natural language commands, thereby modifying the output. It should be noted that the text utilized in this example originates from an unpublished manuscript, which was not previously introduced to ChatGPT. The structured output directive effectively guides the AI model to handle various synthesis parameters derived from text sources, encompassing both published manuscripts and unpublished works.

## S2.4. Interactive Prompt Refinement

> **Human-ChatGPT Interaction**
>
> **Human:** I am going to write a prompt to instruct ChatGPT to help me extract the MOF linker and its amount in the synthesis from the text I provide. Could you please begin by asking me a few questions to gather the necessary information to include in the prompt? Do you understand?
>
> ---
>
> **ChatGPT (GPT-4):** Yes, I understand. I'll ask you a few questions to gather the necessary information for creating the prompt.
>
> 1. What specific text or document are you planning to provide for ChatGPT to extract the MOF linker and its amount from? Is it a research article, a patent, or some other type of document?
> 2. Are there any specific types or classes of MOF linkers that you want ChatGPT to focus on?
> 3. Do you want ChatGPT to provide the information in a particular format or style, such as a table, bullet points, or a summary?
> 4. Are there any additional details or context you'd like ChatGPT to consider when extracting the information?
>
> ---
>
> **Human:**
> 1. This will be a section from a research article.
> 2. No, you should extract the linker as long as it is involved in the reaction to form the MOF.
> 3. Yes, please make it a table and ensure that the table has a fixed header for each output.
> 4. If multiple linkers are presented, please separate them by semicolons within the same cell. Do the same for their amount and ensure that the order is consistent.
>
> ---
>
> **ChatGPT (GPT-4):** Thank you for providing the necessary information. Based on your responses, I'll create a prompt for you to instruct ChatGPT to help you extract MOF linker and its amount in the synthesis from a section of a research article. Here's the prompt:
>
> *"Please help me extract the MOF linker and its amount used in the synthesis from the following text taken from a research article. Present the information in a table format with fixed headers: 'MOF Linker' and 'Amount'. If multiple linkers are present, separate them using a semicolon (;) in the same cell. Do the same for their corresponding amounts, ensuring the order is consistent between linkers and amounts."*

**Figure S15.** A conversation with ChatGPT (GPT-4) for the interactive prompt refinement process, highlighting the initial engagement with GPT-4 to develop a guiding prompt, followed by iterative discussions and refinements to generate a testable prompt for subsequent evaluation with another instance of ChatGPT (GPT-4) model.

> **Human-ChatGPT Interaction**
>
> **Human:** *Please help me extract the MOF linker and its amount used in the synthesis from the following text taken from a research article. Present the information in a table format with fixed headers: 'MOF Linker' and 'Amount'. If multiple linkers are present, separate them using a semicolon (;) in the same cell. Do the same for their corresponding amounts, ensuring the order is consistent between linkers and amounts*
>
> Context:
> *ZIF-20: Zn(Pur)2·(DMF)0.75(H2O)1.5. A solid mixture of zinc nitrate tetrahydrate Zn(NO3)2·4H2O (65 mg, 0. 25 mmol, EM Science) and purine (150 mg, 1.25 mmol, Fluka or Sigma) was dissolved in 5 ml DMF (Fisher) in a 20 ml vial to obtain a slightly turbid solution. The vial was tightly capped and heated at 65 ◦C in an isothermal oven for 3 days to yield pale yellow octahedral crystals.*
>
> **ChatGPT:**
>
> | MOF Linker | Amount |
> |---|---|
> | Purine | 1.25 mmol |
> | Zn(NO3)2·4H2O | 0.25 mmol |

**Figure S16.** A conversation illustrating the evaluation of a preliminarily devised prompt. The process involves integrating the prompt within a given context and assessing the outcomes. It was observed that the initial prompt did not account for potential metal sources, leading to erroneous classification of these sources as linkers in the synthetic process.[8]

> **Human-ChatGPT Interaction**
>
> *---Continued from previous conversation shown in Figure S15---*
>
> **ChatGPT (GPT-4):** **previous prompt**
>
> ---
>
> **Human:** When I use this prompt, I sometimes find that the metal source, such as "Zn(NO3)2·4H2O", is mistakenly counted as a MOF linker. In addition, there is no header. Could you please revise the prompt to fix this issue?
>
> ---
>
> **ChatGPT (GPT-4):** I apologize for the confusion. I have revised the prompt to explicitly mention that metal sources should not be considered as MOF linkers. Here's the updated prompt:
>
> *"Please help me extract the organic MOF linker and its amount used in the synthesis from the following text taken from a research article. Exclude any metal sources or inorganic components, such as Co(NO3)2·4H2O, AlCl3, and HfOCl2, focusing only on the organic linkers. Present the information in a table format with fixed headers: 'MOF Linker' and 'Amount'. If multiple linkers are present, separate them using a semicolon (;) in the same cell. Do the same for their corresponding amounts, ensuring the order is consistent between linkers and amounts."*

**Figure S17.** A continued conversation following on from Figure S15 with ChatGPT (GPT-4), discussing the evaluation outcomes of a preliminary prompt, providing feedback on identified inaccuracies, and suggesting areas of potential improvement. ChatGPT responds by generating a refined prompt, thereby addressing the initially identified issues.

> **Human-ChatGPT Interaction**
>
> **Human:** *Please help me extract the organic MOF linker and its amount used in the synthesis from the following text taken from a research article. Exclude any metal sources or inorganic components, such as Co(NO3)2·4H2O, AlCl3, and HfOCl2, focusing only on the organic linkers. Present the information in a table format with fixed headers: 'MOF Linker' and 'Amount'. If multiple linkers are present, separate them using a semicolon (;) in the same cell. Do the same for their corresponding amounts, ensuring the order is consistent between linkers and amounts.*
>
> Context:
> *ZIF-20: Zn(Pur)2·(DMF)0.75(H2O)1.5. A solid mixture of zinc nitrate tetrahydrate Zn(NO3)2·4H2O (65 mg, 0. 25 mmol, EM Science) and purine (150 mg, 1.25 mmol, Fluka or Sigma) was dissolved in 5 ml DMF (Fisher) in a 20 ml vial to obtain a slightly turbid solution. The vial was tightly capped and heated at 65 ◦C in an isothermal oven for 3 days to yield pale yellow octahedral crystals.*
>
> ---
>
> **ChatGPT:**
>
> | MOF Linker | Amount |
> |---|---|
> | purine | 150 mg (1.25 mmol) |

**Figure S18.** A new conversation representing the ongoing evaluation of the updated prompt from Figure S17. The previously highlighted issue regarding the metal source was rectified, but a new challenge, specifically related to the formatting of the linker amount, surfaced during the process.

> **Human-ChatGPT Interaction**
>
> *---Continued from previous conversation shown in Figure S17---*
>
> **ChatGPT (GPT-4):** **revised prompt**
>
> ---
>
> **Human:** Upon examination, it has come to my attention that the weight of the linker is sometimes expressed as "150 mg (1.25 mmol)," which presents both molar (mol) and mass (g) units. To ensure consistency, I kindly request that when indicating the amount of the linker, the use of mol units is preferred over g units. Moreover, it is important to display only one unit of measurement, either mol or g, when presenting the amount. Please note that the terms g and mol encompass all units of measurement, such as mg, g, kg, µg, mmol, mol, and µmol. Thank you.
>
> ---
>
> **ChatGPT (GPT-4):** Thank you for the clarification. I have updated the prompt to prioritize the mole unit over the mass unit when extracting the amount of the MOF linker. Here's the revised prompt:
>
> *"Please help me extract the organic MOF linker and its amount used in the synthesis from the following text taken from a research article. Exclude any metal sources or inorganic components, such as Co(NO3)2·4H2O, AlCl3, and HfOCl2, focusing only on the organic linkers. When extracting the amount, prioritize the mole unit (e.g., mmol, mol, µmol) over the mass unit (e.g., mg, g, kg, µg) and display only one of them. Present the information in a table format with fixed headers: 'MOF Linker' and 'Amount'. If multiple linkers are present, separate them using a semicolon (;) in the same cell. Do the same for their*

**Figure S19.** An ongoing dialogue with ChatGPT (GPT-4), detailing observed issues related to formatting and initiating a request for further refinement.

> **Human-ChatGPT Interaction**
>
> **Human:** *Please help me extract the organic MOF linker and its amount used in the synthesis from the following text taken from a research article. Exclude any metal sources or inorganic components, such as Co(NO3)2·4H2O, AlCl3, and HfOCl2, focusing only on the organic linkers. When extracting the amount, prioritize the mole unit (e.g., mmol, mol, μmol) over the mass unit (e.g., mg, g, kg, μg) and display only one of them. Present the information in a table format with fixed headers: 'MOF Linker' and 'Amount'. If multiple linkers are present, separate them using a semicolon (;) in the same cell. Do the same for their corresponding amounts, ensuring the order is consistent between linkers and amounts.*
>
> Context:
> *ZIF-20: Zn(Pur)2·(DMF)0.75(H2O)1.5. A solid mixture of zinc nitrate tetrahydrate Zn(NO3)2·4H2O (65 mg, 0. 25 mmol, EM Science) and purine (150 mg, 1.25 mmol, Fluka or Sigma) was dissolved in 5 ml DMF (Fisher) in a 20 ml vial to obtain a slightly turbid solution. The vial was tightly capped and heated at 65 ◦C in an isothermal oven for 3 days to yield pale yellow octahedral crystals.*
>
> ---
>
> **ChatGPT:**
>
> | MOF Linker | Amount |
> |---|---|
> | purine | 1.25 mmol |

**Figure S20.** A conversation with ChatGPT, examining the outcomes of the further refined prompt shown in Figure S19. All previously identified issues were found to be effectively addressed.

> **Human-ChatGPT Interaction**
>
> **Human:** *revised prompt shown in Figure S19*
> Context:
> MOF-905. A solvent mixture of H3BTAC (0.045 g, 0.16 mmol), H2BDC (0.048 g, 0.29 mmol) and Zn(NO3)2·6H2O (0.26 g, 0.87 mmol) in 18 mL DMF and 1.8 mL ethanol was placed in a 20 mL screw-capped scintillation vial, which was heated at 85 °C for 1 d.
>
> **ChatGPT:**
>
> | MOF Linker | Amount |
> |---|---|
> | H3BTAC; H2BDC | 0.16 mmol; 0.29 mmol |
>
> **Human:** *revised prompt shown in Figure S19*
> Context:
> Solvothermal synthesis of MOF-LA2-1:
> In a 4-mL scintillation vial, the linker H2PZVDC (91.0 mg, 0.5 mmol, 1 equiv.) was dissolved in N,N-dimethylformamide (DMF) (0.6 mL) upon sonication. An aqueous solution of AlCl3·6H2O (2.4 mL, 0.2 M, 1 equiv.) was added dropwise, and the resulting mixture was heated in a 120 °C oven for 24 hours.
>
> **ChatGPT:**
>
> | MOF Linker | Amount |
> |---|---|
> | H2PZVDC | 0.5 mmol |
>
> **Human:** *revised prompt shown in Figure S19*
> Context:
> [Al(C6H2O4S)(OH)] · xH2O, CAU-23. 4.30 g H2TDC (25 mmol) was mixed with 2.0 g (50 mmol) sodium hydroxide in 100 mL distilled water until a clear solution of Na2TDC was obtained. After adding 18.75 mL of aqueous aluminum chloride solution (1 mol/L, 18.75 mmol), and 12.5 mL of aqueous sodium aluminate solution (0.5 mol/L, 6.25 mmol), the slurry was stirred under reflux conditions for 6 h, and then filtered off and dried at 100 °C for 4 h.
>
> **ChatGPT:**
>
> | MOF Linker | Amount |
> |---|---|
> | H2TDC | 25 mmol |

**Figure S21.** An extended discussion with ChatGPT, continually testing the refined prompt across various scenarios. The consistent outcomes highlight the robustness of the finalized prompt. The interactive prompt engineering process is deemed completed once user satisfaction is achieved. If new concerns or issues arise, the process loops back to further refine the prompt through continued dialogue.

# Section S3. Text Mining with ChatGPT API

## S3.1. Summarizing Synthesis Conditions with ChatGPT

> **ChatGPT Processing Demo**
>
> **Prompt:** Answer the question as truthfully as possible using the provided context. Please summarize the following details in a table: compound name or chemical formula (if the name is not provided), metal source, metal amount, organic linker(s), linker amount, modulator, modulator amount or volume, solvent(s), solvent volume(s), reaction temperature, and reaction time. If any information is not provided or you are unsure, use "N/A".
>
> Please focus on extracting experimental conditions from only the MOF synthesis and ignore information related to organic linker synthesis, MOF postsynthetic modification or metalation, high throughput (HT) experiment details, or catalysis reactions.
>
> If multiple conditions are provided for the same compound, use multiple rows to represent them. If multiple units or components are provided for the same factor (e.g. g and mol for the weight, multiple linker or metals, multiple temperature and reaction time, mixed solvents, etc.), include them in the same cell and separate by comma.
>
> The table should have 11 columns, all in lowercase:| compound name | metal source | metal amount | linker | linker amount | modulator | modulator amount or volume | solvent | solvent volume | reaction temperature | reaction time |
>
> ---
>
> **Input:**
> *Synthesis of MOF-808. In a 100 mL media bottle were dissolved 1,3,5-benzenetricarboxylic acid (210 mg) and ZrOCl2·8H2O (970 mg) in a solution containing DMF (30 mL) and formic acid (30 mL). The bottle was sealed and heated in a 100 °C isothermal oven for a day. White powder was collected by centrifugation (8000 rpm, 3 min), washed with DMF three times (60 mL × 3) over a 24 h period, and then washed with acetone three times (60 mL × 3) over a 24 h period.*
>
> ---
>
> **Output:**
>
> | compound name | metal source | metal amount | linker | linker amount | modulator | modulator amount or volume | solvent | solvent volume | reaction temperature | reaction time |
> |---|---|---|---|---|---|---|---|---|---|---|
> | MOF-808 | ZrOCl2·8H2O | 970 mg | 1,3,5-benzenetricarboxylic acid | 210 mg | formic acid | 30 mL | DMF | 30 mL | 100 °C | 24 h |

**Figure S22.** An illustration of prompt components, designed to guide ChatGPT in summarizing synthesis conditions, with reference to the three fundamental principles of chemistry prompt engineering. This illustration includes an example input and output table for context. It should be noted that this is a demonstration and various prompt versions could be constructed based on specific research needs.

> **ChatGPT Processing Demo**
>
> **Prompt:** Given the following text describing the synthesis of MOFs, please extract the MOF compound name, synthesis equipment/reaction vessel, reaction type, and reaction product morphology. Present the information in a table format with a header and a new row for each compound. If the compound name, equipment, or reaction results are not mentioned in the text, use "N/A." Determine the reaction type based on the following keywords and prioritize them in this order: Diffusion, Microwave, Conventional, Solvothermal.
>
> *Example 1:*
> Input: Synthesis of Zn(2-mbim)2:the mixture of Zn(CH3CO2)2·2(H2O), 2-methylbenzimdazole in DMF was sealed in a 20 mL vial and heated to 120 oC for 2 days, and then cooled to room-temperature. The pale yellow polyhedral crystals were obtained.
> Output:
> MOF Compound | Equipment | Reaction Type | Reaction Results
> Zn(2-mbim)2 | 20 mL vial | Solvothermal | pale yellow polyhedral crystals
>
> *Example 2:*
> Input: The synthesis of HKUST-1 and MOF-199 was performed in a 20 mL Teflon-lined autoclave. Blue octahedral crystals were obtained for HKUST-1, while MOF-199 yielded a green crystalline powder.
> Output:
> MOF Compound | Equipment | Reaction Type | Reaction Results
> HKUST-1 | 20 mL Teflon-lined autoclave | Solvothermal | Blue octahedral crystals
> MOF-199 | 20 mL Teflon-lined autoclave | Solvothermal | Green crystalline powder
>
> *Example 3:*
> Input: MOF-313. In a 1 L glass round bottom flask, a mixture of H2PylDC (0.2 mol) and NaOH (0.6 mol) was dissolved. The resulting solution was stirred for 10 minutes until all the solids were completely dissolved. Afterward, the reaction mixture was heated to 120°C. After refluxing for 6 hours, the white crystalline powder was collected.
> Output:
> MOF Compound | Equipment | Reaction Type | Reaction Results
> MOF-313 | 1 L glass round bottom flask | Conventional | white crystalline powder
>
> *Example 4:*
> Input: H3BTB and Bi(NO3)3·5H2O were mixed into 30 ml microwave glass reaction vessel. The reaction mixture homogenized and heated to 120 °C for 20 min in a microwave synthesizer. The mixture was stirred with a magnetic stirring bar during the reaction. The yellow precipitated of CAU-7 was filtered, washed with methanol.
> Output:
> MOF Compound | Equipment | Reaction Type | Reaction Results
> CAU-7 | 30 ml microwave glass reaction vessel | Microwave | yellow precipitated
>
> Please provide the extracted information for the following text:
>
> ---
>
> **Input:**
> *Synthesis of ZIF-1001: A mixture of Zn(NO3)2·4H2O (1.2 mmol), HbTZ (1.68 mmol), HbIM (0.75 mmol) was dissolved in DMF (12 mL) under ultrasound. The solution was capped in a 20-ml glass vial and heated at 100 °C for 48 h. White prism-shaped crystals were collected and washed with DMF (6×10 ml). (Yield: 30% based on Zn).*
>
> ---
>
> **Output:**
>
> | MOF Compound | Equipment | Reaction Type | Reaction Results |
> |---|---|---|---|
> | ZIF-1001 | 20-ml glass vial | Solvothermal | White prism-shaped crystals |

**Figure S23.** An illustration of prompt components, designed to guide ChatGPT in summarizing reaction equipment and reaction results using the few-shot prompt strategy. This illustration includes an example input and output table for context.

## S3.2. Classifying Research Article Sections with ChatGPT

> **ChatGPT Processing Demo**
>
> **Prompt:** Determine whether the provided context includes a comprehensive MOF synthesis with explicit reactant quantities or solvent volumes, and answer with either Yes or No.
>
> *Context: In a 4-mL scintillation vial, the linker H2PZVDC (91.0 mg, 0.5 mmol, 1 equiv.) was dissolved in N,N-dimethylformamide (DMF) (0.6 mL) upon sonication. An aqueous solution of AlCl3·6H2O (2.4 mL, 0.2 M) was added dropwise, and the resulting mixture was heated in a 120 °C oven for 24 hours.*
> *Question: Does the section contain a comprehensive MOF synthesis with explicit reactant quantities or solvent volumes?*
> *Answer: Yes.*
>
> *Context: A 0.150 M solution of imidazole in DMF and a 0.075M solution of Zn(NO3)2·4H2O in DMF were used as stock solutions, and heated in an 85 °C isothermal oven for 3 days.*
> *Question: Does the section contain a comprehensive MOF synthesis with explicit reactant quantities or solvent volumes?*
> *Answer: Yes.*
>
> *Context: Solvothermal reactions of Co(NO3)·6H2O, Hatz, and L1/L2 in a 2:2:1 molar ratio in DMF solvent at 180 °C for 24 h yielded two crystalline products, 1 and 2, respectively.*
> *Question: Does the section contain a comprehensive MOF synthesis with explicit reactant quantities or solvent volumes?*
> *Answer: No.*
>
> *Context: A 22.9% weight loss was observed from 115 to 350 °C, which corresponds to the loss of one DEF molecule per formula unit (calcd: 23.5%).*
> *Question: Does the section contain a comprehensive MOF synthesis with explicit reactant quantities or solvent volumes?*
> *Answer: No.*
>
> ---
>
> **Input #1:**
> *In a 100 mL media bottle were dissolved 1,3,5-benzenetricarboxylic acid (210 mg) and ZrOCl2·8H2O (970 mg) in a solution containing DMF (30 mL) and formic acid (30 mL).*
>
> **Output #1:**
> Yes.
>
> **Input #2:**
> *Single crystal X-ray analyses were performed at room temperature on a Siemens SMART platform diffractometer outfitted with an APEX II area detector and monochromatized Mo Kα radiation.*
>
> **Output #2:**
> No.

**Figure S24.** An illustration of use of few-shot prompts to guide ChatGPT in classifying synthesis paragraphs. Two example inputs and their respective outputs are provided for clarification.

## S3.3. Filtering Text using OpenAI Embeddings

> **Embedding Demo**
>
> **Prompt Embedding (ada-002):** Identify the experimental section or synthesis method. This section should cover essential information such as the compound name (e.g., MOF-5, ZIF-1, Cu(Bpdc), compound 1, etc.), metal source (e.g., ZrCl4, CuCl2, AlCl3, zinc nitrate, iron acetate, etc.), organic linker (e.g., terephthalate acid, H2BDC, H2PZDC, H4Por, etc.), amount (e.g., 25mg, 1.02g, 100mmol, 0.2mol, etc.), solvent (e.g., N,N Dimethylformamide, DMF, DCM, DEF, NMP, water, EtOH, etc.), solvent volume (e.g., 12mL, 100mL, 1L, 0.1mL, etc.), reaction temperature (e.g., 120°C, 293K, 100C, room temperature, reflux, etc.), and reaction time (e.g., 120h, 1 day, 1d, 1h, 0.5h, 30min, a week, etc.).
>
> **Input #1:**
> *The synthesis of Zr-DTDC was performed by adding ZrCl 4 (0.466 g), H2DTDC (1.025 g), and hydrochloric acid (0.33 ml) into DMF (12 ml) at room temperature. The slurry was transferred to 20 ml Teflon-lined steel autoclave. The autoclave was placed in an oven with 2 °C/min hearting up to 220 °C, then held at 220 °C for 16 h.*
>
> **Output #1:**
> 0.8771 (Yes)
>
> **Input #2:**
> *Furukawa, Nakeun Ko, Yong Bok Go, Naoki Aratani, Sang Beom Choi, Eunwoo Choi, A. Özgür Yazaydin, Randall Q. Snurr, Michael O'Keeffe, Jaheon Ki.*
>
> **Output #2:**
> 0.6651 (No)
>
> **Input #3:**
> *Thus, a fixed bed was packed with activated ZIF-204 and subsequently subjected to a binary gas mixture containing CO 2(35%, v/v) and CH 4(65%, v/v) at room temperature. It is worthwhile to note that the composition of this binary gas mixture was chosen to simulate the typical volumetric percentage of CO 2 and CH4 found in biogas sources produced from the decomposition of organic matter.*
>
> **Output #3:**
> 0.7904 (No)

**Figure S25.** An illustration of the prompt to be converted into embeddings to facilitate the search and filter process for synthesis paragraphs. This is achieved by determining the semantic similarity between the context and the prompt, assessed via cosine similarity scores ranging between 0 and 1. A high score (denoted as "Yes" in the output) implies the paragraph is relevant and retained, while a low score (denoted as "No") leads to the exclusion of the paragraph. As with previous figures, multiple prompt versions could be used, and the details are largely dependent on the specific requirements of the study.

## S3.4. Batch Text Processing with ChatGPT API

In the preceding sections, we have demonstrated how fixed prompts with context derived from research articles can guide ChatGPT in performing non-dialogue text processing tasks such as summarization and classification.

The ChatGPT API offers a notable advantage over web-based interactions with the model. This advantage lies in its ability to iterate over multiple text inputs using a programming construct like a 'for' loop. This approach enables concurrent processing of a multitude of requests, thus enhancing efficiency for large-scale tasks.

**ChatGPT API**

```python
import openai

response = openai.ChatCompletion.create(
  model="gpt-3.5-turbo",
  messages=[
        {"role": "user", "content": "What is the synthesis condition of MOF-5?"}
    ]
)

print(response.choices[0].message.content)
```
---
The synthesis condition of MOF-5 is typically conducted under solvothermal conditions, where a mixture of zinc nitrate hexahydrate and terephthalic acid is dissolved in N,N-dimethylformamide (DMF) and heated to a temperature of 120-150°C for several hours. The resulting solution is then cooled and the solid MOF-5 material is collected via filtration and washing.

**Figure S26.** An illustration of Python code utilizing the ChatGPT API to ask the question on the synthesis condition of MOF-5 and displaying the subsequent output. This output aligns closely with responses generated via the web-based ChatGPT interface.

```
ChatGPT API

mofs = ["MOF-5", "MOF-520", "MOF-999999", "CAU-10", "ZIF-8"]

for mof_name in mofs:
    question = f"Has {mof_name} been published before? Answer with Yes or No."
    response = openai.ChatCompletion.create(
        model="gpt-3.5-turbo",
        messages=[
            {"role": "user", "content": question}
        ]
    )

    print(f'Question: {question}')
    print(f'Answer: {response.choices[0].message.content}')
------------------------------------------------------------------
Question: Has MOF-5 been published before? Answer with Yes or No.
Answer: Yes.

Question: Has MOF-520 been published before? Answer with Yes or No.
Answer: Yes.

Question: Has MOF-999999 been published before? Answer with Yes or No.
Answer: No.

Question: Has CAU-10 been published before? Answer with Yes or No.
Answer: Yes.

Question: Has ZIF-8 been published before? Answer with Yes or No.
Answer: Yes.
```

**Figure S27.** Python code demonstrating the iterative question-asking functionality made possible through the use of a 'for' loop. The ChatGPT API enables efficient batch processing of text, allowing a pre-compiled series of inputs to be queried rapidly, which otherwise is a process significantly more time-consuming on the web-based ChatGPT due to manual input requirements.

```
contexts = [
    "Context: In a 100 mL media bottle ……",
    "Context: In a 4-mL scintillation vial, the linker ……",
    "Context: In a 20 mL scintillation vial, H2BDH (98 mg, 0.5 mmol) was ……"
    #full context not shown for brevity
]

prompt = "Answer the question as truthfully as possible ……"

for context in contexts:
    message = {
        "role": "user",
        "content": f"{prompt}... {context}...",
    }
    response = openai.ChatCompletion.create(
        model="gpt-3.5-turbo",
        messages=[message],
    )

    print(f'Answer: {response.choices[0].message.content}')
------------------------------------------------------------------
Answer: | Compound Name/Chemical Formula | Metal Source | Metal Amount | Organic Linker(s) |
Linker Amount | Modulator | Modulator Amount/Volume | Solvent(s) | Solvent Volume(s) |
Reaction Temperature | Reaction Time |
|--------------------------------|--------------|--------------|-------------------|---------------|----------|-------------------------|------------|-------------------|------------------|---------------|
| MOF-808                        | ZrOCl2·8H2O  | 970 mg       | 1,3,5-benzenetricarboxylic acid | 210 mg  | Formic acid | 30 mL            | DMF        | 30 mL             | 100 °C           | 1 day         |

Answer: | Compound Name/Chemical Formula | Metal Source | Metal Amount | Organic Linker(s) |
Linker Amount | Modulator | Modulator Amount/Volume | Solvent(s) | Solvent Volume(s) |
Reaction Temperature | Reaction Time |
|--------------------------------|--------------|--------------|-------------------|---------------|----------|-------------------------|------------|-------------------|------------------|---------------|
| MOF-LA2-1                      | AlCl3·6H2O   | 0.48 mmol    | H2PZVDC           | 91.0 mg, 0.5 mmol, 1 equiv. | N/A | N/A          | DMF        | 0.6 mL            | 120 °C           | 24 hours      |

Answer: | Compound Name/Formula | Metal Source | Metal Amount | Organic Linker(s) | Linker Amount | Modulator | Modulator Amount/Volume | Solvent(s) | Solvent Volume(s) | Reaction Temperature | Reaction Time |
|-----------------------|--------------|--------------|-------------------|---------------|----------|-------------------------|------------|-------------------|------------------|---------------|
| MOF-419               | Yttrium (III) nitrate hexahydrate | 191 mg, 0.5 mmol | H2BDH | 98
```

**Figure S28.** Illustration of performing batch text summarization using a 'for' loop in Python to engage the ChatGPT API. Preloaded paragraphs serve as the context, with the Python code assembling the prompts and context into a series of inputs. Once the model generates an answer, it is collected, and the process moves on to the next question automatically. The responses from the API are string-based; therefore, to represent answers in a tabular format, "|" and "-" symbols are used. To maintain clarity and brevity, full prompts and context details are not shown here. Please refer to Figure S22 for complete prompt information.

```python
contexts = [
    "Context: Introduction  A wide range of metal centers ……",
    "Context: In a 100 mL media bottle were ……",
    "Context: Results and Discussion 1 consists of isolated ……",
    "Context: 4. Conclusions A three-dimensional ……",
    #full context not shown for brevity
]

prompt ="""Determine whether the provided context includes a comprehensive ……"""

question= """Question: Does the section contain a comprehensive MOF synthesis with explicit
reactant quantities or solvent volumes?
Answer: """

for context in contexts:
    message = {
        "role": "user",
        "content": f"{prompt}... {context}...{question}",
    }
    response = openai.ChatCompletion.create(
        model="gpt-3.5-turbo",
        messages=[message],
    )

    print(f'Answer: {response.choices[0].message.content}')

---------------------------------------------------------------
Answer: No.

Answer: Yes.

Answer: No.

Answer: No.
```

**Figure S29.** Illustration of performing batch synthesis paragraph classification using a 'for' loop in Python to engage the ChatGPT API. As with Process 1, preloaded paragraphs provide context, and the Python code prepares the prompts and context as inputs. Answers are collected automatically and the process continues to the subsequent question. In this example, ChatGPT has been directed to respond with "Yes" or "No" for classification purposes, following the instructions provided in the fixed prompt. To ensure clarity and brevity, full prompts and context details are omitted here. Please refer to Figure S23 for complete prompt information.

## S3.5. Generating Python Code with ChatGPT

Paper Parsing

As the first step towards enabling ChatGPT to autonomously parse and read academic papers, it is necessary to convert the PDF files into text strings, which can subsequently be incorporated into the prompt as context and processed using the ChatGPT API. This conversion can be achieved by employing a Python function, as illustrated in the following figures.

---

**ChatGPT Code Synthesis**

**Human:** Write a Python code to extract text from multiple research article PDF files. For each page in each file, divide the text into four sections and record their page and section numbers (1-4). Count the tokens for each section and construct a dataframe with the following columns: 'file name', 'page number', section number', 'content', and 'tokens'. The input will be a list of PDF file names as strings, and the output should be a dataframe.

---

**ChatGPT (GPT-4):**

```python
import PyPDF2
import pandas as pd
from nltk.tokenize import word_tokenize
import nltk
nltk.download('punkt') # Download the Punkt tokenizer for tokenizing

def get_txt_from_pdf(pdf_files):
    data = [] # Create an empty list to store the data
    for pdf in pdf_files: # Iterate over the PDF
        with open(pdf, 'rb') as pdf_content: # Fetch the PDF content from the pdf
            pdf_reader = PyPDF2.PdfReader(pdf_content) # Create a PDF reader object
            for page_num in range(pdf_reader.getNumPages()): # Iterate over all the pages
                page = pdf_reader.getPage(page_num) # Extract the text from the current page
                page_text = page.extract_text()
                words = page_text.split() # Split the page text into individual words
                page_text_join = ' '.join(words) # Join the words
                page_len = len(page_text_join)
                div_len = page_len // 4 # Divide the page into 4 parts
                page_parts = [page_text_join[i*div_len:(i+1)*div_len] for i in range(4)]
                for i, page_part in enumerate(page_parts): # Append the data to the list
                    data.append({
                        'file name': pdf,
                        'page number': page_num + 1,
                        'page section': i+1,
                        'content': page_part,
                        'tokens': len(word_tokenize(page_part))
                    })
    return pd.DataFrame(data)
```

**Figure S30.** Demonstration of the interaction with ChatGPT (GPT-4) to generate a Python function capable of parsing academic papers. The function extracts text from multiple PDF files, dividing each page into four sections and recording their respective page and section numbers.

It should be noted that while the majority of the time ChatGPT generates code with fluent and seemingly correct syntax, the code might not always execute as expected. In such instances, ChatGPT can be interactively tasked with debugging by providing the error messages received when executing the code.

---

**ChatGPT Code Synthesis**

*---Continued from previous conversation shown in Figure S30---*

**Human:** DeprecationError: reader.getNumPages is deprecated and was removed in PyPDF2 3.0.0. Use len(reader.pages) instead.

---

**ChatGPT (GPT-4):**
Apologies for the confusion, and thank you for pointing out the change.

You're right, in the recent versions of PyPDF2, getNumPages has been replaced with len(reader.pages). Here's the corrected code:
…
```
    for page_num in range(len(pdf_reader.pages)):  # Iterate over all the pages
```
…

---

**Human:** DeprecationError: reader.getPage(pageNumber) is deprecated and was removed in PyPDF2 3.0.0. Use reader.pages[page_number] instead.

---

**ChatGPT (GPT-4):**
Apologies for the confusion, and thank you for the correction. Here's the updated code:
…
```
    page = pdf_reader.pages[page_num]  # Extract the text from the current page
```
…

---

**Figure S31.** Illustration of the debugging process, where human simply copies and pastes the error message and ChatGPT (GPT-4) assists in revising code that had issues during execution. Please note that for brevity, only the revised sections of the code are shown.

Furthermore, users can direct the code implementation using narrative language to articulate the request. This feature allows the user to specify the use of targeted models or functions.

> **ChatGPT Code Synthesis**
>
> *---Continued from previous conversation shown in Figure S31---*
>
> **Human:** Revise the code to use OpenAI's tiktoken "cl100k_base" to count for tokens
>
> ---
>
> **ChatGPT (GPT-4):**
> To count tokens using the OpenAI's tiktoken library, you first need to install it via pip:
>
> ```
> pip install tiktoken
> ```
>
> *Here's how you can modify the function to use tiktoken:*
>
> ```python
> import PyPDF2
> import pandas as pd
> import tiktoken
>
> def count_tokens(text):
>     encoding = tiktoken.get_encoding("cl100k_base")
>     num_tokens = len(encoding.encode(text))
>     return num_tokens
>
> def get_txt_from_pdf(pdf_files):
>     data = [] # Create an empty list to store the data
>     for pdf in pdf_files: # Iterate over the PDF
>         with open(pdf, 'rb') as pdf_content: # Fetch the PDF content from the pdf
>             pdf_reader = PyPDF2.PdfReader(pdf_content) # Create a PDF reader object
>             for page_num in range(len(pdf_reader.pages)): # Iterate over all the pages
>                 page = pdf_reader.pages[page_num] # Extract the text from the current page
>                 page_text = page.extract_text()
>                 words = page_text.split() # Split the page text into individual words
>                 page_text_join = ' '.join(words) # Join the words
>                 page_len = len(page_text_join)
>                 div_len = page_len // 4 # Divide the page into 4 parts
>                 page_parts = [page_text_join[i*div_len:(i+1)*div_len] for i in range(4)]
>                 for i, page_part in enumerate(page_parts): # Append the data to the list
>                     data.append({
>                         'file name': pdf,
>                         'page number': page_num + 1,
>                         'page section': i+1,
>                         'content': page_part,
>                         'tokens': count_tokens(page_part)
>                     })
>     return pd.DataFrame(data)
> ```

**Figure S32.** Illustration of how ChatGPT (GPT-4) can be directed to modify code according to specific user requests. The fully revised and correct code after interactive refinement was also shown, capable of proficiently parsing academic papers.

Text Encoding

Upon obtaining text segments from a research paper and storing them in a DataFrame, the execution of Process 3 requires OpenAI embeddings of each segments. This can be achieved by instructing ChatGPT about the DataFrame's contents and providing it with an exemplary code snippet derived from OpenAI's official documentation, which illustrates how to engage the OpenAI model for embedding conversion.

> **ChatGPT Code Synthesis**
>
> **Human:** Task: Write a function called add_embedding
>
> You need to implement a Python function called add_embedding that takes a dataframe as input and returns a new dataframe with an additional column called 'embedding'. The 'embedding' column will contain the embeddings generated from the strings in the 'content' column.
>
> To generate the embeddings, you can use the following code:
>
> ```python
> response = openai.Embedding.create(input="Your text string goes here",
>                                    model="text-embedding-ada-002")
> embeddings = response['data'][0]['embedding']
> ```
>
> ---
>
> **ChatGPT (GPT-4):**
>
> ```python
> import pandas as pd
> import openai
>
> # Make sure to set your OpenAI key
> openai.api_key = 'your-api-key'
>
> def add_embedding(df):
>     # Define a function to get embeddings
>     def get_embedding(text):
>         response = openai.Embedding.create(
>             input=text,
>             model="text-embedding-ada-002"
>         )
>         return response['data'][0]['embedding']
>
>     # Apply the function to the 'content' column
>     df['embedding'] = df['content'].apply(get_embedding)
>
>     return df
> ```

**Figure S33.** Demonstration of GPT-4's auto-generation of a function to append embeddings to all research paper segments in a DataFrame. This process is enabled by describing the Dataframe's contents and demonstrating the use of an example code from OpenAI's official documentation. Please note that the placeholder 'your-api-key' should be replaced with the user's actual OpenAI API key.

To circumvent the need for recurring embedding conversions for the same paper pool, the embeddings can be stored locally for subsequent reloading. This process can be facilitated by informing ChatGPT about the requirement and requesting it to generate an additional code that allows for the conversion between the DataFrame and locally saved CSV files storing the embeddings. Moreover, the function should be adapted to ensure that if the DataFrame already contains embeddings, no further actions are taken.

---

**ChatGPT Code Synthesis**

*---Continued from previous conversation shown in Figure S33---*

**Human:** Please revised the code and add a function to store a dataframe with embeddings locally as a CSV file. Also, write a function to convert the CSV file back to a dataframe. The next time the dataframe is loaded to add embeddings function, first check if it already has a column called 'embedding'. If the column already exists, there is no need to add it again.

---

**ChatGPT (GPT-4):**

```python
import pandas as pd
import openai

# Make sure to set your OpenAI key
openai.api_key = 'your-api-key'

def add_embedding(df):
    # If the dataframe already has 'embedding' column, return as is
    if 'embedding' in df.columns:
        return df

    # Define a function to get embeddings
    def get_embedding(text):
        response = openai.Embedding.create(
            input=text,
            model="text-embedding-ada-002"
        )
        return response['data'][0]['embedding']

    # Apply the function to the 'content' column
    df['embedding'] = df['content'].apply(get_embedding)

    return df

def save_df_to_csv(df, filename):
    df.to_csv(filename, index=False)

def load_df_from_csv(filename):
    df = pd.read_csv(filename)
    return df
```

---

**Figure S34.** Depiction of code modifications and implementations by ChatGPT (GPT-4) to append functions for the storage and loading of embeddings.

Nonetheless, the manner in which ChatGPT stores and loads embeddings is rudimentary, leading to type inconsistency issues when attempting to utilize reloaded embeddings from the CSV files. This arises from the storage of embeddings as text in the CSV files, and upon reloading, ChatGPT, unaware of the embeddings' type, simply reloads the CSV files. This issue can be readily resolved by informing ChatGPT about the accurate form of embeddings and requesting it to revise the function accordingly.

---

**ChatGPT Code Synthesis**

*---Continued from previous conversation shown in Figure S34---*

**Human:** I noticed that df['embedding'] is a string instead of a list of floats after loading it from a CSV. Please modify the code to rectify this issue.

---

**ChatGPT (GPT-4):**
You're right. When data is saved to a CSV file, the list of floats in the 'embedding' column is converted into a string representation of the list. When this data is read back into a DataFrame, the 'embedding' column consists of strings that need to be converted back into lists of floats.

We can achieve this by modifying the load_df_from_csv function to include conversion of the 'embedding' column back to lists of floats using the ast.literal_eval function.

Here's the modified code:

```python
def load_df_from_csv(filename):
    df = pd.read_csv(filename)

    # Convert the string representation of list back to list of floats
    df['embedding'] = df['embedding'].apply(ast.literal_eval)

    return df
```

---

**Figure S35.** Demonstration of code modifications by ChatGPT (GPT-4) following the user's issue description. It is noteworthy that the user is not required to suggest a solution for ChatGPT to write; instead, ChatGPT can deduce the issue based on the situation description provided by the user, explain the underlying problem, and modify the code accordingly.

Process 3 and Similarity Search

After having a dataframe with segmented paper content and their respective embeddings, we can proceed to develop Process 3. The primary objective of Process 3 lies in search and filtering operations, as illustrated in the conversation below.

> **ChatGPT Code Synthesis**
>
> **Human:** Your task is to write Python code to construct a function called Model_3. The function takes a dataframe called df as input, which contains columns named 'content' and 'embeddings'. Your job is to first convert a fixed prompt, which is a text string used for search purposes, into OpenAI embeddings using the following code template:
>
> ```
> response = openai.Embedding.create(input="Your text string goes here",
>                                    model="text-embedding-ada-002")
> embeddings = response['data'][0]['embedding']
> ```
>
> Next, you need to compare the converted embedding of the fixed prompt with all embeddings in the dataframe. Calculate the cosine similarity for each comparison and select the top 10 neighbors. Drop the remaining rows to obtain the filtered dataframe. Finally, you should pass the filtered dataframe to a new function called Model_2 for further processing. To calculate the cosine similarity and select the top neighbors, you can write two additional functions: add_similarity and select_top_neighbors.
>
> ---
>
> **ChatGPT (GPT-4):**
> ```python
> import openai
> import pandas as pd
> import numpy as np
> from sklearn.metrics.pairwise import cosine_similarity
>
> def add_similarity(df, given_embedding):
>     def calculate_similarity(row_embedding):
>         return cosine_similarity([row_embedding], [given_embedding])[0][0]
>     df['similarity'] = df['embeddings'].apply(calculate_similarity)
>     return df
>
> def select_top_neighbors(df):
>     df = df.sort_values('similarity', ascending=False)
>     top_neighbors = df.head(10)
>     return top_neighbors
>
> def Model_3(df):
>     response = openai.Embedding.create(
>         input="Your fixed text string goes here",model="text-embedding-ada-002")
>     given_embedding = response['data'][0]['embedding']
>     df = add_similarity(df, given_embedding)
>     filtered_df = select_top_neighbors(df)
>     return Model_2(filtered_df) # Assuming Model_2 is previously defined
> ```

**Figure S36.** Illustration of Process 3 configuration, guided by code generated by ChatGPT (GPT-4). For information regarding the fixed prompt, please refer to Figure S25.

The fine tuning of the functions can be achieved by providing more detailed instructions to ChatGPT (GPT-4) and requesting its revision. For instance, we aimed to filter less relevant content for each paper, therefore necessitating the selection of the top 10 sections based on each unique file name.

> **ChatGPT Code Synthesis**
>
> *---Continued from previous conversation shown in Figure S36---*
>
> **Human:** There is another column called 'file name'. Please modify your code accordingly to select the top neighbors for each different file name.
>
> **ChatGPT (GPT-4):** If you want to select the top neighbors for each different file name, you can modify the select_top_neighbors function to group by the 'file name' column before applying the sorting and selection operation. Here is the updated code:
>
> ```python
> def select_top_neighbors(df):
>     df = df.sort_values('similarity', ascending=False)
>     top_neighbors = df.head(10)
>     return top_neighbors
> ```

**Figure S37.** Example of function optimization according to user requirements, facilitated by ChatGPT (GPT-4).

Process 2 and Classification

Upon setting up Process 3, we can shift our attention to the configuration of Process 2. The approach mirrors the workflow of Process 3: the user provides a detailed description of the function's name, input, and expected output to ChatGPT (GPT-4). It's crucial to inform ChatGPT (GPT-4) about the standard Python code needed to call the ChatGPT API, which can be obtained from OpenAI's official documentation.

---

**ChatGPT Code Synthesis**

**Human:** Your task involves writing Python code to construct a function named Model_2. This function should accept a dataframe, denoted as 'df', which includes a column labeled 'content'. Start by constructing a prompt that combines "given few-shot prompt" with 'content', followed by:

"""*Question: Does the section contain a comprehensive MOF synthesis with explicit reactant quantities or solvent volumes? Answer:* """

Then, you will pass this prompt to the ChatGPT API using the following code:

```
response = openai.ChatCompletion.create(model="gpt-3.5-turbo",messages=[{"role": "user", "content": "<your_prompt>"}])
Answer = response.choices[0].message.content
```

You will assign the first choice of the response content to a variable, 'Answer' The answers will then be appended to the dataframe as a new column, named 'classification'.
Next, add a mask to remove rows with the value of 'No' in the 'classification' column. Finally, pass the resultant dataframe to a function named model_1 for further processing.

---

**ChatGPT (GPT-4):**

```python
import openai
import pandas as pd

def Model_2(df):
    df['classification'] = ''  # Add a new column 'classification' to the dataframe
    for idx, row in df.iterrows():# Iterate over each row of the dataframe
        prompt = f"given few-shot prompt {row['content']}\n\n\
            Question: Does the section contain a comprehensive MOF synthesis with explicit
            reactant quantities or solvent volumes?\n\Answer: " # Construct the prompt
        response = openai.ChatCompletion.create(
            model="gpt-3.5-turbo",messages=[{"role": "user", "content": prompt}])
        df.loc[idx, 'classification'] = response.choices[0].message.content

    # Filter the dataframe to remove rows with 'No' in the 'classification' column
    df = df[df['classification'] != 'No']

    # Pass the dataframe to the model_1 function for further processing
    return model_1(df)
```

**Figure S38.** Depiction of Process 2 setup, aided by code generated by ChatGPT (GPT-4). For details concerning the fixed few-shot prompt, please refer to Figure S24.

Process 1 and Summarization

Similarly, Process 1 can be constructed following the same approach. We note that two sources can provide the input dataframe for Process 1. The first source is the dataframe generated by Process 2 after classification, which contains only the text segments classified as synthesis paragraphs, each having a few hundred tokens. The second source is a dataframe composed of longer, human-selected synthesis paragraphs for each paper, with lengths varying from a few hundred to many thousand tokens. Given the 4096-token limitation, we must consider the total token count of synthesis paragraph context plus the fixed token. Hence, we established a cutoff at 3000 tokens to segment the input content.

> **ChatGPT Code Synthesis**
>
> **Human:** Your task is to write Python code to construct a function named Model_1. This function should accept a dataframe, df, which contains a column named 'content'. Initially, apply the count_tokens() function to determine the number of tokens for each row in 'content'. If the token count exceeds 3000 max_tokens, split the content at each 3000-token boundary.
> Then, for each string in the 'content' column, concatenate "Context: " + 'content' + "Given tubulation prompt". Following this, pass the resulting prompt to the ChatGPT API using the code:
>
> *response = openai.ChatCompletion.create(model="gpt-3.5-turbo",messages=[{"role": "user", "content": "<your_prompt>"}])*
> *Answer = response.choices[0].message.content*
>
> Assign the first choice of the response content to a variable, 'Answer'. Append the 'Answer' to the dataframe as a new column, named 'summarized'. Finally, return the updated dataframe.
>
> ---
>
> **ChatGPT (GPT-4):**
> ```python
> def Model_1(df):
>     max_tokens = 3000
>     df['tokens'] = df['content'].apply(count_tokens)
>     summarized = []
>     for i, row in df.iterrows():
>         content = row['content']
>         token_count = row['tokens']
>         if token_count > max_tokens: # Split the content into chunks of max_tokens
>             content_chunks = [' '.join(content.split()[i:i+max_tokens]) for i in range(0, token_count, max_tokens)]
>         else:
>             content_chunks = [content]
>         for chunk in content_chunks:
>             prompt = f"Context: {chunk} Given tubulation prompt"
>             response = openai.ChatCompletion.create(model="gpt-3.5-turbo", messages=[{"role": "user", "content": prompt}])
>             answer = response['choices'][0]['message']['content']
>             summarized.append(answer)
>     df['summarized'] = pd.Series(summarized)
>     return df
> ```

**Figure S39.** Demonstration of Process 1 establishment, supported by code produced by ChatGPT (GPT-4). For specifics of the fixed prompt, please refer to Figure S22 and Figure S22.

Text Parsing

As previously illustrated in Figure S26, responses from the ChatGPT API present tabular information as text strings filled with '|' symbols, rendering them unreadable. To address this, we enlisted ChatGPT to write a Python function to extract this data, restructure it into a clear, readable format, and convert it into a dataframe. This was achieved by providing ChatGPT with an exemplary output from the 'summarized' column in Process 1, which stores the API responses, and instructs it to reassemble the table into a dataframe and compile all the synthesis parameters.

**ChatGPT Code Synthesis**

**Human:** Your task is to write Python code to construct a function named 'tabulate_condition'. This function should accept a dataframe, 'df', which contains a column named 'summarized'. This column consists of text strings of tabulated data that could take the following format:
```
| Header 1 | Header 2 | Header 3 | …… | Header 11 |
|---------------|-------------|-------------|……|-------------|
| Data 1-1 | Data 1-2 | Data 1-3 | …… | Data 1-11 |
| Data 2-1 | Data 2-2 | Data 2-3 | …… | Data 2-11 |
```
Here, the "……" represent repeating units. Please note that the text strings can have one, two, or more rows of data, as shown in the exemplary input above. In rare cases, they may have no rows of data and consequently, no header. In such instances, no "|" symbol will be found in the text string. These text strings should be dropped.

The function should iterate over all cells in the 'summarized' column, remove the header, extract the data, merge all the data, and then reinsert a single header. The data will always consist of 11 columns: 'compound name', 'metal source', 'metal amount', 'linker', 'linker amount', 'modulator', 'modulator amount or volume', 'solvent', 'solvent volume', 'reaction temperature', and 'reaction time'. The final dataframe returned by the function will consist of these 11 columns.

---

**ChatGPT (GPT-4):**
```python
def tabulate_condition(df):
    columns = ['compound name', 'metal source', 'metal amount', 'linker', 'linker amount', 'modulator', 'modulator amount or volume', 'solvent', 'solvent volume', 'reaction temperature', 'reaction time'] # Initialize an empty dataframe with the desired columns
    result_df = pd.DataFrame(columns=columns)
    for index, row in df.iterrows():
        summarized = row['summarized']
        if "|" in summarized: # Check if the "|" symbol is present in the text string
            # Split the text string into lines and remove the header
            lines = summarized.strip().split("\n")[2:]
            # Iterate through the lines and extract the data
            for line in lines:
                data = [x.strip() for x in line.split("|")[1:-1]]
                if len(data) == len(columns):
                    result_df = result_df.append(pd.DataFrame([data], columns=columns), ignore_index=True)
    return result_df
```

**Figure S40.** Demonstration of instruction for ChatGPT (GPT-4) to generate a Python function for converting string-based tabular data from the 'summarized' column of a dataframe into a structured Dataframe format.

Moreover, given the exemplary input already provided, we further directed ChatGPT to generate code to create a sample dataframe. This dataframe was used to call and test the function that was written by ChatGPT, enabling us to evaluate the function's output and ensure its accuracy and effectiveness.

---

**ChatGPT Code Synthesis**

*---Continued from previous conversation shown in Figure S40---*

**Human:** Please help me write code to test this function tabulate_condition.

---

**ChatGPT (GPT-4):**
Sure! Here is an example of how you can test this function:

```
# Test the function with a sample dataframe
sample_data = {'summarized': [
    """
    | compound name | metal source | metal amount | linker | linker amount | modulator | modulator amount or volume | solvent | solvent volume | reaction temperature | reaction time |
    |---------------|--------------|--------------|--------|---------------|-----------|----------------------------|---------|----------------|----------------------|---------------|
    | Data 1-1      | Data 1-2     | Data 1-3     | Data 1-4 | Data 1-5    | Data 1-6  | Data 1-7                   | Data 1-8 | Data 1-9      | Data 1-10            | Data 1-11     |
    """,
    """Other text""",
    """
    | compound name | metal source | metal amount | linker | linker amount | modulator | modulator amount or volume | solvent | solvent volume | reaction temperature | reaction time |
    |---------------|--------------|--------------|--------|---------------|-----------|----------------------------|---------|----------------|----------------------|---------------|
    | Data 5-1      | Data 5-2     | Data 5-3     | Data 5-4 | Data 5-5    | Data 5-6  | Data 5-7                   | Data 5-8 | Data 5-9      | Data 5-10            | Data 5-11     |
    """
]}

sample_df = pd.DataFrame(sample_data)
result = tabulate_condition(sample_df)

print(result)
```

**Output:**

| compound name | metal source | metal amount | linker | linker amount | modulator | modulator amount or volume | solvent | solvent volume | reaction temperature | reaction time |
|---|---|---|---|---|---|---|---|---|---|---|
| Data 1-1 | Data 1-2 | Data 1-3 | Data 1-4 | Data 1-5 | Data 1-6 | Data 1-7 | Data 1-8 | Data 1-9 | Data 1-10 | Data 1-11 |
| Data 5-1 | Data 5-2 | Data 5-3 | Data 5-4 | Data 5-5 | Data 5-6 | Data 5-7 | Data 5-8 | Data 5-9 | Data 5-10 | Data 5-11 |

**Figure S41.** Demonstration of testing the function generated by ChatGPT (GPT-4) through instructing ChatGPT to build up a sample dataframe.

## Section S4. ChatGPT-Assisted Chemistry Data Processing Cookbook

Reaction Time and Temperature Conversion

To handle basic entity resolution and data unification tasks, such as unit and format unification, we can either import the data into the ChatGPT API or, more conveniently, copy and paste the data into the web-based ChatGPT platform. Each specific task requires a preconfigured prompt, developed based on the three fundamental principles discussed in Supplementary Information Section S2. The prompts are refined iteratively, as demonstrated in Figures S13−S19, to enhance performance. The final versions of these prompts are displayed in the top box of each figure below. For easier proofreading, we instructed ChatGPT to include the original input in the first column of the output. It should be noted that the examples in this section were executed by ChatGPT based on the GPT-4 model. While both the GPT-4 and GPT-3.5 models perform adequately for such simple conversions, GPT-4 was chosen due to its superior handling of empty cells when processing large data inputs. Based on token limitations, each ChatGPT operation can process between 50 to 100 data points, typically completing in about 10 seconds.

**ChatGPT Data Processing**

**Prompt:** Please convert the following reaction times into hours and generate a two-column table. The first column should contain the original text input, and the second column should contain the converted hours without units. The input will consist of various formats like days, weeks, hours, minutes, and seconds, with some entries possibly containing multiple time values. In cases with multiple time values, please use the longest time value. If "N/A," a blank space, or irrelevant information with no indication of the time value is present, treat it as N/A and convert it to the text string "N/A. Treat "overnight" as 12 hours.

**Input:**

```
"""
24h
18 hours
1 week
150 sec
20 min
one day
overnight
3 days
"""
```

**Output:**

| Original Input | Converted Hours |
|---|---|
| 24h | 24 |
| 18 hours | 18 |
| 1 week | 168 |
| 150 sec | 0.0417 |
| 20 min | 0.3333 |
| one day | 24 |
| overnight | 12 |
| 3 days | 72 |

**Figure S42.** Illustration of a custom prompt instructing ChatGPT (GPT-4) to automate time conversion. This process can be performed iteratively until all data in the dataset is converted. Since the conversion uniformizes the unit to hours, we instructed ChatGPT to exclude the unit (hour) in the second column for simplicity.

Similarly, reaction temperatures can also be standardized using a pre-configured prompt. It's worth noting that ChatGPT could be guided to write Python code, Excel functions, or even use NLP methods for these conversions. However, the most remarkable feature of ChatGPT lies in its ability to accommodate a wide range of situations that would otherwise require extensive rule-based coding. This includes variations in phrases with equivalent meanings such as "overnight," "for one night," and "Over the course of a night," as well as special characters and incorrect representations of units that are easily recognized and understood by ChatGPT with a single sentence instruction.

---

**ChatGPT Data Processing**

**Prompt:** Please convert the given reaction temperatures into numbers without units in °C and generate a two-column table, where the first column shows the original temperature and the second column displays the converted temperature in °C without units. The input may include various units such as °C, K, and possibly incorrect representations of °C (for example, represented with any special character or alphanumeric character before 'C'). If "N/A," a blank space, new line, or irrelevant information with no indication of the temperature value is present, treat it as N/A and convert it to the text string "N/A". In cases with multiple temperature values or ramping instructions, use the highest temperature value and disregard ramping time. When encountering temperatures in Kelvin (K), convert them to Celsius (°C) before adding to the table. Consider "room temperature" or "rt" or "ambient" as 25 °C. Exclude "°C" from the second column of the table. Ensure that the output is formatted as a two-column table.

---

**Input:**

```
"""
120 °C
room temperature
ambient
383 K
298 K
120 8C
160 C
ramp of 10 ??C per minute to
120 ??C
160 oC
reflux
reflux at 120 C
microwave (700W)
100 uC
85??C
65C
"""
```

**Output:**

| Original Temperature | Converted Temperature (°C) |
|---|---|
| 120 °C | 120 |
| room temperature | 25 |
| ambient | 25 |
| 383 K | 110 |
| 298 K | 25 |
| 120 8C | 120 |
| 160 C | 160 |
| ramp of 10 ??C per minute to 120 ??C | 120 |
| 160 oC | 160 |
| reflux | N/A |
| reflux at 120 C | 120 |
| microwave (700W) | N/A |
| 100 uC | 100 |
| 85??C | 85 |
| 65C | 65 |

**Figure S43.** Demonstration of a custom prompt instructing ChatGPT (GPT-4) to automate temperature conversion.

Chemical Formula Conversion and Molecular Weight Calculation

Compared to unifying reaction temperature and reaction time data, the process of standardizing compound notation often necessitates intricate rule-based codes, making it a time-consuming task. Special characters can further complicate this process. However, the implementation of ChatGPT, equipped with carefully designed prompts, significantly simplifies this task. As depicted in Figure S42, ChatGPT successfully converts a variety of chemical formulae representations into standard notation in a matter of seconds.

**ChatGPT Data Processing**

**Prompt:** Given a list of metal salts compound names in various formats, clean, standardize, and unify these names by identifying the correct chemical formula for each compound and converting them into their standard notation (e.g., AlCl3·6H2O, ZnCl4). Then, create a two-column table to display the results, with the left column containing the original text and the right column showing the converted text in the standard format. The input may include compound names written in different ways, such as using common names, abbreviations, or Roman numerals. If "N/A," a new line, or only a metal (cation) is mentioned without any associated anion, treat it as N/A and convert it to the text string "N/A".

**Input:**

```
"""
Zirconium oxynitrate hydrate
al2so43 18h2o
Zn(Ac)2 ·2H2O
ZnCO3??2Zn(OH)2??H2O
ZnNO3•6(H2O)
stock solution
iron chloride anhydrous
zn(no3)2??4h2o
In(NO3)3??5H2O and
Zn(NO3)2??4H2O
Manganese(II) chloride,
Magnesium chloride
cupric sulfate pentahydrate
"""
```

**Output:**

| Original Text | Converted Text |
|---|---|
| Zirconium oxynitrate hydrate | Zr(NO3)4·nH2O |
| al2so43 18h2o | Al2(SO4)3·18H2O |
| Zn(Ac)2 ·2H2O | Zn(C2H3O2)2·2H2O |
| ZnCO3??2Zn(OH)2??H2O | Zn5(CO3)2(OH)6 |
| ZnNO3•6(H2O) | Zn(NO3)2·6H2O |
| stock solution | N/A |
| iron chloride anhydrous | FeCl3 |
| zn(no3)2??4h2o | Zn(NO3)2·4H2O |
| In(NO3)3??5H2O and Zn(NO3)2??4H2O | In(NO3)3·5H2O, Zn(NO3)2·4H2O |
| Manganese(II) chloride, Magnesium chloride | MnCl2, MgCl2 |
| cupric sulfate pentahydrate | CuSO4·5H2O |

**Figure S44.** Utilization of ChatGPT (GPT-4) to standardize the notation of metal compounds.

The molecular weight of compounds can be calculated in a similar manner. With the standard notation established, it is a matter of extending the prompt to request ChatGPT to generate Python code to perform these calculations.

> **ChatGPT Code Synthesis**
>
> *---Continued from previous data processing shown in Figure S44 ---*
>
> **Human:** Could you write a Python script that calculates the molecular weights of given compounds and adds this information as a new column in a DataFrame? If a compound name includes an undetermined number of water molecules (denoted as 'nH2O'), please exclude this portion during the calculation. Additionally, if multiple compounds are listed in a single entry, please separate them with a semicolon (';').
>
> ---
>
> **ChatGPT (GPT-4):**
> ```python
> !pip install mendeleev pandas
>
> import re
> import pandas as pd
> from mendeleev import element
>
> def get_molar_mass(compound):
>     # Ignore any undefined number of water molecules
>     compound = compound.replace('nH2O', '')
>     elements = re.findall(r'([A-Z][a-z]*)(\d*)', compound)
>     mass = 0
>     for e in elements:
>         symbol, quantity = e
>         quantity = int(quantity) if quantity else 1
>         mass += element(symbol).atomic_weight * quantity
>     return mass
>
> data = {
>     'Original Compound': ['Zirconium oxynitrate hydrate', 'al2so43 18h2o', 'Zn(Ac)2·2H2O',
> 'ZnCO3??2Zn(OH)2??H2O', 'ZnNO3•6(H2O)', 'stock solution', 'iron chloride anhydrous',
> 'zn(no3)2??4h2o', 'In(NO3)3??5H2O and Zn(NO3)2??4H2O', 'Manganese(II) chloride, Magnesium
> chloride','cupric sulfate'],
>     'Standard Notation': ['Zr(NO3)4·nH2O', 'Al2(SO4)3·18H2O', 'Zn(C2H3O2)2·2H2O',
> 'Zn5(CO3)2(OH)6', 'Zn(NO3)2·6H2O', 'N/A', 'FeCl3', 'Zn(NO3)2·4H2O', 'In(NO3)3·5H2O;
> Zn(NO3)2·4H2O', 'MnCl2; MgCl2', 'CuSO4·5H2O']}
>
> df = pd.DataFrame(data)
>
> # Handling multiple compounds and calculating molar mass
> df['Molar Mass'] = df['Standard Notation'].apply(lambda x: ';
> '.join([str(get_molar_mass(i.strip())) if i != 'N/A' else 'N/A' for i in x.split(';')]))
>
> print(result)
> ```

**Figure S45.** Instruction of ChatGPT (GPT-4) to generate Python code to calculate the molecular weight of given compounds. The conversation was continued from the context set up in Figure S44, and ChatGPT was able to know the format of the standard notation.

After executing the ChatGPT-provided code in a Python environment, the resulting DataFrame is presented in Table S1. The code efficiently calculates the molecular weight of the given compounds in standard notation. Subsequently, we instructed ChatGPT to modify the code to create a function, thereby facilitating batch molecular weight calculations for a DataFrame containing standard notations.

**Table S1.** Molecular weights of various compounds calculated using the code provided by ChatGPT.

| Original Compound | Standard Notation | Molar Mass |
|---|---|---|
| Zirconium oxynitrate hydrate | Zr(NO3)4·nH2O | 153.228 |
| al2so43 18h2o | Al2(SO4)3·18H2O | 168.034 |
| Zn(Ac)2 ·2H2O | Zn(C2H3O2)2·2H2O | 142.439 |
| ZnCO3??2Zn(OH)2??H2O | Zn5(CO3)2(OH)6 | 403.915 |
| ZnNO3•6(H2O) | Zn(NO3)2·6H2O | 145.399 |
| stock solution | N/A | N/A |
| iron chloride anhydrous | FeCl3 | 162.195 |
| zn(no3)2??4h2o | Zn(NO3)2·4H2O | 145.399 |
| In(NO3)3??5H2O and Zn(NO3)2??4H2O | In(NO3)3·5H2O, Zn(NO3)2·4H2O | 194.837; 145.399 |
| Manganese(II) chloride, Magnesium chloride | MnCl2, MgCl2 | 125.838; 95.205 |
| cupric sulfate | CuSO4·5H2O | 177.617 |

**ChatGPT Code Synthesis**

*---Continued from previous conversation shown in Figure S45---*

**Human:** I have a DataFrame that includes a 'Standard Notation' column. Could you provide a Python script that uses this DataFrame as input and adds an additional 'Molar Mass' column to it? Each row of the 'Molar Mass' column should correspond to the molecular weight of the compound in the same row of the 'Standard Notation' column.

**ChatGPT (GPT-4):**
```python
import re
from mendeleev import element
import pandas as pd

def add_molar_mass_column(df):
    # Assuming 'df' is your DataFrame and it has a column named 'Standard Notation'
    df['Molar Mass'] = df['Standard Notation'].apply(lambda x: '; '.join([str(get_molar_mass(i.strip())) if i != 'N/A' else 'N/A' for i in x.split(';')]))
    return df
```

**Figure S46.** Instruction of ChatGPT (GPT-4) to modify the Python code for batch processing of molecular weight calculations for compounds listed in a DataFrame.
49

Reaction Outcome Classification

We demonstrate here that ChatGPT can assist in classifying synthesis outcomes by crystalline state. As shown in Figure S21, direct synthesis results such as "colorless crystals" "red solid", and "microcrystalline powder" can be extracted and summarized. Now, we engage ChatGPT to categorize these descriptions, stored in a Dataframe. We note that in literature we select, there are no synthesis condition that yield amorphous, but rather only crystalline MOFs, with difference on whether or not they are single crystals or just microcrystalline powder or solid. Consequently, this MOF crystalline state classification will have two categories, one is "single-crystal" (SC), which includes all cases where the author clearly says that crystals form, and the other is "polycrystalline" (P), where the author mentions solid, powder, or did not mention the single-crystal.

**ChatGPT Data Processing**

**Prompt:** Please classify the provided descriptions of materials based on their crystalline structure into either single-crystal (SC) or polycrystalline (P). The descriptions should be parsed and classified as per the following rules: If the description includes words like "powder", "solid", "microcrystalline", "crystalline product", or "N/A", classify it as "P" (polycrystalline). If the description includes words like "single-crystal" or "crystal", classify it as "SC" (single-crystal).
Please present the results in a two-column table. The first column should contain the original description, and the second column should display the classification result "SC" or "P". Ensure that each row in the table represents one description and its corresponding classification.

**Input:**
```
"""
red solid
Green block crystals
colorless rodlike crystals
pink precipitate
yellow crystalline product
purple hexagonal platelike crystals
large yellow crystalline product
Colorless rod-shaped crystals
colorless rhombus-shaped block crystals
Off-white powdered product
Tetragonal colorless crystals
Sky-blue powdered product
Yellow needle crystals
fine powdered MOF
dark yellow cubic crystals
white microcrystalline product
Needlelike colorless crystals
"""
```

**Output:**

| Original Description | Classification |
|---|---|
| red solid | P |
| Green block crystals | SC |
| colorless rodlike crystals | SC |
| pink precipitate | P |
| yellow crystalline product | P |
| purple hexagonal platelike crystals | SC |
| large yellow crystalline product | P |
| Colorless rod-shaped crystals | SC |
| colorless rhombus-shaped block crystals | SC |
| Off-white powdered product | P |
| Tetragonal colorless crystals | SC |
| Sky-blue powdered product | P |
| Yellow needle crystals | SC |
| fine powdered MOF | P |
| dark yellow cubic crystals | SC |
| white microcrystalline product | P |
| Needlelike colorless crystals | SC |

**Figure S47.** Utilization of GPT-4 in classifying synthesized MOFs by crystalline morphology.

Unifying Linker Representations

The lack of a consistent representation for organic linkers in MOF synthesis often results in data confusion, with full names, abbreviations, and special characters all being used interchangeably. By standardizing linker representations using a format like the SMILES (Simplified Molecular Input Line Entry System) code, we can facilitate better data exchange. However, this standardization process can be challenging and often requires human intervention, especially for cases involving misspelled names or complex linker structures that have not yet been named. Nevertheless, we have developed a method for instructing ChatGPT to write code that handles the majority of relatively simple cases, accelerating the process while reducing manual effort.

Our approach begins by converting abbreviations to full names. We do this by instructing ChatGPT to implement Process 3, which prompts the system to search for the full name of the linker. The paragraph with the highest similarity score is then presented to ChatGPT, which is then prompted to provide the full name of the abbreviation.

---

**Embedding Demo**

**Prompt Embedding (ada-002):** Provide the full name of linker ({*Input_Linker_Abbreviation*}) or denoted as {*Input_Linker_Abbreviation*} in chemicals, abstract, introduction or experimental section.

**Input #1:**
*H2FDC*

**Output #1:**
…The linker 2,5-furanedicarboxylic acid (H2FDC) is considered to be a biorenewable organic building unit 27,28 and hence considered a promising alternative to terephthalic or isophthalic acid…

**Input #2:**
*H3BTCB*

**Output #2:**
Section 2. Ligand Synthesis. The organic linker H3L1 (4,4',4''-((benzene -1,3,5 -tricarbonyl)tris (azanediyl))tribenzoic acid, H3BTCB ) was synthesized according to the literature procedure……

**Input #3:**
*2-nIM.*

**Output #3:**
Experimental measurements and methods used: General Remarks: 2-nitroimidazole (2-nIM) was purchased from ……

---

**Figure S48.** An illustration of how an embedding model (OpenAI ada-002) is employed to derive the full names of linkers from their abbreviations. Please note that the outputs are abridged for clarity.

Once the full names of the linkers are obtained, we instruct ChatGPT to fetch the corresponding SMILES code and PubChem ID using an API. Given that a single compound may have multiple synonymous names, we consolidate the names sharing the same PubChem ID and SMILES code to create a dictionary. This cleaned dataset replaces all names and abbreviations recorded in the dictionary with their SMILES codes. The compounds that cannot be fetched are manually reviewed. While it is possible for ChatGPT to provide the SMILES code given the standard full name of a compound, this method is limited to common compounds and is at risk of generating inaccurate information. Hence, we recommend using the API to retrieve this information.

---

**ChatGPT Code Synthesis**

**Human:** I have a dataframe with two columns: "full name" and "abbreviation". Please write Python code to create a function. This function will take the dataframe as input and will search for each "full name" value (which are strings) using the PubChem API, PubChemPy, to retrieve its PubChem ID and canonical SMILES code. These two pieces of information will then be appended as two additional columns. If the information cannot be found on PubChem, the function should assign the value "N/A" to the corresponding cells. The function should then merge the rows based on unique PubChem CID values. In other words, it should create a new dataframe with four columns: "full name", "abbreviation", "CID", and "SMILES code". If there are rows with the same CID, they should be merged. The resulting merged row should have full names and abbreviations separated by ";", with just one "CID" and "SMILES" code. The function should not merge items with "N/A" in their CID column. Finally, the function should return this new dataframe.

---

**ChatGPT (GPT-4):**

```python
import pandas as pd
import pubchempy as pcp

def get_pubchem_info(df):
    df['CID'] = 'N/A'
    df['SMILES code'] = 'N/A'
    for i, row in df.iterrows():
        try:
            compound = pcp.get_compounds(row['full name'], 'name')
            if compound:
                df.at[i, 'CID'], df.at[i, 'SMILES code'] = compound[0].cid, compound[0].isomeric_smiles
        except:
            continue
    filtered_df = df[df['CID'] != 'N/A']

    return filtered_df.groupby('CID').agg({
        'full name': ';'.join,
        'abbreviation': ';'.join,
        'SMILES code': 'first'
    }).reset_index()
```

**Figure S49.** Demonstration of using ChatGPT (GPT-4) to generate code that retrieves CID and SMILES codes from PubChem and merges dataframes based on unique compound identifiers.

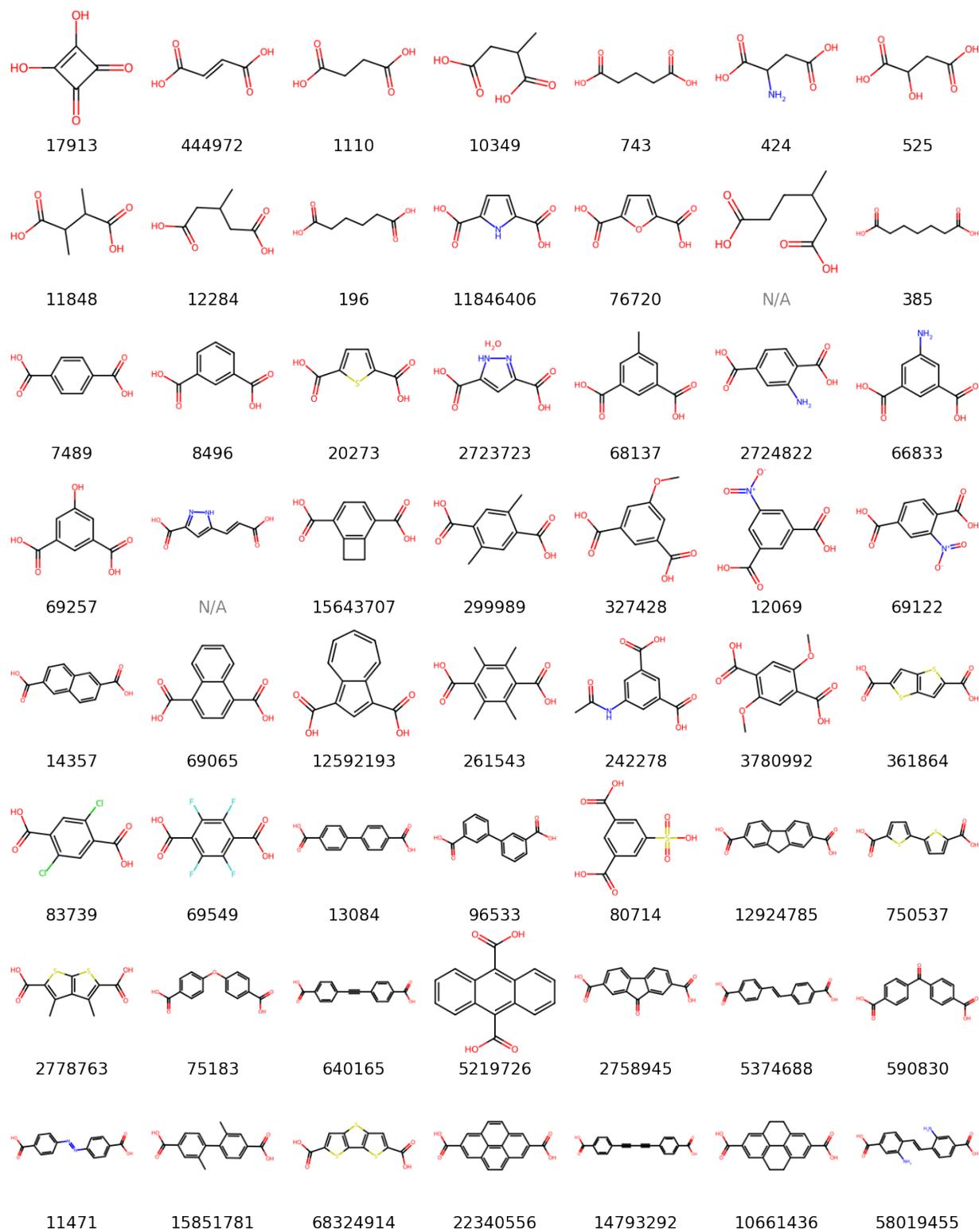

**Figure S50.** MOF linkers obtained from text mining and their CID numbers.

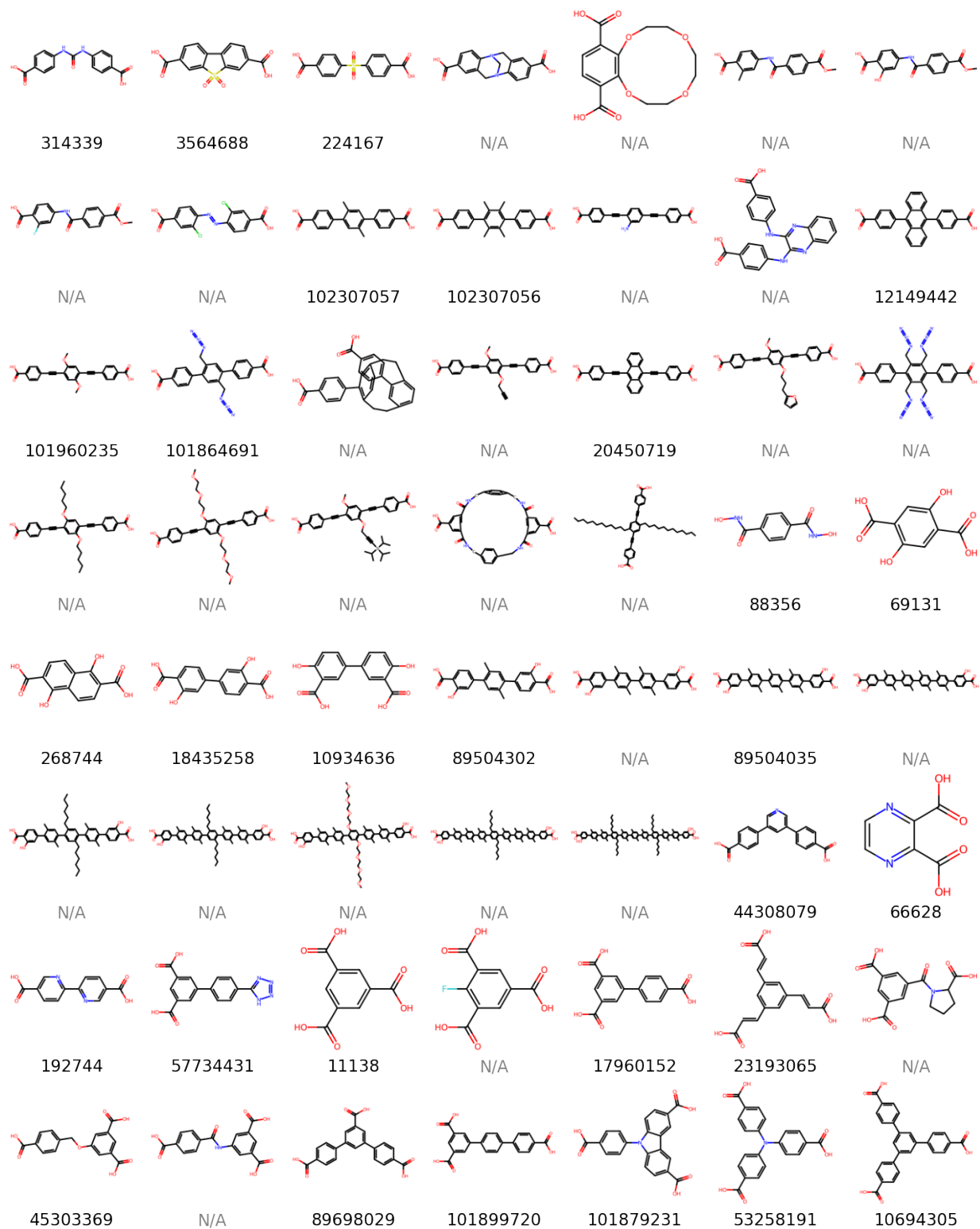

**Figure S51.** MOF linkers obtained from text mining and their CID numbers.

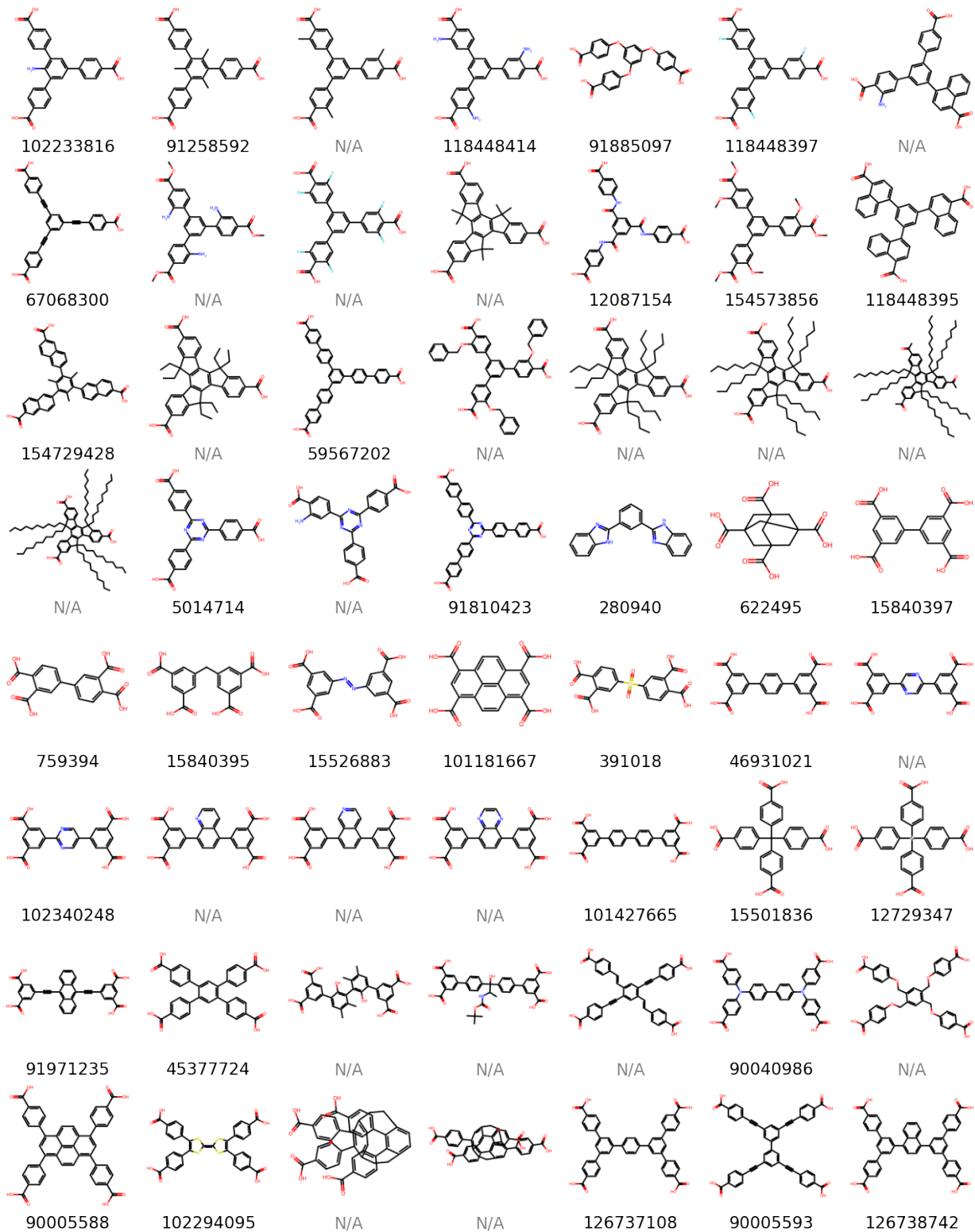

**Figure S52.** MOF linkers obtained from text mining and their CID numbers.

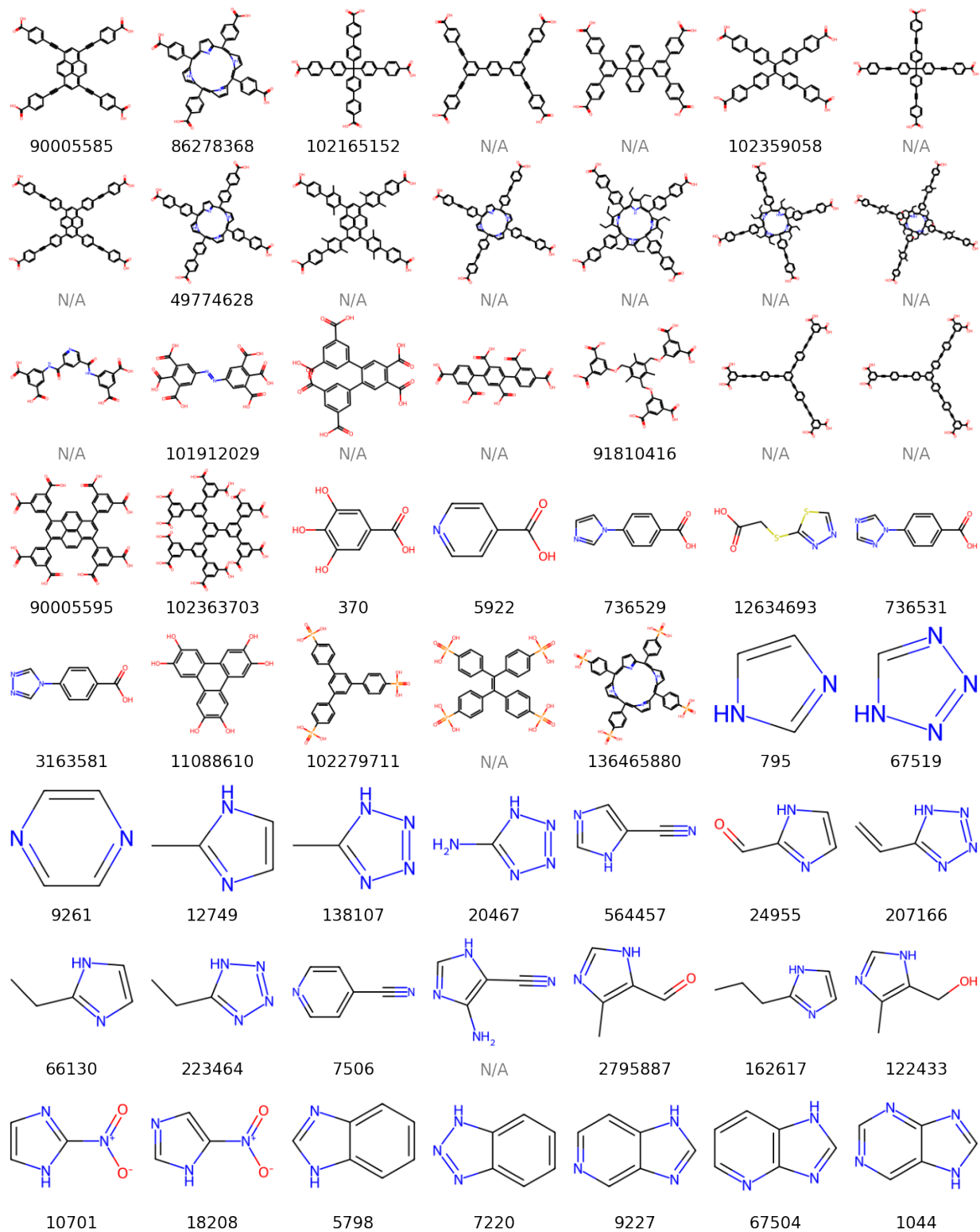

**Figure S53.** MOF linkers obtained from text mining and their CID numbers.

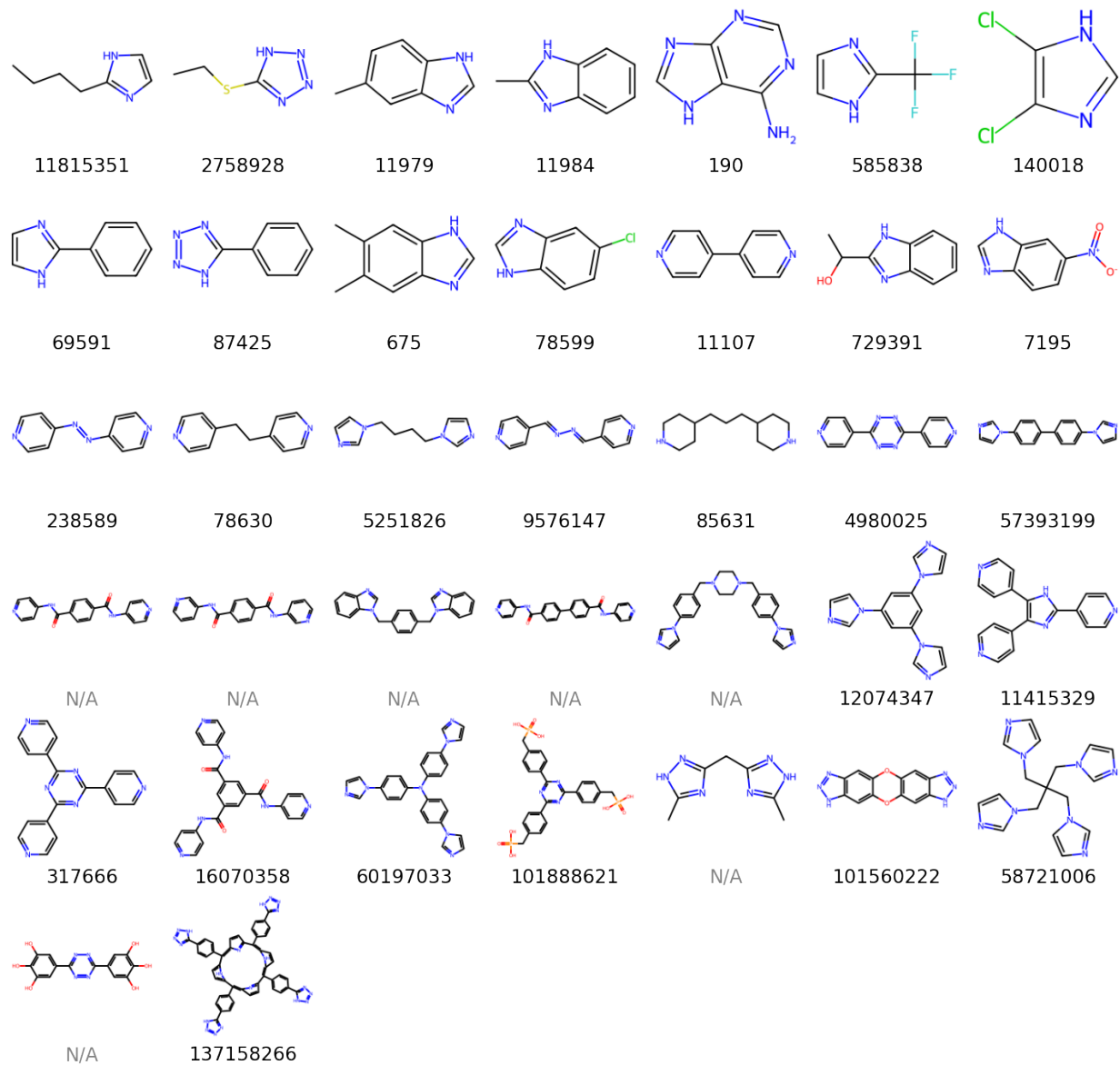

**Figure S54.** MOF linkers obtained from text mining and their CID numbers.

Linguistic Versatility in Synthesis Condition Summarization

We showcase ChatGPT's capability to process research papers written in languages other than English with minor modifications to the prompt. As evidenced, ChatGPT accurately identifies each synthesis parameter and tabulates them in English - a task that proves significantly challenging for traditional English-based NLP methods.[9-12]

**ChatGPT Processing Demo**

**Prompt:** Please provide a truthful response based on the given context. Translate and summarize the following details into an English table: compound name or chemical formula (if the name is not mentioned), metal source, metal quantity, organic linker(s), amount of linker, modulator, volume or quantity of modulator, solvent(s), volume of solvent(s), reaction temperature, and reaction duration. If any of the data is not provided or you are uncertain, please fill in "N/A".

Your attention should be on extracting only the experimental conditions from the MOF synthesis, disregarding information associated with the organic linker synthesis, MOF postsynthetic modifications or metalation, high throughput (HT) experiment details, or catalytic reactions.

If there are various conditions mentioned for the same compound, represent them in multiple rows. If there are multiple units or components provided for the same factor (such as g and mol for weight, multiple linkers or metals, varied temperature and reaction duration, mixed solvents, etc.), include them in a single cell, separating each by a comma.

The table should have 11 columns, all in lowercase:| compound name | metal source | metal amount | linker | linker amount | modulator | modulator amount or volume | solvent | solvent volume | reaction temperature | reaction time |

**Input:**
*Synthese: Die solvothermale Synthese von DUT-9 kann sowohl in DMFals auch in DEF durchgefhrt werden. Die Kristallstruktur von DUT-9 wurde anhand eines in DMF-Lösung gewachsenen Kristalls bestimmt. Weitere analytische Untersuchungen erfolgten mit Material, das aus Synthesen in DEF-Lösung stammt. In einer typischen Synthese werden 180 mg H3btb (0.41 mmol) und 366 mg Ni-(NO3)2·6H2O (1.26 mmol) in ein Pyrex-Röhrchen gegeben und in 10.5 mL DEF gelöst. Die Lösung wird 20 h auf 120°C erhitzt. Dieerhaltenen, hellgrünen Kristalle von DUT-9 werden durch Filtration unter Argon von der überstehenden Lösung getrennt, zweimal mit frischem DEF gewaschen und im Argonstrom bei Raumtemperatur getrocknet. Ausbeute: 412 mg (54% bezogen auf die eingesetzte Menge an H3btb).*

**Output:**

| compound name | metal source | metal amount | linker | linker amount | modulator | modulator amount or volume | solvent | solvent volume | reaction temperature | reaction time |
|---|---|---|---|---|---|---|---|---|---|---|
| DUT-9 | Ni(NO3)2·6H2O | 366 mg, 1.26 mmol | H3btb | 180 mg, 0.41 mmol | N/A | N/A | DEF | 10.5 mL | 120 °C | 20 hours |

**Figure S55.** Demonstration of ChatGPT's proficiency in interpreting and summarizing synthesis condition from German text.

> **ChatGPT Processing Demo**
>
> **Prompt:**
> ---Same prompt as shown in previous demo in Figure S55---
>
> ---
>
> **Input:**
>
> 单金属 Mn-MOF-74 的制备
>
> 单金属 Mn-MOF-74 的制备参考了已报道的文献[25]。在 250 mL 反应瓶中加入 MnCl2·4H2O(1 098 mg) 和 2,5-二羟基对苯二甲酸(DOBDC,333 mg),再加入 150 mL N,N-二甲基甲酰胺(DMF)-乙醇-水(体积比为 15:1:1) 混合溶剂,超声 20 min 得到澄清溶液后,置入已经预热至 135 ℃的烘箱中反应 24 h。从烘箱中取出自然冷却后,倾倒出反应母液,得到棕黑色晶体材料。晶体用 DMF 洗涤 3 次(3×20 mL) 后,再用甲醇进行溶剂置换 3 d(3×3×20 mL)。甲醇置换过后的 Mn-MOF-74 材料先于室温下真空脱溶剂 6 h,之后缓慢升温至 250 ℃真空条件下脱溶剂处理 12 h。
>
> ---
>
> **Output:**
>
> | compound name | metal source | metal amount | linker | linker amount | modulator | modulator amount or volume | solvent | solvent volume | reaction temperature | reaction time |
> |---|---|---|---|---|---|---|---|---|---|---|
> | Mn-MOF-74 | MnCl2·4H2O | 1,098 mg | 2,5-dihydroxyterephthalic acid | 333 mg | N/A | N/A | DMF-ethanol-water | 150 mL | 135 ℃ | 24 hours |

**Figure S56.** Demonstration of ChatGPT's proficiency in interpreting and summarizing synthesis condition from Chinese text.

> **ChatGPT Processing Demo**
>
> **Prompt:**
> ---Same prompt as shown in previous demo in Figure S55---
>
> ---
>
> **Input:**
> *MOF-5 の合成：テフロンで口を巻いた 100 mL のガラス瓶に 49 mL の DMF と 1 mL の水を加える。その混合液に Zn(NO3)2 · 6H2O (0.45 g, 1.5 mmol) とテレフタル酸 H2(bdc) (0.083 g, 0.50 mmol) を溶かす。ガラス瓶を閉め、それを 100 °C のオーブンで 7 時間加熱する。加熱後、ガラス瓶を室温に冷ます。これ以降の操作は、不活性雰囲気下のグローブテントで、無水溶媒を用いて行う。混合物の上澄み液をデカンテーションにより取り除き、新しい無水 DMF 50 mL を加えて、固体を 8 時間浸す。この DMF 洗浄を 3 回行ったら、同様に無水ジクロロメタン DCM による洗浄を 3 回行う。最後の DCM 洗浄の後、無色の結晶をろ過により回収し、150 °C で加熱真空引きすることで孔内の溶媒を除去する。*
>
> ---
>
> **Output:**
>
> | compound name | metal source | metal amount | linker | linker amount | modulator | modulator amount or volume | solvent | solvent volume | reaction temperature | reaction time |
> |---|---|---|---|---|---|---|---|---|---|---|
> | MOF-5 | Zn(NO3)2 · 6H2O | 0.45 g, 1.5 mmol | H2(bdc) | 0.083 g, 0.50 mmol | N/A | N/A | DMF, water | 49 mL, 1 mL | 100 °C | 7 hours |

**Figure S57.** Demonstration of ChatGPT's proficiency in interpreting and summarizing synthesis condition from Japanese text.

> **ChatGPT Processing Demo**
>
> **Prompt:**
>
> *---Same prompt as shown in previous demo in Figure S55---*
>
> ---
>
> **Input:**
> 2.1 금속 유기 골격체(MOFs):
> MIL-53(Al), MIL-53-NH2(Al) 합성
> MIL-53 의 합성은 Al(NO3)3·9H2O(1.3 g)와 H2BDC(0.288 g), 5 mL H2O 를 sonication 에 30 분 정도 놓아 섞은 후 autoclave 에 넣어, Box furnace 에서 분당 5 oC 씩 220 °C 까지 승온 한 다음 220 °C 에서 약 72 시간 반응을 진행 시켰다. 그 후 centrifugation 으로 수득 한 후 DI water 로 2 회, N,N-Dimethyl formamide(DMF)로 1 회 washing 작업을 진행하였다. 색이 하얀 페이스트가 얻어지면 DMF 25 ml 를 autoclave 에 넣어 분당 5 °C 씩 150 °C 까지 승온 후 150 oC 에서 약 15 시간 동안 H2BDC 를 충분히 제거 하였다. 이 과정을 통해 하얀색 색을 띄는 파우더 형태 의 MIL-53 을 얻게 되었다.
> MIL-53-NH2 의 합성은 AlCl3·6H2O(0.5 g)와 H2BDCNH2(0.38 g), 5 mL H2O 를 autoclave 에 넣어, Box furnace 에서 분당 5 °C 씩 150 °C 까지 승온 한 다음 그 온도에 서 약 5 시간 반응을 진행시킨다. 색이 노란 페이스트가 얻어지면 DMF 25 ml 를 autoclave 에 넣어 분당 5 °C 씩 150 °C 까지 승온 후 150 °C 에서 약 15 시간 동안 H2BDC 를 충분히 제거하였다. 추가 DMF 제거를 위해서 150 °C 12 시 간 동안 box furnace 에 놓아 MIL-53-NH2 를 얻게 되었다.
>
> ---
>
> **Output:**
>
> | compound name | metal source | metal amount | linker | linker amount | modulator | modulator amount or volume | solvent | solvent volume | reaction temperature | reaction time |
> |---|---|---|---|---|---|---|---|---|---|---|
> | MIL-53(Al) | Al(NO3)3·9H2O | 1.3 g | H2BDC | 0.288 g | N/A | N/A | H2O, DMF | 5 mL, 25 mL | 220 °C | 72 hours |
> | MIL-53-NH2(Al) | AlCl3·6H2O | 0.5 g | H2BDCNH2 | 0.38 g | N/A | N/A | H2O, DMF | 5 mL, 25 mL | 220 °C | 72 hours |

**Figure S58.** Demonstration of ChatGPT's proficiency in interpreting and summarizing synthesis condition from Korean text.

Multidimensional Physicochemical Information Summarization

While the primary focus of this paper is to automate the process of reading research papers, identifying and summarizing synthesis procedures using a three-tiered ChatGPT model, the versatility of the large language model extends to diverse tasks. The adaptable nature of ChatGPT enables the search for various information types, such as sorption behaviors, BET surface area, gas uptake, thermal stability, decomposition temperature, and even chemical stability. Additionally, it can extract structural information such as MOF topology and crystal parameters. These tasks would typically require a specialist to manually establish a training set or write specific patterns - an approach that lacks flexibility when the search target changes. By slightly modifying the structure of the search input and the summarization prompt, ChatGPT can efficiently accomplish these tasks.

Furthermore, we demonstrate the ability to search for specific information from the same pool of papers without the need to process all the text from the papers again, saving significant time. This is achieved by converting all papers into embeddings, which can easily be reloaded. As a demonstration, we design a prompt to search for the decomposition temperature obtained from TGA plots for the compounds reported in the papers, changing the search target from synthesis parameters to decomposition temperature.

> **Embedding Demo**
>
> **Prompt Embedding (ada-002):** Identify the section discussing thermogravimetric analysis (TGA) and thermal stability. This section typically includes information about weight-loss steps (e.g., 20%, 30%, 29.5%) and a decomposition temperature range (e.g., 450°C, 515°C) or a plateau.
>
> ---
>
> **Article Input #1:**
> *doi.org/10.1021/jacs.1c04946*
>
> ---
>
> **Search Output #1:**
> …Figure S24. TGA plot of the as-synthesized 3W-ROD-2-CH3, <177 °C, loss of free/surface water/ DMF; >380 °C, framework degradation…
>
> ---
>
> **Article Input #2:**
> *doi/10.1073/pnas.0602439103*
>
> ---
>
> **Search Output #2:**
> …The TGA trace for ZIF-8 showed a gradual weight-loss step of 28.3% (25–450°C), corresponding to partial loss of guest species [N,N- dimethylformamide (DMF) and 3H2O; calcd. 35.9%], followed by a plateau (450–550°C). More impressively, the TGA trace for ZIF-11 revealed a sharp weight-loss step of 22.8% (25–250°C), corresponding to the escape of all N,N- diethylformamide (DEF) solvent molecules trapped in the pores (0.9 DEF; calcd. 23.3%), despite the fact that DEF is actually much larger than the aperture of ZIF-11 in size. The TGA trace of ZIF-11 also showed a long plateau in the temperature range 250–550°C, indicating its high thermostability in the absence of guest molecules.…
>
> ---
>
> **Article Input #3:**
> *doi.org/10.1021/ja512973b*
>
> ---
>
> **Search Output #3:**
> … The results revealed that all the materials are stable at elevated temperature, having decomposition temperatures around 470 °C (see SI section S8.). The TGA of the solvated samples all show significant weight loss due to guest solvent molecules in the cavity prior to 200 °C, consistent with the highly porous nature of these materials. ……

**Figure S59.** An illustration of ChatGPT's adaptability, where the prompt is modified to target TGA data. The text parsed from the research article is not displayed for clarity, and the DOI of the paper is provided instead. The output displays one of the top similarity results, with the remainder omitted for brevity.

Once the embedding search yields a selection of paragraphs with embeddings exhibiting high similarity to the prompt's embedding, and these paragraphs potentially contain TGA information, we can adjust the prompt to facilitate summarization of this data.

---

**ChatGPT Processing Demo**

**Prompt:** Answer the question as truthfully as possible using the provided context. Please summarize the following details in a table: compound name and decomposition temperature. If any information is not provided or you are unsure, use "N/A".

The table should have 2 columns, all in lowercase:| compound name | decomposition temperature |

**Input:**

   ---*Full text from Search Output #1, #2 and #3 in previous Embedding Demo shown in Figure S59*---

**Output:**

| compound name | decomposition temperature |
|---|---|
| 3w-rod-2-oh | >375 °C |
| 3w-rod-1 | >375 °C |
| 3w-rod-2-f | >371 °C |
| 3w-rod-2-ch3 | >380 °C |
| zif-8 | 450-550°C |
| zif-11 | 250-550°C |
| nu-1101 | ~470°C |
| nu-1102 | ~470°C |

---

**Figure S60.** A designed prompt to guide ChatGPT in summarizing TGA decomposition temperatures, applying the three principles of chemistry prompt engineering. The figure includes a sample input and output table for context.

Similar adaptability can be exhibited by modifying the search and summary modules to extract BET surface area data. By appropriately adjusting the prompt and the summarization request, ChatGPT can effectively extract and summarize critical data from the literature.

> **Embedding Demo**
>
> **Prompt Embedding (ada-002):** Identify the section discussing nitrogen (N2) sorption, argon sorption, Brunauer-Emmett-Teller (BET) surface area, Langmuir surface area, and porosity. This section typically reports values such as 1000 m2/g, 100 cm3/g STP, and includes pore diameter or pore size expressed in units of Ångströms (Å).
>
> ---
>
> **Article Input #1:**
> *dx.doi.org/10.1021/ic301961q*
>
> ---
>
> **Search Output #1:**
> …The nitrogen sorption experiment clearly yields a type-I-isotherm, proving the microporosity of CAU-8 (Figure 9). The specific surface area according to the Brunauer–Emmett–Teller (BET)-method is S BET= 600 m2/g, and the observed micropore volume is VMIC= 0.23 cm3/g, calculated from the amount adsorbed at p/p0= 0.5. The maximum uptake of hydro- gen at 77 K and 1 bar is 1.04 wt %. …
>
> ---
>
> **Article Input #2:**
> *dx.doi.org/10.1021/acs.cgd.0c00258*
>
> ---
>
> **Search Output #2:**
> … permanent porosity of ZTIF-8 was confirmed by the reversible N2 sorption measurements at 77 K, which showed type I adsorption isotherm behavior (Figure 2 a). The Langmuir and BET surface areas were 1981 m2/g and 1430 m2/g for ZTIF-8, respectively. A single data point at relative pressure at 0.98 gives a pore volume of 0.705 for ZTIF-8 by the Horvath –Kawazoe equation. Furthermore, the single component sorption isotherms for CO 2were measured at 1 atm and 273 and 288 K ….
>
> ---
>
> **Article Input #3:**
> *dx.doi.org/10.1021/ja4050828 |*
>
> ---
>
> **Search Output #3:**
> … The N2 adsorption isotherm of NU-1000 is best described as type IVc (Figure 3); NU-1000 has a Brunauer–Emmett–Teller (BET) surface area of 2320 m2 g−1 and a total pore volume of 1.4 cm3 g−1. The experimentally measured surface area and total pore volumes are in excellent agreement with the theoretical values of 2280 m2 g−1 and 1.4 cm3 g−1 obtained from grand canonical Monte Carlo simulations (GCMC) and subsequent BET analysis (Table S3, SI)…

**Figure S61.** Illustration of ChatGPT's flexibility in changing the search target to sorption properties. The parsed text from the research article is not displayed for clarity, and the DOI of the paper is provided instead. The output displays one of the top similarity results, with the remainder omitted for brevity.

> **ChatGPT Processing Demo**
>
> **Prompt:** Answer the question as truthfully as possible using the provided context. Please summarize the following details in a table: compound name, pore volume, and BET surface area. If any information is not provided or you are unsure, use "N/A".
>
> The table should have 2 columns, all in lowercase:| compound name | pore volume | surface area |
>
> ---
>
> **Input:**
>
> ---Full text from Search Output #1, #2 and #3 in previous Embedding Demo shown in Figure S61---
>
> ---
>
> **Output:**
>
> | compound name | pore volume | surface area |
> |---|---|---|
> | CAU-8 | 0.23 cm3/g | 600 m2/g |
> | ZTIF-8 | 0.705 cm3/g | 1430 m2/g |
> | NU-1000 | 1.4 cm3/g | 2320 m2/g |

**Figure S62.** A prompt designed to instruct ChatGPT to summarize BET surface area and pore volume, with reference to the principles of chemistry prompt engineering. This includes an example input and output table for context, indicating that various prompt versions can be constructed based on specific research requirements.

**Section S5. Prediction Modeling**

To train our model, an initial step of data curation was undertaken. Rare cases in which the synthesis condition contained more than three metals or linkers were pruned and dropped. This was done to manage the complexity of the model, as one-hot encoding for such multi-component systems would introduce a large number of additional features, significantly increasing the model's dimensionality. Furthermore, instances with more than three metals or linkers were relatively rare and could act as outliers, potentially disturbing the learning process. After comparing the quality of the text-mined synthesis conditions by different processes, as shown in Figure 5c, we chose the results from Process 1 for training due to the fewest errors presented that could potentially impact the model. Consequently, data curation based on the output from Process 1 resulted in 764 samples that were used for model training.

Six sets of chemical descriptors were designed in alignment with the extracted synthesis parameters: these pertain to the metal node(s), linker(s), modulator(s), solvent(s), their respective molar ratios, and the reaction condition(s). The metal ions were described by several atomic and chemical properties, including valency, atomic radius[13], electron affinity[14], ionization potential, and electronegativity[15]. For the organic linkers, apart from Molecular Quantum Numbers (MQNs) that encode structural features in atomistic, molecular, and topological spaces,[16, 17] a set of empirical descriptors were also employed. These were based on counts of defined motifs such as carboxylate and phosphate groups (Figure S48−S52).

All solvents and modulators extracted were categorized into eight classes based on the recommendations from ChatGPT, each assigned a number from 1 to 8 and this assignment was made based on ranking the frequency of the compounds within each group (Table S2). These categories were represented by one-hot encodings. Molecular weights were also incorporated as descriptors for the linker(s), modulator(s), and solvent(s) sets. When multiple metals and organic linkers were present in the synthesis, the descriptors were calculated by taking a molar weighted average of the individual components. This approach was also employed to obtain the categorical encoders for multiple solvents and modulators used in combination. Here, the normalized molar fraction was entered into the cell where the corresponding solvent or modulator category was present, while all other entries were zero. In instances where solvents or modulators were absent in the synthesis parameters, arrays of zeros were used.

The RF models were trained using Scikit-Learn's *RandomForestClassifier* implementation for varying train size on 80% random split of the curated data. We used grid search to determine the optimal hyperparameters for our model, specifically the number of tree estimators and the minimum samples required for a leaf split. Model performance was evaluated using cross-validation and the metrics used for assessing the model's predictive power included class-weighted accuracy, precision, recall, and F1 score on the test set and the held out set. Feature permutation importance, quantified by the percent decrease in model accuracy by permutating one feature at a time, was used to identify which descriptors were the most influential in predicting the crystalline state outcome of a given synthesis condition.

**Table S2.** Classification of solvent and modulator groups.

| Solvent and Modulator Class | Assigned Number for Solvent Class | Compound Name |
|---|---|---|
| Acids | 8 | acetic anhydride; hydrofluoric acid; hydrochloric acid; tetrafluoroboric acid; formic acid; acetic acid; trifluoroacetic acid; benzoic acid; biphenyl-4-carboxylic acid; 4-nitrobenzoic acid; 2-fluorobenzoic acid; octanoic acid; nonanoic acid; phosphoric acid; nitric acid; sulfuric acid |
| Alcohols | 3 | methanol; ethanol; 1-propanol; 2-propanol; ethylene glycol; 2-amino-1-butanol; 3-amino-1-propanol; 1-butanol; 3-methylphenol; phenylmethanol |
| Amides, Sulfur-containing, and Cyclic Ethers | 1 | 1,4-dioxane; acetone; 1,3-dimethyl-2-imidazolidinone; 1-cyclohexyl-2-pyrrolidone; dimethylformamide; diethylformamide; 1-methyl-2-pyrrolidone; dimethyl sulfoxide; *N,N*-dimethylacetamide; *N*-methylformamide; tetrahydrofuran; 2-imidazolidinone |
| Amines and Ammonium Compounds | 6 | ammonia; methylamine; dimethylamine; triethylamine; tetrabutylammonium hydroxide; tetramethylammonium bromide; tetraethylammonium hydroxide; ammonium fluoride; 1-ethyl-3-methylimidazolium tetrafluoroborate; 1-ethyl-3-methylimidazolium chloride |
| Base | 7 | sodium hydroxide; sodium azide; lithium hydroxide; potassium hydroxide; sodium fluoride |
| Heterocyclic Compounds | 5 | 2-(1-hydroxyethyl)-1h-benzimidazole; 1,4-diazabicyclo[2.2.2]octane; 4,4′-bipyridine; pyrazine; piperazine; morpholine; pyridine; s-triazine; meso-tetra(n-methyl-4-pyridyl) porphine tetratosylate |
| Hydrocarbons and Derivatives | 4 | hexadecyltributylphosphonium bromide; benzene; toluene; chlorobenzene; p-xylene; acetonitrile; dichloromethane |
| Water and Derivatives | 2 | water; hydrogen peroxide |

Molar ratios were calculated from the total molar amount in the event of multiple species for each set. For the reaction conditions, four categories were identified: vapor diffusion, solvothermal, conventional, and microwave-assisted reaction. These were classified using ChatGPT (Figure S21). With regards to the crystalline state outcome, if the reaction results contained a description of (single) crystal(s), it was classified as the single-crystal (SC). If it included words like microcrystalline product, powder, solid, or no description of product morphology was given, it was classified as polycrystalline (P).

The full descriptor set include the following components: 'temperature', 'time', 'synthesis_method', 'metal_ionenergy', 'metal_affinity', 'metal_radii', 'metal_electronegativity', 'metal_valence', 'n_carboxylate', 'n_N_donnor', 'n_phosphate', 'n_chelating', 'linker_MW', 'solvent_MW', 'modulator_MW', 'linker_metal_ratio', 'solvent_metal_ratio', 'modulator_metal_ratio', 'linker_MQNs1', 'linker_MQNs2', 'linker_MQNs3', 'linker_MQNs4', 'linker_MQNs6', 'linker_MQNs7', 'linker_MQNs8', 'linker_MQNs9', 'linker_MQNs10', 'linker_MQNs11', 'linker_MQNs12', 'linker_MQNs13', 'linker_MQNs14', 'linker_MQNs15', 'linker_MQNs16', 'linker_MQNs17', 'linker_MQNs19', 'linker_MQNs20', 'linker_MQNs21', 'linker_MQNs22', 'linker_MQNs23', 'linker_MQNs24', 'linker_MQNs25', 'linker_MQNs26', 'linker_MQNs27', 'linker_MQNs28', 'linker_MQNs29', 'linker_MQNs30', 'linker_MQNs31', 'linker_MQNs32', 'linker_MQNs34', 'linker_MQNs35', 'linker_MQNs36', 'linker_MQNs37', 'linker_MQNs40', 'linker_MQNs41', 'linker_MQNs42', 'solvent_type1', 'solvent_type2', 'solvent_type3', 'solvent_type4', 'solvent_type5', 'solvent_type6', 'solvent_type7', 'solvent_type8', 'modulator_type1', 'modulator_type2', 'modulator_type3', 'modulator_type5', 'modulator_type6', 'modulator_type7', 'modulator_type8'. In order to extract the most relevant features and to reduce model complexity, a recursive feature elimination (REF) with 5-fold cross validation was performed to yield 26 descriptors from the initial 70 after the down-selection (Figure S61).

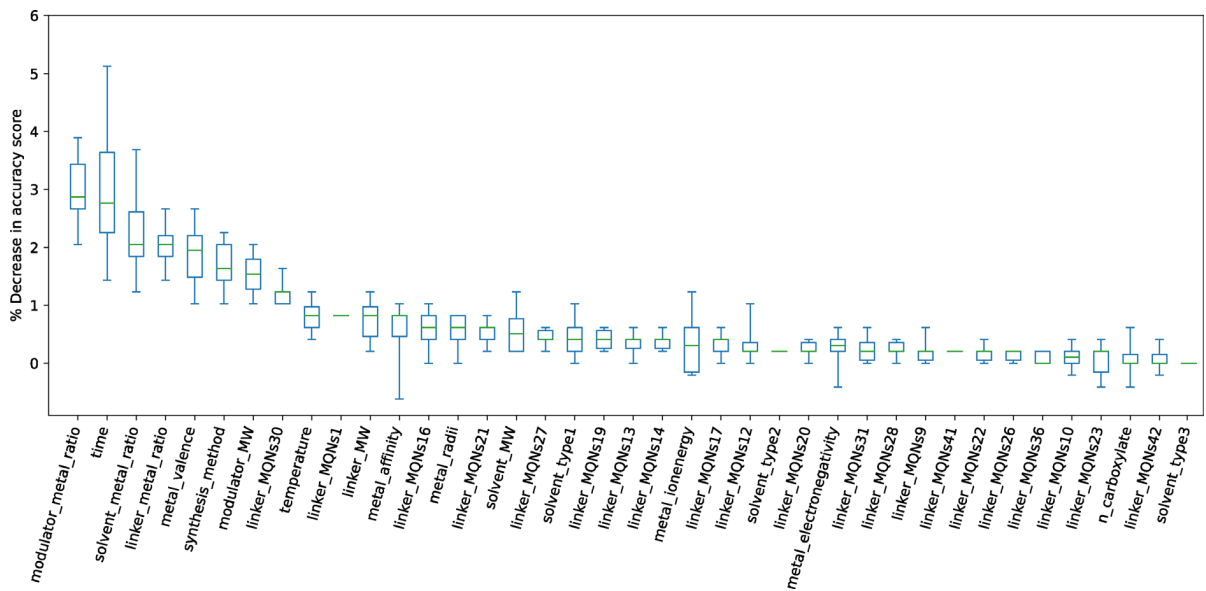

**Figure S63.** Percent decrease in accuracy by permutating features in descriptors set after REF over 10 runs. The boxes for each descriptor extend from the first to the third quartile, with a green line indicating the median. The whiskers span from the minimum to the maximum values of the data.

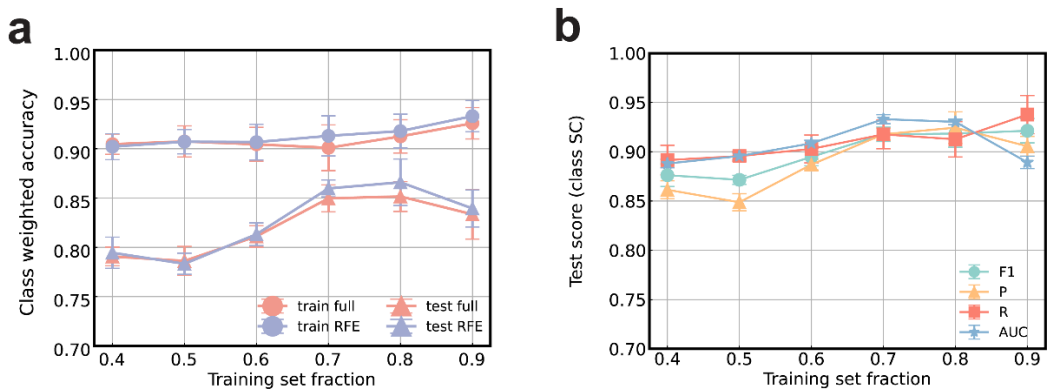

**Figure S64.** Performance of the classification models in predicting the crystalline state of MOFs from synthesis on the train and test set for varying training set ratio to the data excluding the held out set. (a) Learning curves of the classifier model with $1\sigma$ standard deviation error bars. (b) Model performance evaluation through the F1 Score, Precision, Recall, and Area Under the Curve metrics.

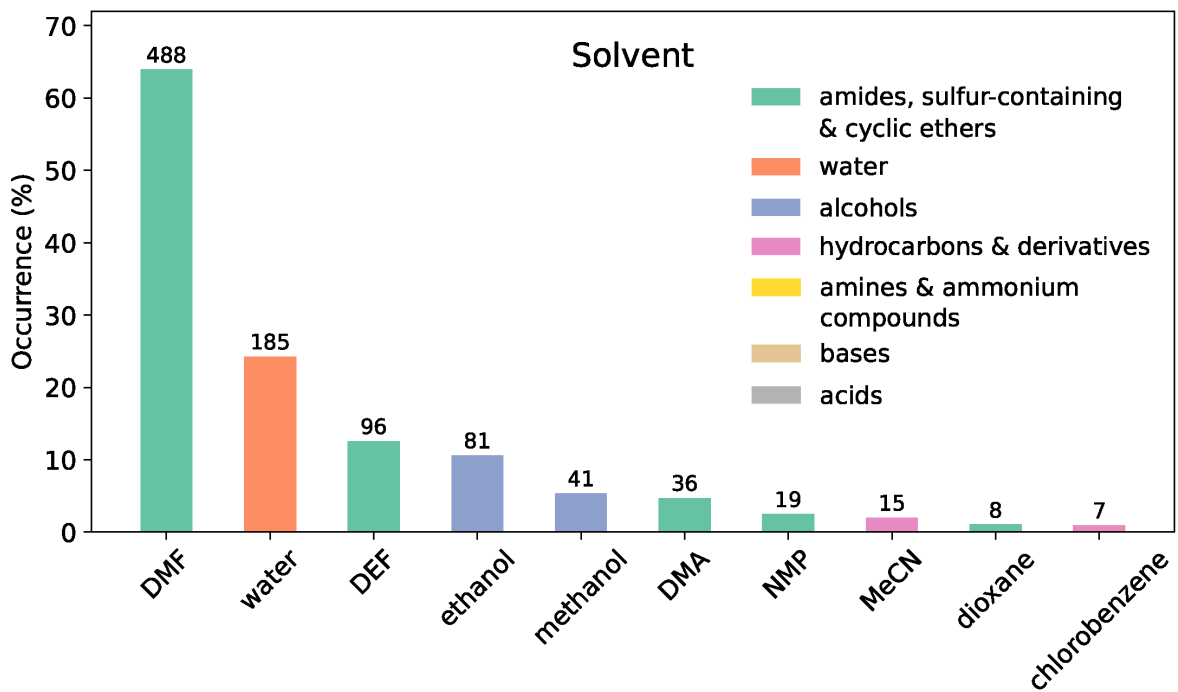

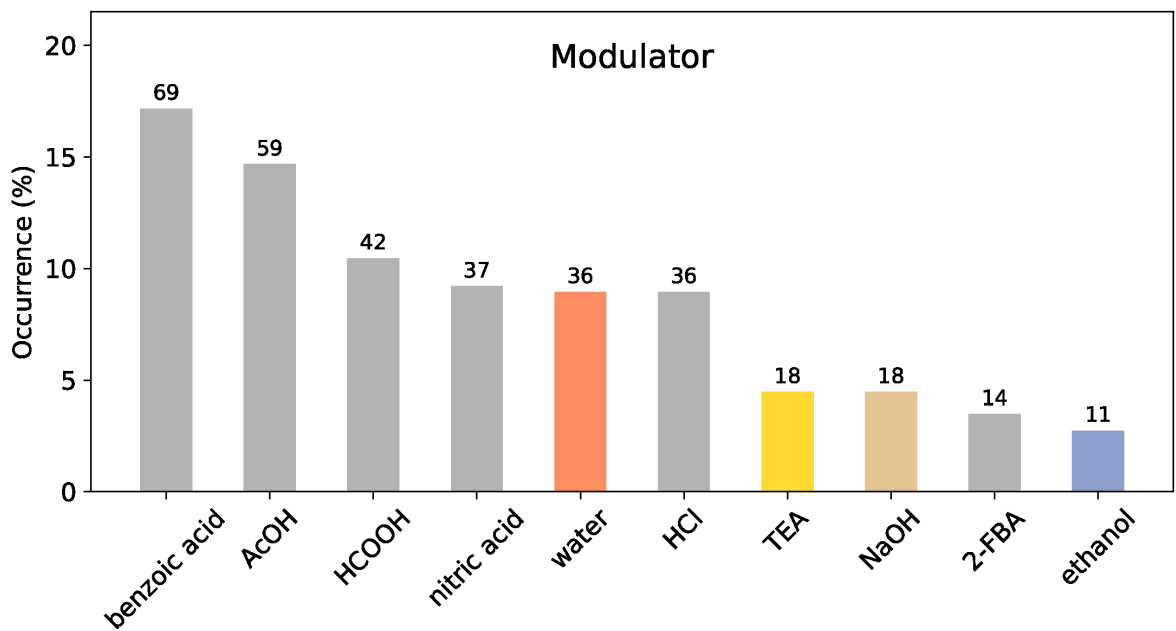

**Figure S65.** Frequency analysis of the synthesis condition dataset. In total, 35 unique solvent compounds and 44 unique modulator compounds were identified, and 10 most frequently occurring solvents and modulators from the extracted synthesis parameters were shown. Percent occurrence of solvents were calculated out of 763 experiments with solvent parameters; those of modulators were calculated out of 402 experiments with modulator parameters.

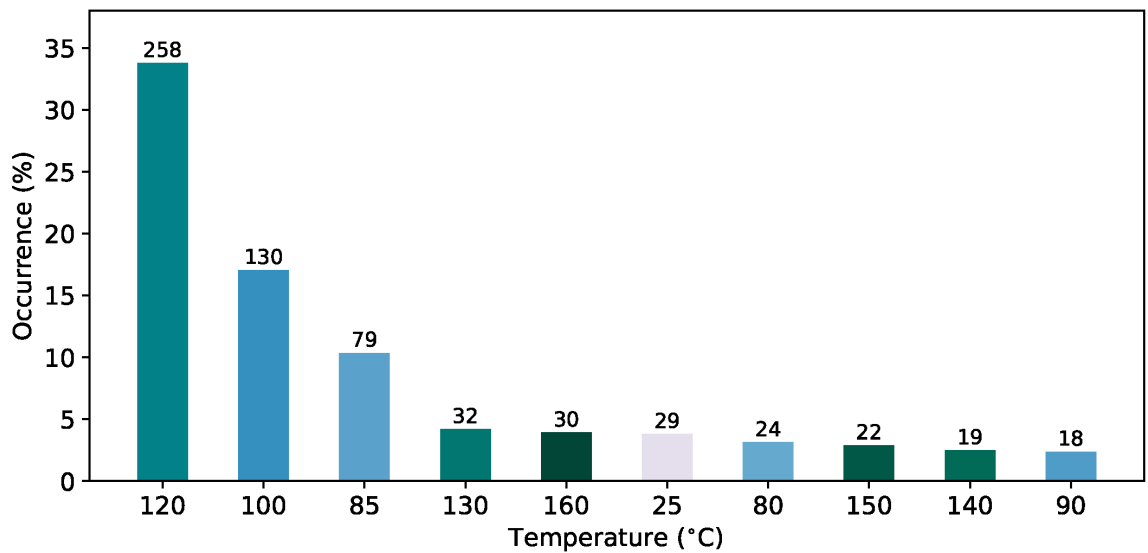

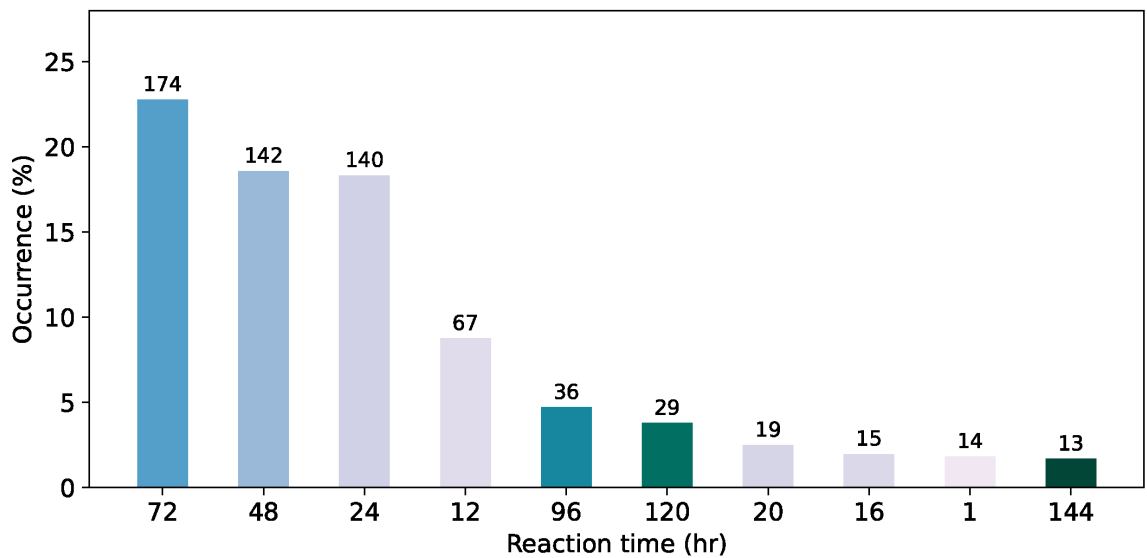

**Figure S66.** Frequency analysis of the synthesis condition dataset. 10 most frequently occurring reaction conditions, out of 30 unique reaction temperature and 48 unique reaction time, from the extracted synthesis parameters were shown.

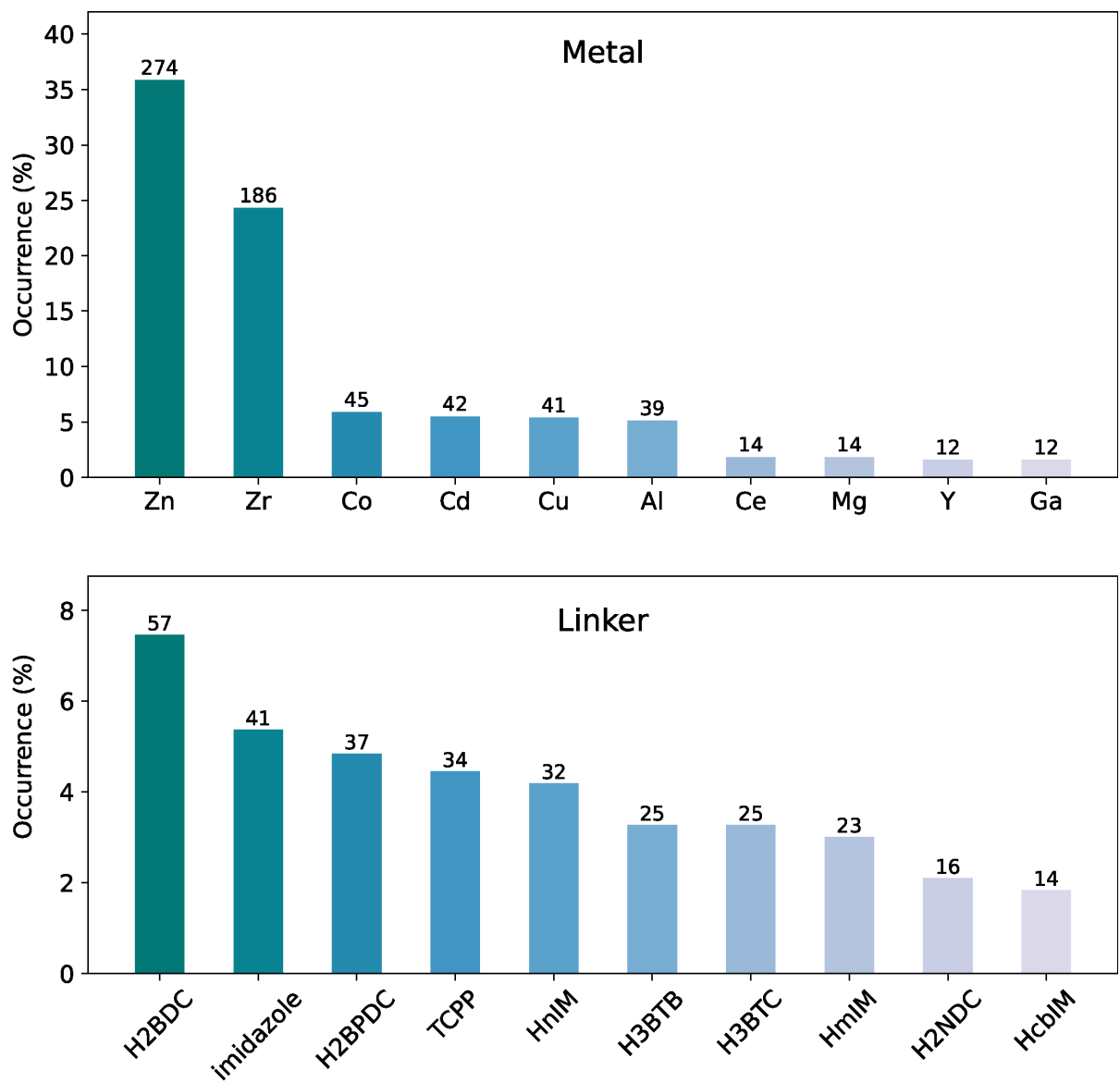

**Figure S67.** Frequency analysis of the synthesis condition dataset. 10 most frequently occurring metal elements and linkers, out of 29 unique metal and 263 unique linker compounds, from the extracted synthesis parameters are shown.

## Section S6. Dataset to Dialogue: The Creation of a MOF Synthesis Chatbot

To enable an automated chatbot drawing upon our dataset acquired from text mining, we initially reformatted the synthesis parameters for each compound into discrete paragraphs. For each paragraph, we also compiled a list of publication data where the compound was reported, such as authors, DOIs, and publication years, retrieved from Web of Science. This approach facilitated the creation of a synthesis and publication information card for each compound. Subsequently, we developed embeddings for the information cards, which form an embedded dataset (Table S3).

**Table S3.** Illustrative information card for MOFs and their respective embeddings

| Synthesis and Paper Information | Embeddings |
|---|---|
| MOF Name: MOF-808<br>Metal Source: ZrOCl2·8H2O<br>Metal Amount: 0.50 mmol<br>Linker: H3BTC (1,3,5-Benzenetricarboxylic acid, CAS number: 554-95-0)<br>Linker Amount: 0.50 mmol<br>Modulator: formic acid<br>Modulator Amount or Volume: 20 mL<br>Solvent: DMF<br>Solvent Volume: 20 mL<br>Reaction Temperature: 100°C<br>Reaction Time: 168 h<br>Reaction Equipment: 60 mL screw capped glass<br>Product Color or Shape: Octahedral colorless crystals<br>Paper DOI: 10.1021/ja500330a<br>Journal: J. Am. Chem. Soc.<br>Publication Year: 2014<br>Publication Date: MAR 19<br>Article Title: Water Adsorption in Porous Metal-Organic Frameworks and Related Materials<br>Author Names: Furukawa, Hiroyasu; Gandara, Felipe; Zhang, Yue-Biao; Jiang, Juncong; Queen, Wendy L.; Hudson, Matthew R.; Yaghi, Omar M. | [0.0009970446117222231,<br>-0.021761000156402588,<br>-0.025494899600744247,<br>-0.027127644047141075,<br>-0.006226510275155306,<br>0.04229075089097023,<br>...,<br>-0.03372553735971451] |
| MOF Name: ZIF-8<br>Metal Source: Zn(NO3)2·4H2O<br>Metal Amount: 0.210 g<br>Linker: H-MeIM (2-methylimidazole, CAS number: 693-98-1)<br>Linker Amount: 0.060 g<br>Modulator: N/A<br>Modulator Amount or Volume: N/A<br>Solvent: DMF<br>Solvent Volume: 18 mL<br>Reaction Temperature: 140°C<br>Reaction Time: 24 h<br>Reaction Equipment: 20-mL vial<br>Product Color or Shape: Colorless polyhedral crystals<br>Paper DOI: 10.1073/pnas.0602439103<br>Journal: Proc. Natl. Acad. Sci. U. S. A.<br>Publication Year: 2006<br>Publication Date: JUL 5<br>Article Title: Exceptional chemical and thermal stability of zeolitic imidazolate frameworks<br>Author Names: Park, Kyo Sung; Ni, Zheng; Cote, Adrien P.; Choi, Jae Yong; Huang, Rudan; Uribe-Romo, Fernando J.; Chae, Hee K.; O'Keeffe, Michael; Yaghi, Omar M. | [-0.0068720560520887375,<br>-0.02060604654252529,<br>-0.03643505275249481,<br>-0.017434848472476006,<br>-0.007826789282262325,<br>0.05133294314146042,<br>...,<br>-0.026907961815595627] |

The system is programmed to navigate to the embedding dataset and locate the most relevant sections based on a user's initial query. This procedure is based on calculating a similarity score between the question and the embeddings and mirrors the foundation of Text Mining Process 3. The highest-ranking entry's information is then dispatched to the ChatGPT Chemistry Assistant's prompt engineering module, which, through the ChatGPT API, crafts responses centered solely around the provided synthesis information. Depending on the user needs, the system can output multiple high similarity scores, such as the top 3 or top 5, provided this does not exceed the token budget (i.e., 4096 tokens for *gpt-3.5-turbo*).

---

**Embedding Demo**

**Query from User:** What is the linker used to synthesis MOF-520?

---

**Search Output:**
MOF Name: MOF-520
Metal Source: Al(NO3)3
Metal Amount: 0.04 mmol
Linker: H3BTB (Benzene-1,3,5-tris(4-benzoic acid), CAS number: 50446-44-1)
Linker Amount: 75 mg
Modulator: formic acid
Modulator Amount or Volume: 1.4 mL
Solvent: DMF
Solvent Volume: 17 mL
Reaction Temperature: 130°C
Reaction Time: 72 h
Reaction Equipment: 20 m L vial
Product Color or Shape: single crystals
Paper DOI: 10.1021/ja501606h
Journal: J. Am. Chem. Soc.
Publication Year: 2014
Publication Date: APR 9
Article Title: High Methane Storage Capacity in Aluminum Metal-Organic Frameworks
Author Names: Gandara, Felipe; Furukawa, Hiroyasu; Lee, Seungkyu; Yaghi, Omar M.

---

**Figure S68.** Illustration of embedding user's initial question to generate context, utilizing the information card of MOF-520 to respond to the query.

To establish a chatbot through the ChatGPT API, we followed a similar methodology to that employed in Processes 1, 2, and 3, specifically using ChatGPT to generate the code. The code takes an input prompt from the search output and a fixed prompt to ensure context-based responses. Furthermore, the function should enable access to prior conversations, maintaining a consistent context based on the synthesis information card. The figure below (Figure S69) displays a representative function that our ChatGPT Chemistry Assistant operates on.

**ChatGPT API**

```python
def chatbot(question, past_user_messages=None, initial_context=None):
    if past_user_messages is None:
        past_user_messages = []
    past_user_messages.append(question) # Store Synthesis and Paper Information Cards
    df_with_emb = pd.read_csv("xxx.csv")# Get Information Cards and Embeddings

    if initial_context is None:
        # Find the context based on the first question
        first_question = past_user_messages[0]
        question_return = openai.Embedding.create(model="text-embedding-ada-002", input=first_question)
        question_emb = question_return['data'][0]['embedding']

        df_with_emb_sim = add_similarity(df_with_emb, question_emb)
        num_paper = 3
        top_n_synthesis_str = top_similar_entries(df_with_emb_sim, num_paper)
        initial_context = top_n_synthesis_str

    message_history = [
        {
            "role": "system",
            "content": "You are a chemistry assistant that specifically handles questions related to MOF synthesis conditions based on the papers you have reviewed. Answer the question using the provided context. If the question is not relevant to the context or the MOF is not mentioned in the context, respond with 'Based on the information available from the MOF paper I have read so far, I cannot provide a reliable answer to this question. Please revise your question.'\n\nContext:\n" + initial_context
        },
    ]

    for user_question in past_user_messages:
        message_history.append({"role": "user", "content": user_question})

    response = openai.ChatCompletion.create(
        model='gpt-3.5-turbo',
        messages=message_history
    )
    answer = response.choices[0].message["content"]
    return answer, initial_context, past_user_messages

def add_similarity(df, given_embedding):
    def calculate_similarity(embedding):
        return cosine_similarity([embedding], [given_embedding])[0][0]
    df['similarity'] = df['embedding'].apply(calculate_similarity)
    return df
```

**Figure S69.** Python code snippet demonstrating the utilization of the ChatGPT API for establishing a literature-grounded chatbot.

We demonstrate the function's capabilities and its potential for building a robust chatbot application through several dialogues between a human user and the ChatGPT Chemistry Assistant, as depicted below. This foundational function could be integrated into an interactive website or mobile application, offering users real-time dialogues with the chatbot. As a user poses a question through the application interface, a POST request is triggered to the server, prompting the chatbot function with the user's query. The function then computes a pertinent response considering the conversation history, which is then relayed back to the user through the front-end.

**ChatGPT API**

```python
first_question = "Do you know MOF-808?"
answer, initial_context, past_user_messages = chatbot(first_question)
print(answer)
```

Yes, I am familiar with MOF-808.

```python
follow_up_question = "Tell me more."
answer, _, past_user_messages = chatbot(follow_up_question, past_user_messages,
initial_context)
print(answer)
```

MOF-808 is a metal-organic framework (MOF) that was first synthesized using zirconium as the metal source and 1,3,5-benzenetricarboxylic acid (H3BTC) as the organic linker. The synthesis was carried out using DMF as the solvent and formic acid as the modulator. The resulting MOF-808 was a white powder and was characterized using various techniques such as X-ray diffraction, thermogravimetric analysis, and nitrogen adsorption measurements. The MOF-808 has been used as a catalyst for selective methane oxidation to methanol.

```python
follow_up_question = "Can you give me the title or doi of the paper that reports MOF-808?"
answer, _, past_user_messages = chatbot(follow_up_question, past_user_messages,
initial_context)
print(answer)
```

Sure! The paper that reports MOF-808 is titled "Bioinspired Metal-Organic Framework Catalysts for Selective Methane Oxidation to Methanol" and has the DOI: 10.1021/jacs.8b11525.

**Figure S70.** Demonstration of a dialogue with the ChatGPT Chemistry Assistant using Python, showing the potential for conversion into an interactive web or mobile application.

Expanding upon this, we demonstrate the potential and versatility of the ChatGPT Chemistry Assistant through a series of queries that reflect its ability to make a dataset more accessible, in addition to responding effectively to synthesis inquiries.

In the first example (Figure S71), the user sought to understand the difference between two MOFs, ZIF-8 and ZTIF-8. The assistant detailed their distinct characteristics, which includes the metal sources, solvents, co-linkers, and synthesis conditions. When the user expressed interest in synthesizing ZIF-8, the assistant underscored the importance of lab safety and suggested familiarizing oneself with essential lab techniques before attempting the MOF synthesis. In response to the user's request for a detailed synthesis procedure for ZIF-8, the assistant provided a step-by-step process while highlighting necessary safety precautions.

In the second example (Figure S72), the assistant showcased its ability to provide information on the synthesis of Zirconium MOFs. It shared two examples of synthesis procedures and explained the concept and role of a modulator. The assistant further elaborated that the outcome might differ if the modulator (acetic acid, in this case) was not included, leading to uncontrolled crystal growth and possibly reducing the yield. When the user inquired about alternative modulators, the assistant explained that the selection of modulators depends on the desired MOF structure and properties, and their choice could significantly impact the outcome.

The third example demonstrated how the assistant can effectively handle misinformation (Figure S73). Initially, the user asked about synthesizing MOF-499, a compound the assistant couldn't find information on. However, when the user corrected the typo to MOF-419, the assistant provided a detailed synthesis procedure. The user also sought to understand the post-synthesis steps, and the assistant mentioned common processes such as activation, doping, and functionalization, though specific steps for MOF-419 were not detailed in the reviewed paper.

The fourth example highlighted the assistant's capability to provide detailed information about the synthesis of single crystals of MOF-801 (Figure S74). The assistant confirmed that single crystals of MOF-801 could be made and explained the significance of obtaining single crystals for accurate structural characterization. It also explained the potential applications of MOF-801 and the difference in synthesis procedures for single crystal and microcrystalline powder forms. When the user asked about other MOFs reported in the paper, the assistant provided information on the reported zirconium MOFs.

In each of these cases, the assistant has shown its utility as an effective tool for helping researchers navigate the complex world of MOF synthesis. By providing contextually aware, focused responses, the assistant makes it easier for researchers to access relevant data and learn more about MOF synthesis, serving as an invaluable tool in the field of chemistry.

> **Human-MOF Chatbot Interaction**
>
> **Human:** What is the difference between ZIF-8 and ZTIF-8?
>
> ---
>
> **MOF Chatbot:**
> ZIF-8 and ZTIF-8 are different metal-organic frameworks. Although both MOFs share a similar ligand, which is 2-methylimidazole, ZIF-8 is synthesized using Zn(NO3)2·4H2O as a metal source and DMF as a solvent, while ZTIF-8 is synthesized using Zn(CH3COO)2·2H2O as a metal source and DMF and EtOH as solvents. Additionally, ZTIF-8 uses 5-methyltetrazole as a co-linker, which is not present in ZIF-8. ZIF-8 is also synthesized at a lower temperature for a longer time compared to ZTIF-8. Their different synthesis conditions lead to different crystal structures and properties.
>
> ---
>
> **Human:** I have never conducted any MOF synthesis before, and I want to make ZIF-8. How should I prepare for it?
>
> ---
>
> **MOF Chatbot:**
> Making MOFs requires knowledge and experience in handling chemicals and laboratory equipment. It also involves following strict safety protocols to avoid any accidents or mishaps. Before attempting to make ZIF-8, it is recommended that you first gain experience in basic laboratory techniques and familiarize yourself with the necessary equipment and safety measures. Additionally, it is essential to carefully read and understand the synthesis procedure from a reliable source, such as a peer-reviewed scientific journal article. It is not advised to attempt to make MOFs without proper knowledge, training, and safety precautions.
>
> ---
>
> **Human:** Could you please provide me with a detailed synthesis procedure for this MOF, highlighting the parts that require caution?
>
> ---
>
> **MOF Chatbot:**
> Sure! Here's a detailed synthesis procedure for ZIF-8:
> 1. Dissolve Zn(NO3)2·6H2O in DMF (20 mL) to create a solution.
> 2. Add 2-methylimidazole (0.5 g) to the solution.
> 3. Mix the solution thoroughly.
> 4. Transfer the solution to a Teflon-lined autoclave (23 mL).
> 5. Seal the autoclave and heat it at 85°C for 24 hours.
> 6. Allow the autoclave to cool to room temperature.
> 7. Remove the resulting white precipitate by centrifugation or filtration.
> 8. Wash the precipitate with solvents such as DMF to remove any impurities.
> 9. Dry the resulting white powder at 60°C under vacuum for 12 hours.
>
> Caution:
> - DMF is toxic and should not be inhaled or ingested. Handle it in a well-ventilated area with protective equipment such as gloves and goggles.
> - The autoclave should be sealed tightly to prevent leakage or explosion during heating.
> - The resulting white powder should be handled with care to avoid inhaling or exposing it to moisture, as it can easily absorb water vapor from the air and decompose.

**Figure S71.** A conversation with the MOF Chatbot based on *gpt-3.5-turbo* model.

> **Human-MOF Chatbot Interaction**
>
> **Human:** How to make Zirconium MOF?
>
> ---
>
> **MOF Chatbot:**
> There are different ways to make Zirconium MOFs, and the specific method may depend on various factors such as the linker and modulator used. However, based on the information available from the papers I have reviewed, I can provide two examples:
>
> 1. Zr-CAU-30: Zr-CAU-30 was synthesized using Ni-H8TPPP (linker) and NaF/NaOH (modulators) in H2O solvent at 160°C for 3 hours under stirring. ZrOCl2·8H2O (metal source) and Ni-H8TPPP were mixed in H2O solvent, and then NaF and NaOH were added to the mixture. The mixture was heated and stirred until the reaction was completed. The product was obtained as a highly porous metal phosphonate.
>
> 2. Zr-CAU-28: Zr-CAU-28 was synthesized using H2FDC (linker) and acetic acid (modulator) in H2O solvent at 80°C for 1 hour under microwave-assisted heating. ZrOCl2·8H2O (metal source) and H2FDC were mixed in H2O solvent, and then acetic acid was added to the mixture. The mixture was heated under microwave irradiation until the reaction was completed. The product was obtained as a white crystalline powder.
>
> It's important to note that these are just two examples of Zirconium MOF synthesis, and the specific conditions may vary depending on the desired MOF structure and properties.
>
> ---
>
> **Human:** Could you explain what a modulator is, and if I choose not to add acetic acid during the synthesis, how might it impact the outcome of my results?
>
> ---
>
> **MOF Chatbot:**
> In MOF synthesis, a modulator is a molecule that can promote or inhibit crystal growth and control the morphology of the MOF crystals. In the synthesis of Zr-CAU-28 as described in the paper, acetic acid was used as a modulator. The addition of acetic acid likely helped to increase the yield of the desired MOF product, as well as promote the formation of smaller and more uniform crystals.
>
> If you choose not to add acetic acid during the synthesis of Zr-CAU-28, the outcome of your results may differ. Without a modulator, the crystal growth may be less controlled, leading to the formation of larger or less uniform crystals. Additionally, the yield of the desired MOF product may be lower. It is important to note, however, that the impact of omitting acetic acid may depend on the specific MOF synthesis conditions and may not always have a significant effect on the final product.
>
> ---
>
> **Human:** That's interesting. In the case of synthesizing other MOFs, have you come across any alternative modulators that could potentially be used in place of acetic acid?
>
> ---
>
> **MOF Chatbot:**
> While the paper did not mention any other modulators used in the synthesis of Zr-CAU-28, other papers have used different modulators for synthesizing Zr-based MOFs. For example, in the synthesis of Zr-CAU-30, the modulators used were NaF and NaOH. However, the choice of modulator may impact the outcome of the results, so it is recommended to refer to other papers and perform some preliminary experiments to determine the most suitable modulator for the specific synthesis protocol and desired MOF properties.

**Figure S72.** A conversation with the MOF Chatbot based on *gpt-3.5-turbo* model.

> **Human-MOF Chatbot Interaction**
>
> **Human:** Can you tell me how to synthesize MOF-499?
>
> **MOF Chatbot:**
> Based on the information available from the MOF papers I have read so far, I cannot provide a reliable answer to this question. Please provide more information, or the specific paper reference, on the synthesis of MOF-499.
>
> **Human:** Oops, my bad. There was a typo. I actually meant MOF-419.
>
> **MOF Chatbot:**
> To synthesize MOF-419, you will need to use Y(NO3)3 (0.5 mmol) as the metal source, H2BDH (benzene-1,4-dihydroxamic acid, CAS number: 20073-80-7) (0.5 mmol) as the linker, and formic acid (100 mmol) as the modulator. The reaction should be carried out in H2O (12 mL) solvent in a 20 mL scintillation vial at 100°C for 12 hours. The product should be obtained as single crystals. If you have any further questions, feel free to ask..
>
> **Human:** Thanks for the explanation. But, how about the post-synthesis steps? What should I do after I get the MOF crystals?
>
> **MOF Chatbot:**
> Unfortunately, the paper I have reviewed did not provide information regarding any post-synthesis steps. However, depending on the intended application of the MOF, various post-synthesis modifications are possible including activation, doping, functionalization, and more. Activating the MOF typically involves removing any solvent or guests trapped within the pores of the MOF through solvent exchange or thermal treatment. Doping and functionalization can be achieved through various methods including deposition, exchange, and covalent attachment. The exact post-synthesis steps required for MOF-419 may depend on the intended application and should be further researched.
>
> **Human:** Could you provide me with the exact title of the paper that discusses MOF-419 so that I can have a check by myself?
>
> **MOF Chatbot:**
> The paper discussing MOF-419 is titled "Metal-Organic Frameworks with Rod Yttrium Secondary Building Units" and was published in the Isr. J. Chem. journal.
>
> **Human:** Thank you. Can you give me a link to get access to the paper?
>
> **MOF Chatbot:**
> Unfortunately, I cannot provide a specific link to access the paper as I do not have information on the journal's access policies. However, you can try searching for the paper by its title "Metal-Organic Frameworks with Rod Yttrium Secondary Building Units" in a search engine or a scientific paper database. Some databases that you can use are Google Scholar, PubMed, and ResearchGate.

**Figure S73.** A conversation with the MOF Chatbot based on *gpt-3.5-turbo* model.

> **Human-MOF Chatbot Interaction**
>
> **Human:** Is it possible to make single crystals of MOF-801?
>
> **MOF Chatbot:** Yes, it is possible to make single crystals of MOF-801. In fact, the paper "Water Adsorption in Porous Metal-Organic Frameworks and Related Materials" (DOI: 10.1021/ja500330a) reports the synthesis and characterization of MOF-801 in its single crystal form (MOF-801-SC). The synthesis was performed using ZrOCl2·8H2O as the metal source, fumaric acid as the linker, formic acid as the modulator, DMF as the solvent, and a reaction temperature of 120°C for 24 h in a 60 mL screw capped glass vessel. The product obtained was octahedral colorless single crystals.
>
> **Human:** What is the application of this MOF and why it is important to make it single crystal? Can I make powdered MOF-801?
>
> **MOF Chatbot:** MOF-801 has potential applications in water adsorption, dehumidification, and delivery of drinking water in remote areas. It is important to make single crystals of MOF-801 because they allow for more accurate structural characterization, which is important for understanding the properties and behavior of the material. Powdered MOF-801 can also be made and studied for its properties, but single crystals provide more detailed information. Therefore, both single crystal and powdered forms of MOF-801 have their own importance and applications.
>
> **Human:** How do they differ in terms of synthesis procedure?
>
> **MOF Chatbot:** Based on the information available from the MOF paper I have read so far, MOF-801 can be synthesized as both a single crystal and a microcrystalline powder form. However, the synthesis procedures are different for each form of MOF-801. The synthesis procedure for MOF-801 single crystal involves the use of ZrOCl2·8H2O as metal source, fumaric acid as linker, formic acid as modulator, DMF as solvent, and a 60 mL screw capped glass for 24 hours at 120°C. On the other hand, MOF-801 microcrystalline powder is synthesized using the same metal source and linker but with different amounts of modulator (benzoic acid) and DMF as solvent.
>
> **Human:** What other MOFs does this paper also report?
>
> **MOF Chatbot:** The paper reports on the water adsorption properties of 23 materials, 20 of which are metal-organic frameworks (MOFs). Among the MOFs are 10 zirconium(IV) MOFs, with the following reported for the first time: MOF-801-SC (single crystal form), MOF-802, MOF-805, MOF-806, MOF-808, MOF-812, and MOF-841. MOF-801-P (microcrystalline powder form) was reported earlier and studied for its water adsorption properties. MOF-812 was only made and structurally characterized but not examined for water adsorption because it is a byproduct of MOF-841 synthesis.
>
> **Human:** Thanks. Last question, where can I find this paper?
>
> **MOF Chatbot:** You can find the paper on the ACS Publications website using the Digital Object Identifier (DOI) associated with each paper. Here are the DOIs for the papers mentioned in the context:
>
> - MOF-801-SC paper: 10.1021/ja500330a
>
> Alternatively, you can search for the paper on the ACS Publications website using the journal title I mentioned earlier.

**Figure S74.** A conversation with the MOF Chatbot based on *gpt-3.5-turbo* model.